\documentclass[useAMS,usenatbib,usegraphicx]{mn2e}
\usepackage{journals}
\usepackage{color}
\usepackage{longtable}
\usepackage{supertabular}
\usepackage{colortbl}
\usepackage{rotating}
\usepackage{lscape}

\begin{document}
\def\apj {ApJ}
\def\aj {AJ}
\def\mnras {MNRAS}

\title[2MASS CG catalogue]{Compact Groups of Galaxies selected by stellar mass:\\ The
  2MASS Compact Group Catalogue} 
\author[D\'iaz-Gim\'enez et al.]{Eugenia D\'iaz-Gim\'enez$^{1,2,3}$, 
Gary A. Mamon$^4$, 
Marcela Pacheco$^{1}$, 
\and Claudia Mendes de Oliveira$^3$ \& M. Victoria Alonso$^{1,2}$ \\
$1$ Instituto de Astronom\'{\i}a Te\'orica y Experimental, IATE, CONICET, Argentina\\
$2$ Observatorio Astron\'omico, Universidad Nacional de C\'ordoba, Laprida 854,
X5000BGR, C\'ordoba, Argentina\\
$3$ Instituto de Astronomia, Geofisica e Ciencias Atmosfericas, IAG, USP. Rua do Mat\~ao 1226, S\~ao Paulo, Brazil\\ 
$4$ Institut d'Astrophysique de Paris (UMR 7095: CNRS \& UPMC), 98 bis Bd Arago, F--75014 Paris, France 
}
\date{\today} \maketitle

%
\begin{abstract}
We present a photometric 
catalogue of compact groups of galaxies (\emph{p2MCG}s) automatically extracted from the 
2MASS extended source catalogue.
A total of $262$ \emph{p2MCG}s are identified, 
following the criteria defined by \cite{Hickson82}, of which 230 survive
visual inspection
(given occasional galaxy fragmentation and blends in the 2MASS parent catalogue).
Only one quarter of these 230 groups were previously known compact groups (CGs).
Among the 144 \emph{p2MCG}s that have all their
galaxies with  
known redshifts, 85 (59\%) have 4 or more accordant galaxies. 
This \emph{v2MCG} sample of velocity-filtered \emph{p2MCG}s constitutes
the largest sample of CGs (with $N\geq4$) catalogued to date, 
with both well-defined selection criteria and velocity filtering, 
and is the first CG sample selected by stellar mass.
It is fairly complete up to $K_{\rm group} \sim 9$ and 
radial velocity of $\sim 6000 \, \rm km \, s^{-1}$.

We compared the properties of the 78 \emph{v2MCG}s with median velocities
greater than $3000 \, \rm km \, s^{-1}$ with the properties of other CG
samples, as well as those (\emph{mvCG}s) extracted from the
semi-analytical model (SAM) of Guo et al. (2011) run on the high-resolution
Millennium-II simulation. 
This \emph{mvCG} sample is
similar (i.e. with 2/3 of physically dense CGs) to those we had previously
extracted on  three other SAMs run on the 
Millennium simulation with 125 times worse spatial and mass resolutions.
The space density of \emph{v2MCG}s within $6000 \, \rm km \, s^{-1}$ is 
$8.0 \times 10^{-5} \,h^3 \,\rm Mpc^{-3}$, i.e. 4 times that
of the Hickson sample (HCG)
up to the same distance and with the same criteria used in this work, but
still 40\% less than that of \emph{mvCG}s.

The \emph{v2MCG} constitutes the first 
group catalogue to show a statistically large first-second ranked 
galaxy magnitude difference (in units of the dispersion of the first-ranked
absolute magnitudes) according to Tremaine-Richstone statistics,
as expected if the first ranked group members tend to be the products of galaxy
mergers, and as confirmed in the \emph{mvCG}s.
The \emph{v2MCG} is also the first observed sample
to show that first-ranked galaxies tend to be
centrally located, again consistent with the predictions obtained from
\emph{mvCG}s. 
We found no significant correlation of group 
apparent elongation and velocity
dispersion in the quartets among the 
\emph{v2MCG}s, and the velocity dispersions of apparently
round quartets are not significantly larger than those of chain-like ones,
 in contrast to what has been previously reported
in HCGs. 

By virtue of its automatic selection with the popular Hickson criteria, its
size, its selection on stellar mass, and its 
statistical 
signs of mergers and centrally located brightest galaxies, 
the \emph{v2MCG} catalogue appears to 
be the laboratory of choice 
to study physically dense groups of 4 or more galaxies of
comparable luminosity.

\end{abstract}

\section{Introduction}

Compact Groups (hereafter, CGs) of at least 4 galaxies of comparable
luminosity are the densest galaxy associations known at present.
The compactness of these groups is so high that the typical projected separations 
between galaxies are of the order of 
their own diameters \citep{HMdOHP92,FK02}, hence their space densities can exceed 
those of the cores of rich clusters. 
The combination of their very high number densities and low velocity
dispersion makes CGs the ideal site of galaxy mergers (\citealp{Mamon92}, see
also \citealp{CCS81,Barnes85,Mamon87,BCL93}).

Since the discovery of Stephan's Quintet \citep{Stephan1877} and Seyfert's
Sextet \citep{Seyfert48}, several surveys of CGs have been undertaken:
\cite{Rose77} and \cite{Hickson82} performed visual identifications of 
CGs on the POSS I photographic plates. 
Thereafter, the new catalogues of CGs used automatic searches: 
from the COSMOS/UKST Southern Galaxy Catalogue \citep{PIM94,Iovino02}, 
the DPOSS catalogue \citep{Iovino03,decarvalho05}, 
and the Sloan Digital Sky Survey (SDSS) photometric catalogue DR1 \citep{Lee+04} and 
DR6 \citep{McCPES09}.
All of the above studies used only 2-dimensional information of the galaxies 
(i.e., angular positions). 
Other CG catalogues were obtained by searches in redshift space,
e.g.: 
\cite{Barton+96} from the the CfA2 catalogue, 
\cite{AT00} from the Las Campanas Redshift Survey,
\cite{FK02} from the  UZC Galaxy Catalogue,
and \cite{Deng+08} from the SDSS-DR6 spectroscopic catalogue.

Since the nearly full spectroscopic followup by \cite{HMdOHP92} of
the original Hickson Compact Groups (\citealp{Hickson82}, hereafter, HCGs), the velocity-filtered
sample of 92 HCGs with at least 3 accordant-redshift members and 69 with at
least 4 has been, by far, the most studied to date
(e.g. \citealp{HKA89} for optical photometry;
\citealp{MdOH91} for galaxy morphologies;
\citealp{Moles+94}, \citealp{delarosa07} and \citealp{tzanavaris10} for star formation rates;
\citealp{CRdCC98} for nuclear activity;
\citealp{delarosa01} and \citealp{torres-flores10} for galaxy scaling relations;
\citealp{VerdesMontenegro+01} and \citealp{borthakur10} for neutral gas content;
\citealp{PBEB96} for hot gas content, etc.).

However, the visual inspection performed by \cite{Hickson82} led to a sample of CGs that is
not reproducible, incomplete and not homogeneous \citep{HKA89,WM89,PIM94,Sulentic97,diaz-mamon10}. 
In particular, using the $z=0$ outputs of semi-analytical models of galaxy
formation run on the Millennium cosmological dark matter simulation \citep{Springel+05}, 
\cite{diaz-mamon10} have shown that the HCG sample is 
typically less than 10\% complete at the median distance of the sample.

The properties of CGs and their member galaxies must be studied using complete and well-defined
observed samples. To achieve this goal, 
we present a new sample of automatically selected CGs extracted from the
largest solid angle 
catalogue at present, 
the 2 Micron All Sky Survey. Using 2MASS has two strong advantages: 1) it
provides us with a full-sky survey and 2) the $K$-band photometry is only weakly
sensitive to both galactic extinction, internal extinction and recent star
formation, and is thus a very good tracer of the stellar mass content of galaxies.
For these reasons, it is ideal to build a CG sample from a wide $K$-band galaxy 
survey such as 2MASS (which has the additional benefit of being all-sky) 
with (nearly) full redshift information available from other sources \citep{Mamon94}.

The layout of this paper is as follows. In Sect.~\ref{2mass}, 
we describe the parent catalogue. In Sect.~\ref{2mcgc}, we present the CG catalogue. 
We perform a cross-identification between the 2MASS-CGs and other samples of groups in 
Sect.~\ref{cross-id}. 
In Sect.~\ref{v-filter}, we present a sample of CG after applying a velocity filtering, 
while we present some general properties of the samples in Sect.~\ref{props}, and summarise 
and discuss our results in Sect.~\ref{conclude}.

Throughout this paper we use a Hubble constant $H_0=
100 \, h \, {\rm km \,s^{-1}\, Mpc^{-1}}$, 
and for all cosmology-dependent calculations, 
we assume a flat cosmological model with 
a non-vanishing cosmological constant: $\Omega_{\rm m}=0.25$ and $\Omega_\Lambda=0.75$.
 
\section{The parent catalogue: 2MASS XSC}
\label{2mass}

\begin{figure*}
{\includegraphics[width=\hsize]{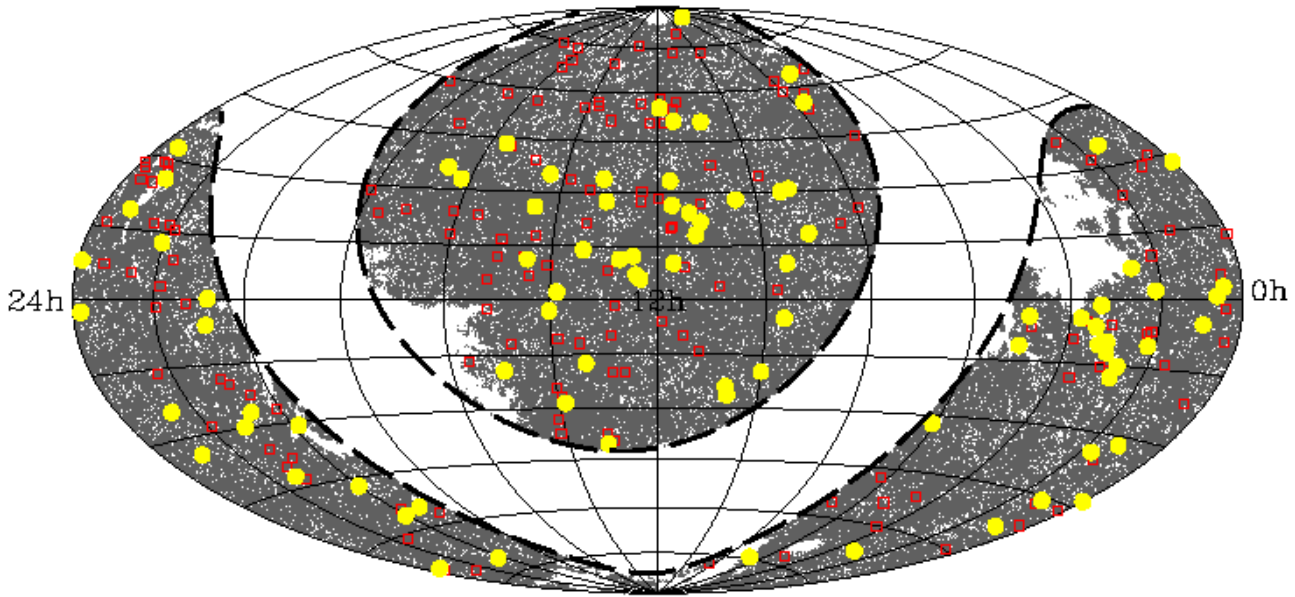}}
\caption{Aitoff projection of galaxies in the 2MASS XSC excluding 
the region around the Galactic Plane (dashed lines) and regions with high galactic extinction (\emph{background points}).
\emph{Open squares} represent the 230 compact groups identified in projection, while \emph{filled circles} are the 85 compact groups after the velocity filter.
\label{aitoff}
}
\end{figure*}

The 2 Micron All Sky Survey (2MASS) \citep{skrutskie06}
 has uniformly scanned the entire sky in 
three near-infrared bands to detect and characterise point sources brighter 
than about 1 mJy in each band, with S/N greater than 10.
2MASS used two highly-automated 1.3-m 
telescopes, one at Mt. Hopkins, AZ, and one at CTIO, Chile. Each telescope was 
equipped with a three-channel camera, each channel consisting of a NICMOS3
256$\times$256 HgCdTe array, capable of imaging a 
8$^\prime$.5 $\times$ 8$^\prime$.5 field at a pixel scale of 2$^{\prime\prime}$ per pixel in the $J$ (1.25 microns), $H$ (1.65 microns), and $K_s$ (2.17 microns) bands.

Our data set was selected from the publicly available full-sky extended
source catalogue (XSC;
\citealp{jarrett00})\footnote{http://irsa.ipac.caltech.edu/cgi-bin/Gator/nph-dd?catalog=fp\_xsc},
which contains over $1.6$ million extended objects brighter than
$K_s=14.3$. We adopted the ``K20 isophotal fiducial elliptical aperture
magnitudes'' and selected galaxies not flagged as artifacts ({\tt cc\_flg != 'a'}) nor close to large
galaxies --- thus avoiding spurious fragments in the envelopes of large
galaxies ({\tt cc\_flg != 'z'}).
There is a strong correlation between dust extinction
and stellar density, which increases exponentially towards the Galactic
Plane.  Stellar density is a contaminant factor of the XSC since the
reliability of separating stars from extended sources is very sensitive to
this quantity \citep{jarrett00}. In order to avoid contamination from stars,
we have constructed a mask for the 2MASS survey using the HEALPix
\citep{gorski05} map with $N_{\rm side}=256$ and excluding those pixels where
the $K_s$-band extinction $A(K_s)=0.367 \, E(B$--$V) > 0.05$ and $|b|<20$,
which reduces galactic contaminant sources to 2\% \citep{maller05}.  
This filtering on galactic extinction reduced the solid angle from
$27\,334\,\rm deg^2$ to $23\,844 \, {\rm deg}^2$.

The raw
magnitudes were corrected for galactic extinction using the reddening map of
\cite{schlegel98}.  We also followed \citeauthor{maller05} and imposed a cut
at $K_{\rm lim}^{\rm 2MASS} =13.57$ in the corrected magnitudes.  
The sky distribution of these galaxies is shown as 
the grey points in Fig.~\ref{aitoff}.
These
restrictions produced a sample of $408\,618$ extended sources which
constitute our parent catalogue.  
\begin{figure}
\centering
{\includegraphics[scale=0.8]{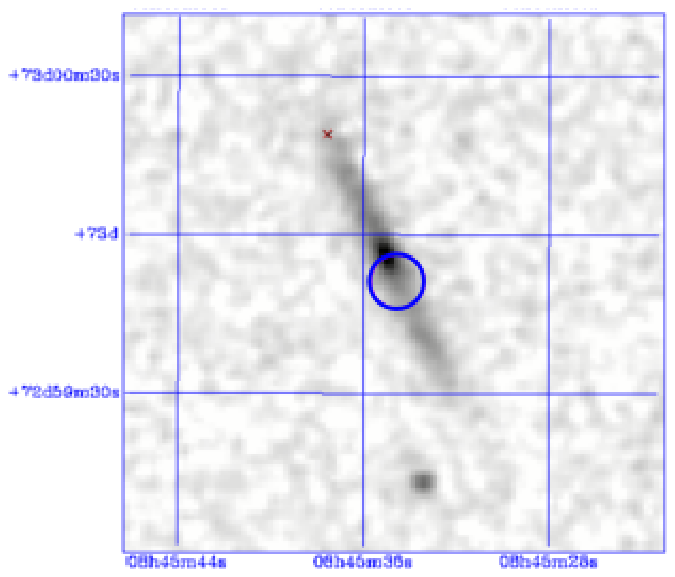}
\includegraphics[scale=0.8]{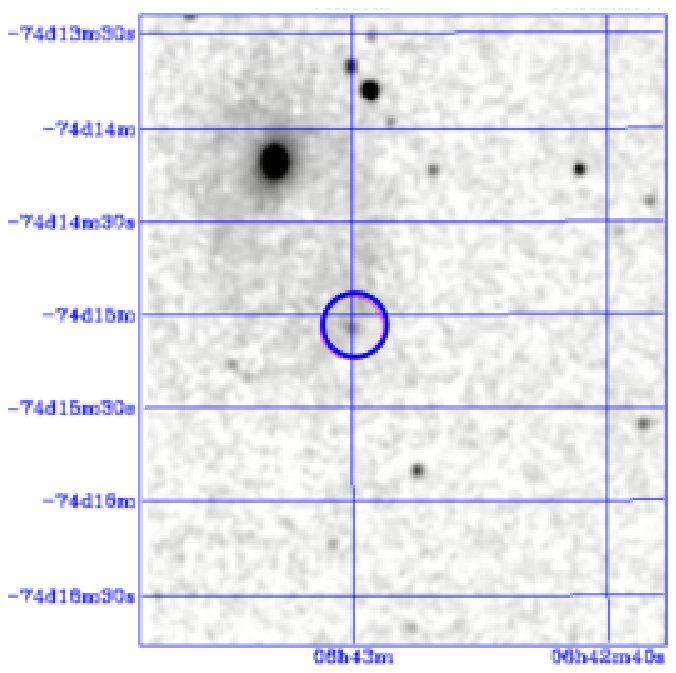}}
\includegraphics[scale=0.8]{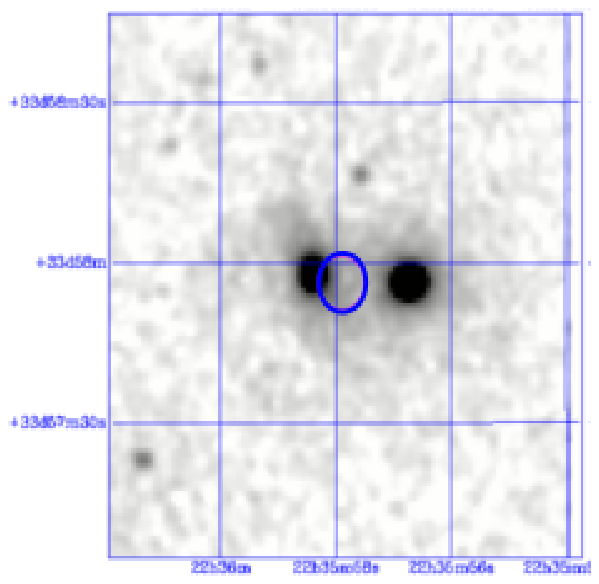}

\caption{Images in the $K_s$-band taken from the Interactive 2MASS Image
  Service showing 2MASS misidentification examples (see
  Table~\ref{2MASS_pofG}). Circles show the position of these objects in the
  2MASS XSC.  From top to bottom: 2MASXJ08453453+7259512,
  2MASXJ06430003-7415042, 
2MASXJ22355791+3357562.
In all but the last image, the large galaxy close to the circles also belongs to 2MASS XSC.
\label{pofG}}
\end{figure}

\section{The 2MASS CG catalogue}
\label{2mcgc}
We identify CGs in projection (\emph{p2MCG}s) by using an automated searching 
algorithm very similar to that defined by \cite{Hickson82} which is 
fully described in \cite{diaz-mamon10}. Briefly, this algorithm 
identifies as \emph{p2MCG}s those systems that satisfy the following criteria:
\begin{itemize}
 \item $4 \le N \le 10$ (population)
\item $\mu_K \le 23.6 \ \rm mag \ arcsec^{-2} $ (compactness)
\item $\theta_{\cal N} > 3\theta_G$ (isolation)
\item $K_{\rm brightest} \le K_{\rm lim}^{\rm 2MASS}-3=10.57$ (flux limit)
\end{itemize}
where 
$N$ is the total number of galaxies whose $K$-band magnitude satisfies 
$K < K_{\rm brightest}+3$, and $K_{\rm brightest}$ is the apparent 
magnitude of the brightest galaxy of the group; 
$\mu_K$ is the mean $K$-band surface brightness, averaged 
over the smallest circle circumscribing the galaxy centres; $\theta_G$ is the 
angular diameter of the smallest circumscribed circle, and $\theta_{\cal N}$ 
is the angular diameter of the largest concentric circle that contains no 
other galaxies within the considered magnitude range or brighter. 
Our compactness criterion is set to match that of the HCG, using a mean
colour transformation of $K=R-2.4$ (see appendix~\ref{mu_k}).

In order to speed up this computationally extensive algorithm, we used the 
subroutines of the HEALPix\footnote{http://healpix.jpl.nasa.gov} package to find 
neighbours within 5 degrees around each galaxy, and the 
STRIPACK\footnote{http://people.sc.fsu.edu/burkardt/f\_src/stripack/stripack.html} 
subroutines to compute the centres  and radii of the
minimum enclosing circles (hereafter CG centres and CG radii, respectively). 
 
Using this algorithm, we found $262$ 
\emph{p2MCG}s in the 2MASS XSC,
containing 1158 
galaxies.
We note, as a curiosity, that $3\pm0.5\%$ (binomial errors) 
of our compact groups with $N>4$ contain a 
compact quartet core that also meets all the CG criteria.
These are, in fact, CGs within CGs. 
Note that this percentage is significantly
lower than the $(6-13)\%$ predicted by \cite{diaz-mamon10} 
from the semi-analytical models (with binomial uncertainty less than 0.5\%).
Following \citeauthor{diaz-mamon10}, we always choose the larger CG.

Using the Aladin interactive sky atlas\footnote{http://aladin.u-strasbg.fr/java/nph-aladin.pl} \citep{aladin00} 
and the Interactive 2MASS image
server\footnote{http://irsa.ipac.caltech.edu/applications/2MASS/IM/inter\-active.html}, 
we performed a visual inspection of all of these \emph{p2MCG}s. We found that 
there were 26 
galaxy misidentifications in the 2MASS XSC: fragments of larger galaxies
(often HII regions) or blends of two galaxies.
In other words, 
since 26 galaxies are misidentifications over a total of 1158, then,
for our purposes,
the 2MASS XSC turned out to be 97.8$\pm$0.4\% reliable.
Figure~\ref{pofG} shows a few examples of these misidentifications.
In Table~\ref{2MASS_pofG}, we list the 26 
objects that belonging to CGs were
incorrectly classified as galaxies by 2MASS, and also are quoted the names
 of their host galaxies. We discarded those CGs of 4 members 
that hosted one of these galaxies. If a misidentified galaxy belonged to 
a CG with more than 4 members, then only this galaxy is discarded, 
and all properties of the CG are recomputed and all the criteria are checked again.
In total, 20 
groups were discarded because of incorrect 2MASS galaxy identifications.

Moreover, 2MASS fails to identify some large galaxies that are close to another 
large galaxy belonging to a CG. 
For instance, galaxy NGC 7578A 
does not appear in the 2MASS XSC, while its pair-neighbour, 
NGC 7578B, 
does. 
The same happened with the following 12 galaxies: 
NGC0414-NED02, 
IC0590-NED02, 
NGC~3750,  
NGC~4783,
NGC~5354,
NGC~4796,
IC~1165~NED02, 
ESO~284-IG~041~NED02, 
ESO596-49, 
LCRSB210329.4-450104, 
NGC~7318A, 
NGC~7318B.
These 13 missing galaxies among 1158 detected ones make the 2MASS XSC  99\%
complete for our purposes.
Given the lack of $K$-band magnitudes for these galaxies, 
we omitted from our sample the 12 CGs containing these 13 galaxies.
 
\begin{table}
\begin{center}
\tabcolsep 3pt
\caption{Objects in the 2MASS XSC that are actually part of larger galaxies\label{2MASS_pofG}}
\begin{tabular}{rlc}
\hline
\hline
\# & Galaxy Name in 2MASS & Main galaxy \\
\hline
1 & 2MASXJ18533628$-$5643133 & 2MASXJ18533694$-$5643078\\
2 & 2MASXJ14080439$-$3318147 & 2MASXJ14080314$-$3318542\\
3 &  2MASXJ22355791+3357562 & galaxy pair \\
4 & 2MASXJ03554380$-$4222233 & 2MASXJ03554474$-$4222024\\
5 & 2MASXJ07271181+8544540 & 2MASXJ07271448+8545162\\
6 & 2MASXJ08453453+7259512 & 2MASXJ08453501+7259560\\
7 & 2MASXJ03582336$-$4428024 & 2MASXJ03582180$-$4427585\\
8 & 2MASXJ10421741$-$0022318 & 2MASXJ10421797$-$0022365\\
9 & 2MASXJ12422507$-$0702456 & 2MASXJ12422554$-$0702364\\
10 & 2MASXJ16013973+2121296 & 2MASXJ16014023+2121106\\
11 & 2MASXJ12040147+2013489 & 2MASXJ12040140+2013559\\
12 & 2MASXJ07222530+4916277 & 2MASXJ07222519+4916427\\ 
13 & 2MASXJ06430003$-$7415042 & 2MASXJ06430596$-$7414103\\
14 & 2MASXJ00364578+2134078 & 2MASXJ00364500+2133594\\
15 & 2MASXJ17465074+2045440 & 2MASXJ17465132+2045400\\
16 & 2MASXJ09054355+1820276 & 2MASXJ09054305+1820226\\
17 & 2MASXJ11282505+0924272 & 2MASXJ11282405+0924279\\
18 & 2MASXJ23223215+1153235 & 2MASXJ23223093+1153332\\
19 & 2MASXJ02142411$-$0722178 & 2MASXJ02142586$-$0722064\\
20 & 2MASXJ12214093+1129448 & 2MASXJ12214230+1130118\\
21 & 2MASXJ13193834$-$1242052 & 2MASXJ13193805$-$1241562\\
22 & 2MASXJ12494210+2653266 & 2MASXJ12494226+2653312\\
23 & 2MASXJ13561035+0514388 & 2MASXJ13560724+0515169\\
24 & 2MASXJ23535429+0757368 & 2MASXJ23535389+0758138\\
25 & 2MASXJ11561045+6031300 & 2MASXJ11561032+6031211\\
26 & 2MASXJ15375266+5923382 & 2MASXJ15375345+5923304\\ 
\hline
\end{tabular}
\end{center}
\end{table}

\begin{figure*}
\centering
{\includegraphics[scale=0.66]{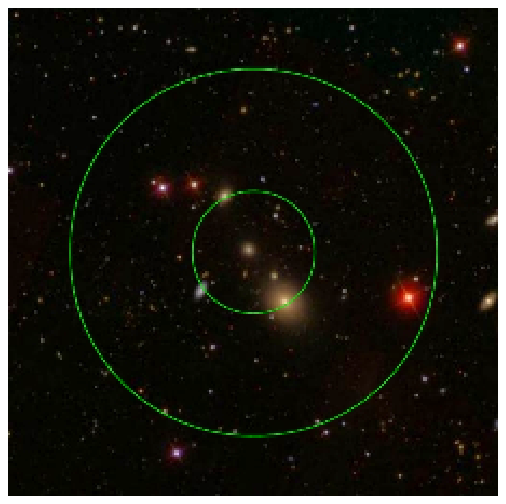}
\includegraphics[scale=0.66]{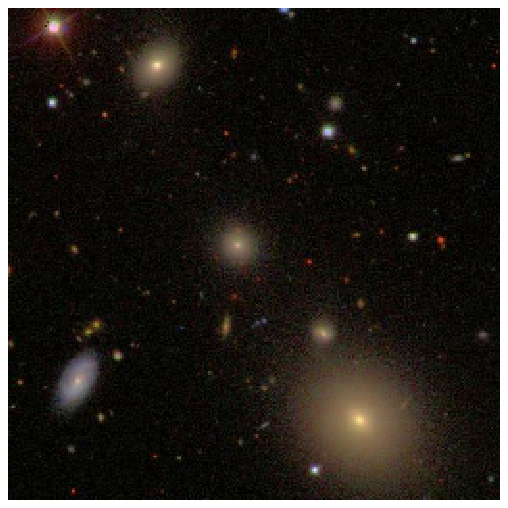}
\hskip 0.3cm
\includegraphics[scale=0.66]{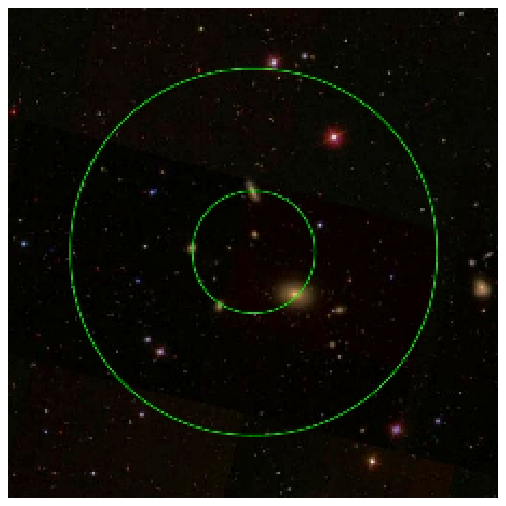}
\includegraphics[scale=0.66]{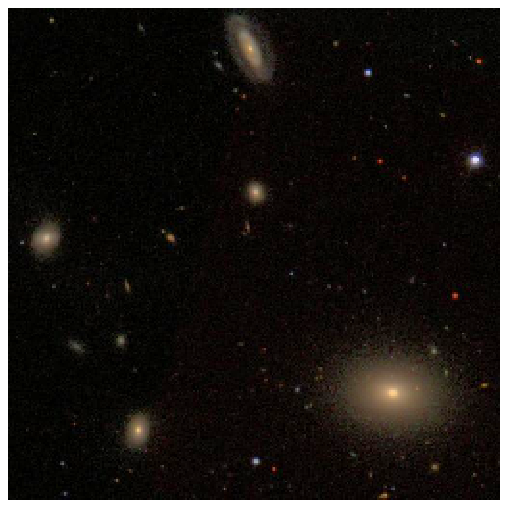}}
\vskip 0.3cm

{\includegraphics[scale=0.66]{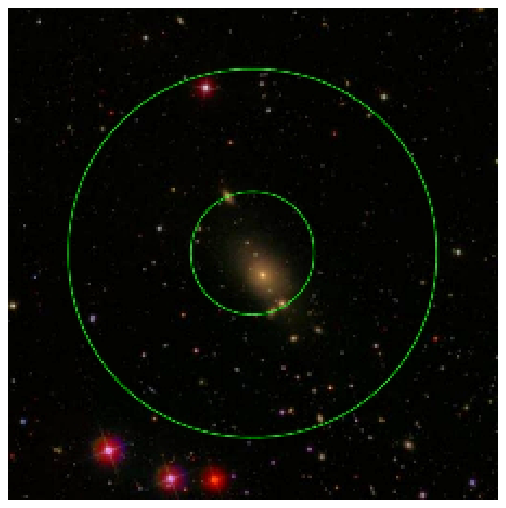}
\includegraphics[scale=0.66]{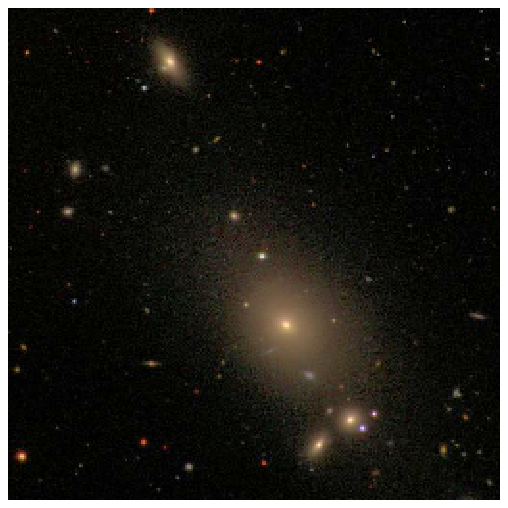}
\hskip 0.3cm
\includegraphics[scale=0.66]{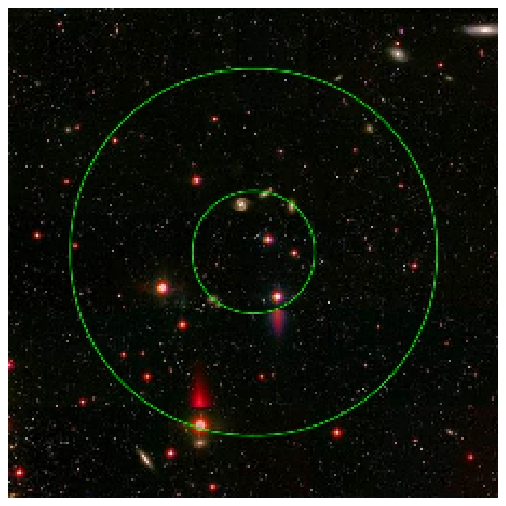}
\includegraphics[scale=0.66]{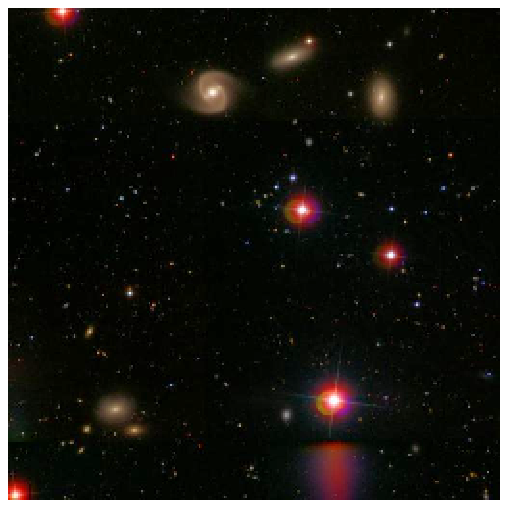}}
\vskip 0.3cm

{\includegraphics[scale=0.66]{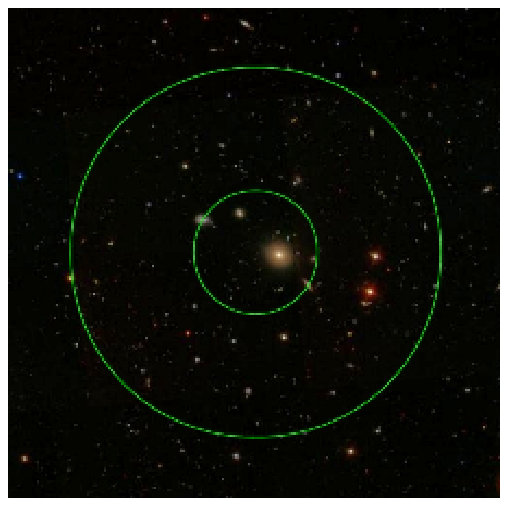}
\includegraphics[scale=0.66]{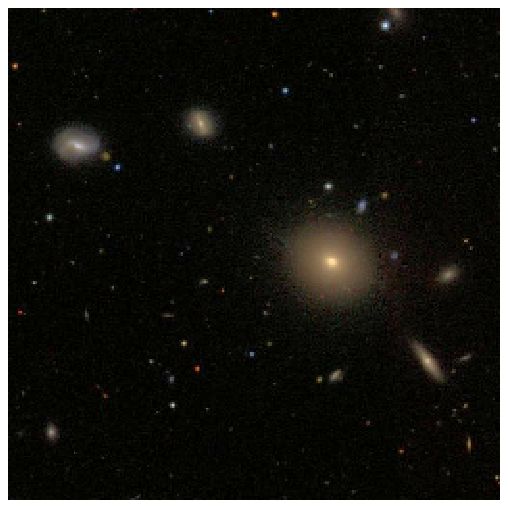}
\hskip 0.3cm
\includegraphics[scale=0.66]{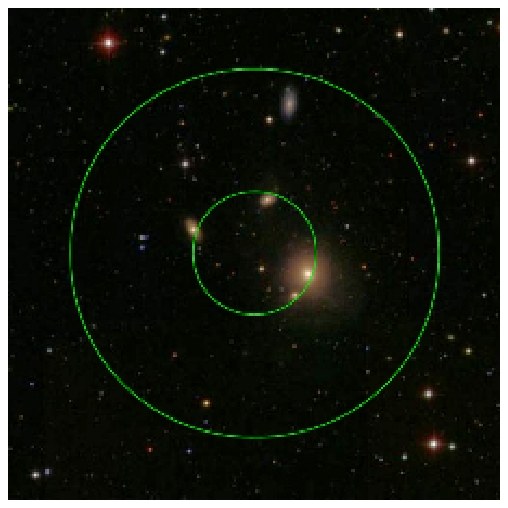}
\includegraphics[scale=0.66]{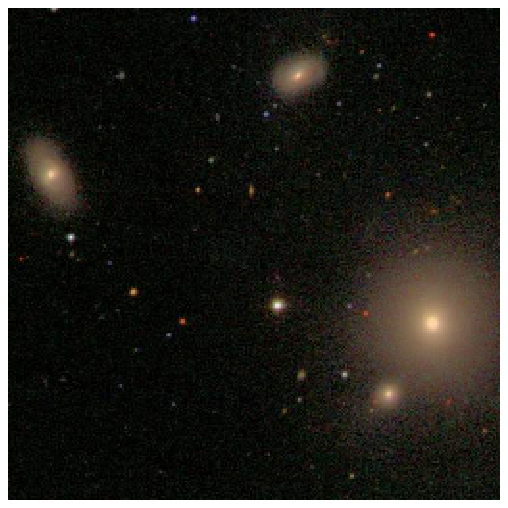}}
\vskip 0.3cm

{\includegraphics[scale=0.66]{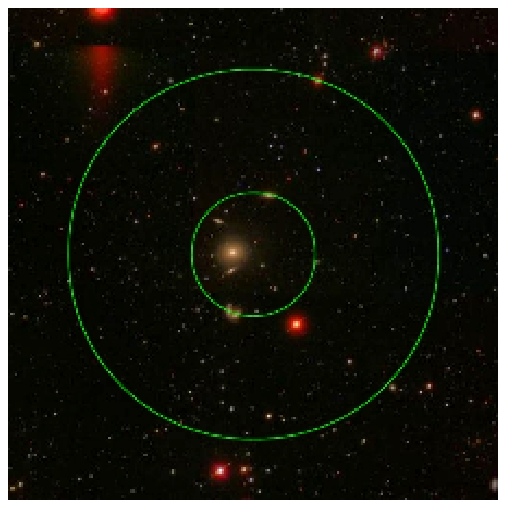}
\includegraphics[scale=0.66]{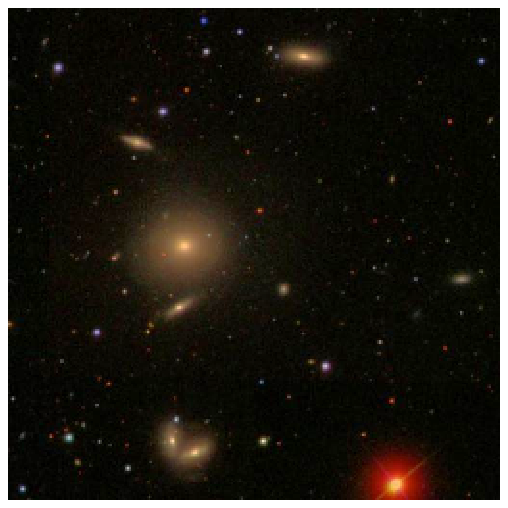}
\hskip 0.3cm
\includegraphics[scale=0.66]{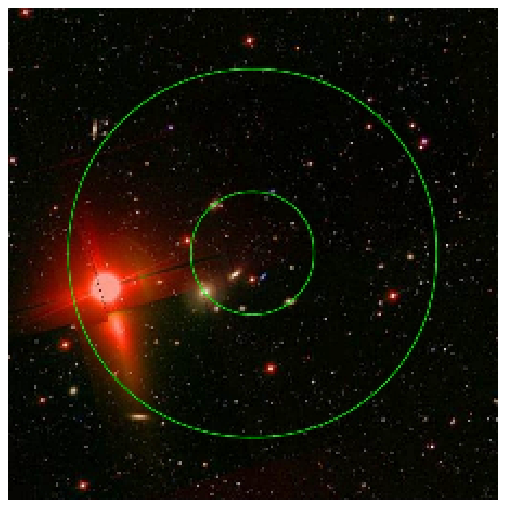}
\includegraphics[scale=0.66]{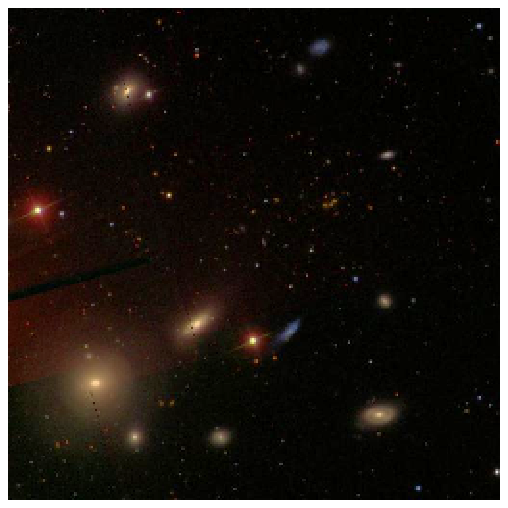}}
\vskip 0.3cm

{\includegraphics[scale=0.66]{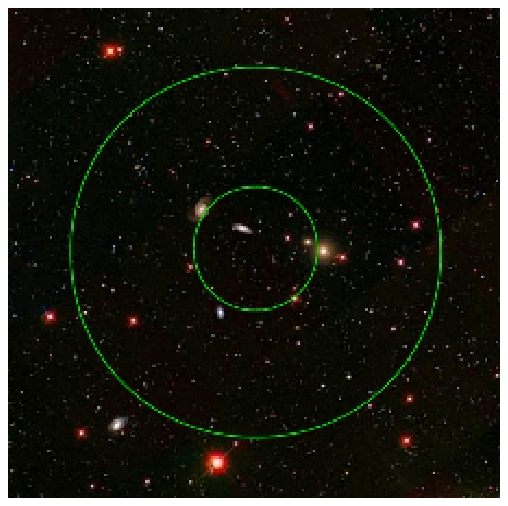}
\includegraphics[scale=0.66]{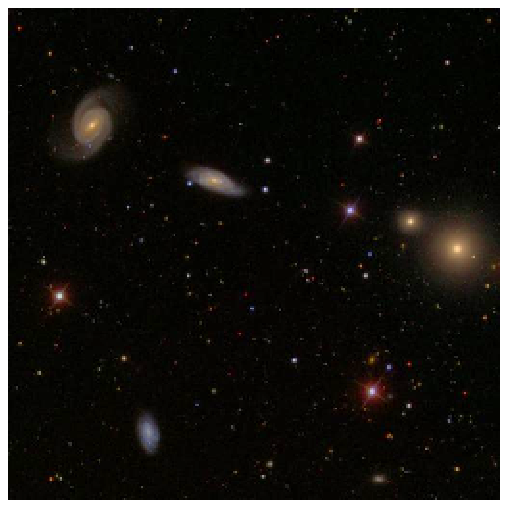}
\hskip 0.3cm
\includegraphics[scale=0.66]{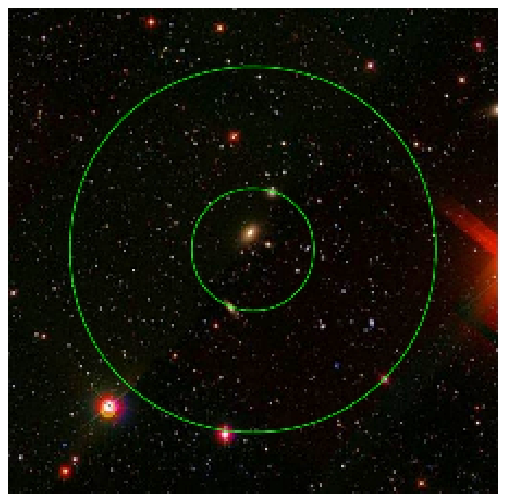}
\includegraphics[scale=0.66]{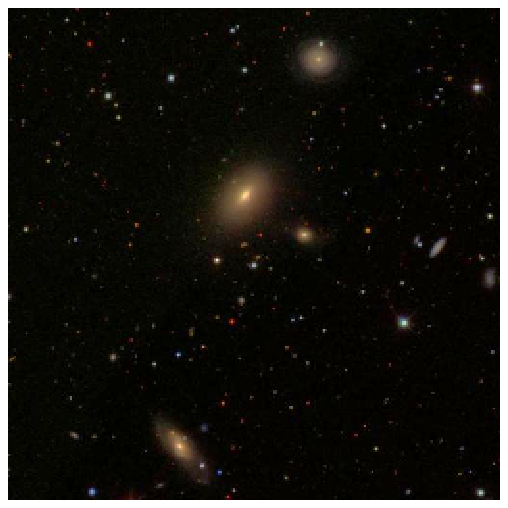}}
\vskip 0.3cm

{\includegraphics[scale=0.66]{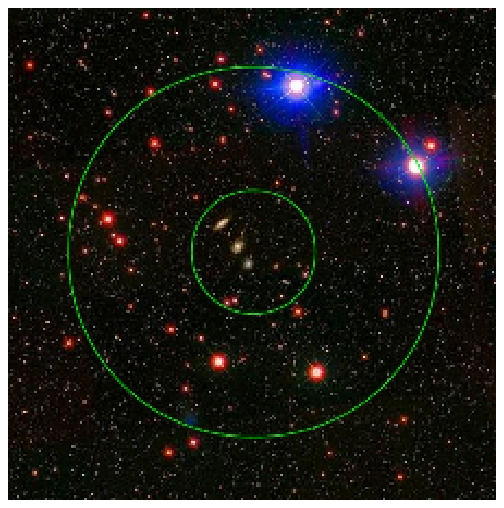}
\includegraphics[scale=0.66]{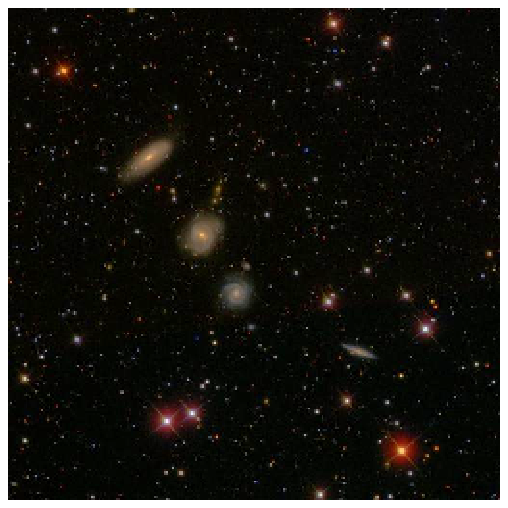}
\hskip 0.3cm
\includegraphics[scale=0.66]{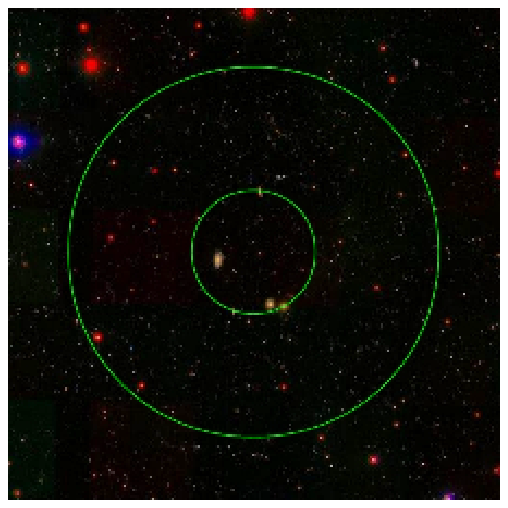}
\includegraphics[scale=0.66]{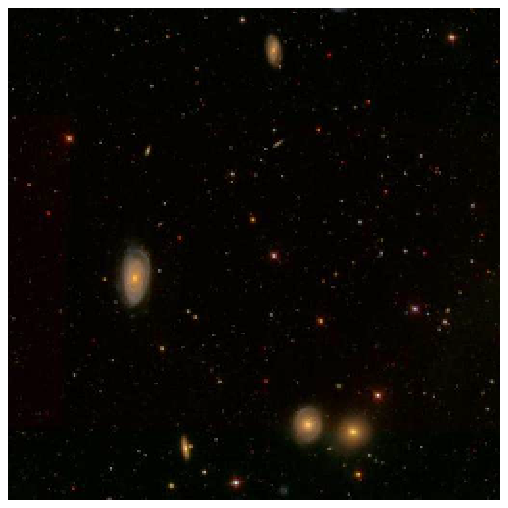}}

\caption{A few examples of \emph{p2MCG}s that lie within the SDSS area, 
none of them are already known groups. There are 2 frames per \emph{p2MCG}:
the \emph{left frames} show concentric circles which correspond to 1 $\theta_G$ and 3 $\theta_G$ (see text).
The \emph{right frames} are zoomed images which show the regions within $\theta_G$, for each group.
According to the notation in 
Table~\ref{v2MASSCG} they are (from left to right and top to bottom): 
32, 36, 40, 50, 52, 57, 59, 62, 64, 66, 74, 85. 
\label{2MASSCG_sdss}}
\end{figure*}

\begin{figure}
\includegraphics[width=\hsize]{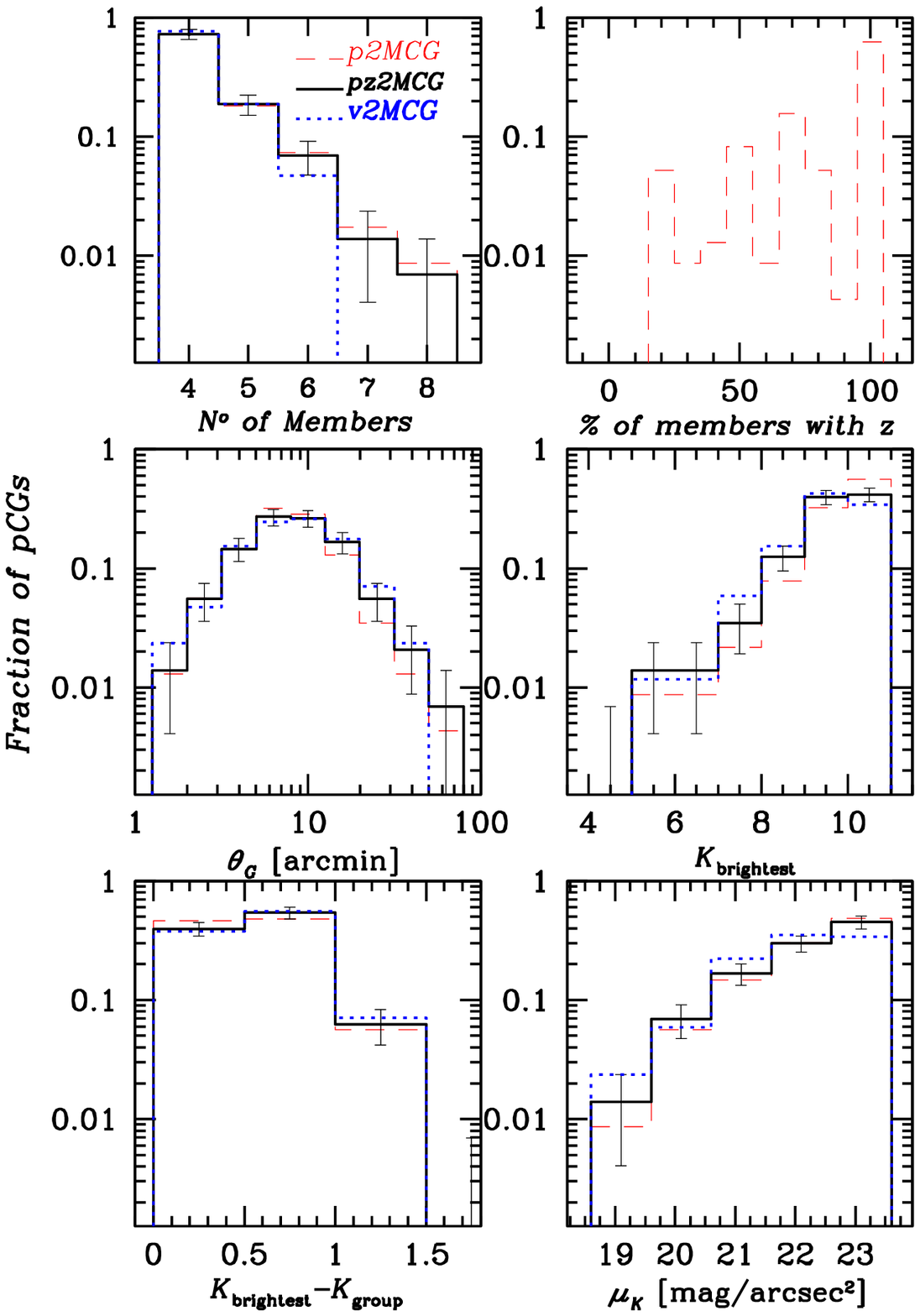}
\caption{Distributions of observable properties for the CGs identified in projection in the 2MASS XSC \label{p2MCG}:
Number of members in the CG (\emph{top left panel}), percentage of members with redshifts available 
(\emph{top right panel}), group angular diameter (\emph{middle left panel}),
$K$-band apparent magnitude of the brightest galaxy member (\emph{middle
  right panel}), difference between the brightest galaxy and the total group
magnitudes (\emph{bottom left panel}) and mean group surface brightness
(\emph{bottom right panel}). 
\emph{Dashed histograms} correspond to the sample of 230 \emph{p2MCG}s,
 \emph{solid histograms} correspond to the sample of 144 
\emph{p2MCG}s that have all the redshifts of their galaxies known 
(\emph{pz2MCG}s), and will be filtered in Sect.~\ref{v-filter}, 
while 
\emph{dotted histograms} correspond to the sample of 85 \emph{v2MCG}s.
Error bars correspond to Poisson errors.
}
\end{figure}

As a result, we identify $230$ 
\emph{p2MCG}s in the 2MASS catalogue.
In Fig.~\ref{aitoff} we show the sky coverage of these groups (\emph{empty squares}). Figure~\ref{2MASSCG_sdss} shows images of a few examples of \emph{p2MCG}s 
that lie in the SDSS area. Some of the observable properties of the \emph{p2MCG}s 
are shown in Fig.~\ref{p2MCG} and their median values are shown in the second column of 
Table~\ref{projected}.

A list of acronyms used to refer different samples to be defined throughout this work is provided in Table~\ref{acronyms}.
\begin{table}
\begin{center}
\tabcolsep 3pt
\caption{List of acronyms used throughout this work \label{acronyms}}
\begin{tabular}{ll}
\hline
CG & general compact groups \\
\emph{p2MCG} & CGs identified in projection from the 2MASS catalogue\\
\emph{pz2MCG} & \emph{p2MCG}s whose galaxies have their radial velocities known \\
\emph{v2MCG} & CGs with 4 or more accordant galaxies (velocity filtered)\\
\emph{mvCG}  & mock velocity-filtered CGs\\
\emph{HCG}  & Hickson Compact Groups\\
\hline
\end{tabular}
\end{center}
\end{table}


\section{Cross-identification}
\label{cross-id}

We compared our sample of CGs to the original HCG sample.
We looked for the $K$-band magnitudes of all the original members of the HCG 
sample in the 2MASS catalogue.
There are $42$ HCGs that lie within the studied area 
(HCG~33 and 34 lie within $20^\circ$ of the Galactic Plane)
and whose brightest galaxy $K$-band magnitude is brighter than 10.57
(fourth criterion). 
However, only $20$ HCGs have been identified with the \emph{p2MCG}s in the 2MASS sample, 
and they are: HCG~4, 7, 10, 15, 16, 21, 22, 23, 25, 40, 42, 51, 
58, 86, 87, 88, 93, 97, 99, 100. While 10 of these 
20 CGs have the exact
same member galaxies, the remaining $10$ have galaxies in common but
 are not exactly the same:
some groups have more galaxies unidentified by Hickson 
while others have fewer. 

We therefore  analysed the reasons why we 
 failed to identify the 22 remaining HCGs among the 42.
First, in HCG 68, HCG 92 (Stephan`s quintet) and HCG 94, the 2MASS XSC
photometric pipeline blends
a pair of galaxies into a single galaxy or only identifies one galaxy of a
pair. This then falls into the category of 
groups discarded due to problematic galaxy identification 
described in the previous section
(in this case, galaxy NGC 5354 for HCG 68, 
galaxies NGC 7318A and NGC 7318B for Stephan's quintet, and 
NGC 7578A for HCG 94).
Second, among the  19 remaining  unidentified  HCGs in our sample, 10
(HCG~5, 56, 57, 61, 65, 74, 90, 91, 96, 98)
have less than 4 members within our adopted 2MASS limit of $K=13.57$, 
i.e, some of their members do not belong to our parent sample. 
Moreover, HCG~57  also fails
the HCG isolation criterion in the $R$ band \citep{Sulentic97},
while HCG~74 and 96 fail the membership criterion in the $R$ band. 
Finally, $9$ of the HCGs  (HCG~11, 19, 30, 41, 44, 48, 53, 62, 67)
fail to meet the $K$-band membership criterion, i.e., have fewer than 4 galaxies
with $K-K_{\rm brightest} < 3$,
one of which (HCG~30) also fails to meet this criterion in the $R$ band.

The visual inspection performed using Aladin images has also provided 
information about other cross-identifications. 
Only $25\%$ of our \emph{p2MCG}s have already been completely or partially 
identified by other authors.

\section{Velocity-filtered compact groups}
\label{v-filter}

\subsection{Velocity filtering}
We searched in the literature for available redshifts for all galaxies in the \emph{p2MCG}
sample, in order to have a sample of concordant groups.
First, we correlated the galaxies in the 2MASS extended source catalogue 
with galaxies in the 2MASS Redshift catalogue (2MRS, \citealp{Huchra+12}).
We have found $561$ 
of our galaxies in \emph{p2MCG}s in the main catalogue of those authors. 
Also, another $280$ 
were present in the ``extra" catalogue presented by the authors. 
Then, we looked for the remaining galaxies in the 
2M++ redshift compilation \citep{2M++_11}. We found $9$ 
of the remaining \emph{p2MCG} galaxies in this catalogue. 
We also looked for available redshifts in the NED for those galaxies in the \emph{p2MCG}s
that do not belong to the 2MRS nor to 2M++. We have found another $19$ 
redshifts of galaxies in \emph{p2MCG}s. 
All in all, we find that $869$ 
out of $1020$ 
galaxies (85\%) already have measurements of their redshifts available. 

A total of 144 (62\%) 
of the \emph{p2MCG}s have \emph{all} their members with
available redshifts, and we hereafter refer to these as \emph{pz2MCG}s.
In 20\% of the \emph{p2MCG}s there is one galaxy without available redshift, 
while in 10\%  (8\%)
of the \emph{p2MCG}s there are two (three or more) galaxies without
redshifts. 

Fig.~\ref{p2MCG} shows that 
the distributions of observable properties of the 
144 \emph{pz2MCG}s are very similar to those for the full sample of \emph{p2MCG}s.
Therefore, our subsample of \emph{pz2MCG}s does not appear biased relative to
the full sample of \emph{p2MCG}s.

Using these 144 \emph{pz2MCG}s, 
we built a sample of velocity-filtered CGs (\emph{v2MCG}s) by following an iterative
procedure (see \citealp{HMdOHP92} and \citealp{diaz-mamon10}). Briefly, 
after computing the median velocity of the group, we discard the galaxy whose
velocity is furthest and at least $\Delta v = 1000 \, \rm km \, s^{-1}$ from the median.
We recompute the velocity median of the remaining galaxies and iterate until
all, and at least 4, galaxies lie within $\Delta v$ from the new median. We
then check that the brightest remaining galaxy is brighter than 10.57, and that
$\mu_K \leq 23.6\,\rm mag\,arcsec^{-2}$. If not, we discard the
group. 

\begin{figure*}
\includegraphics[width=0.93\hsize]{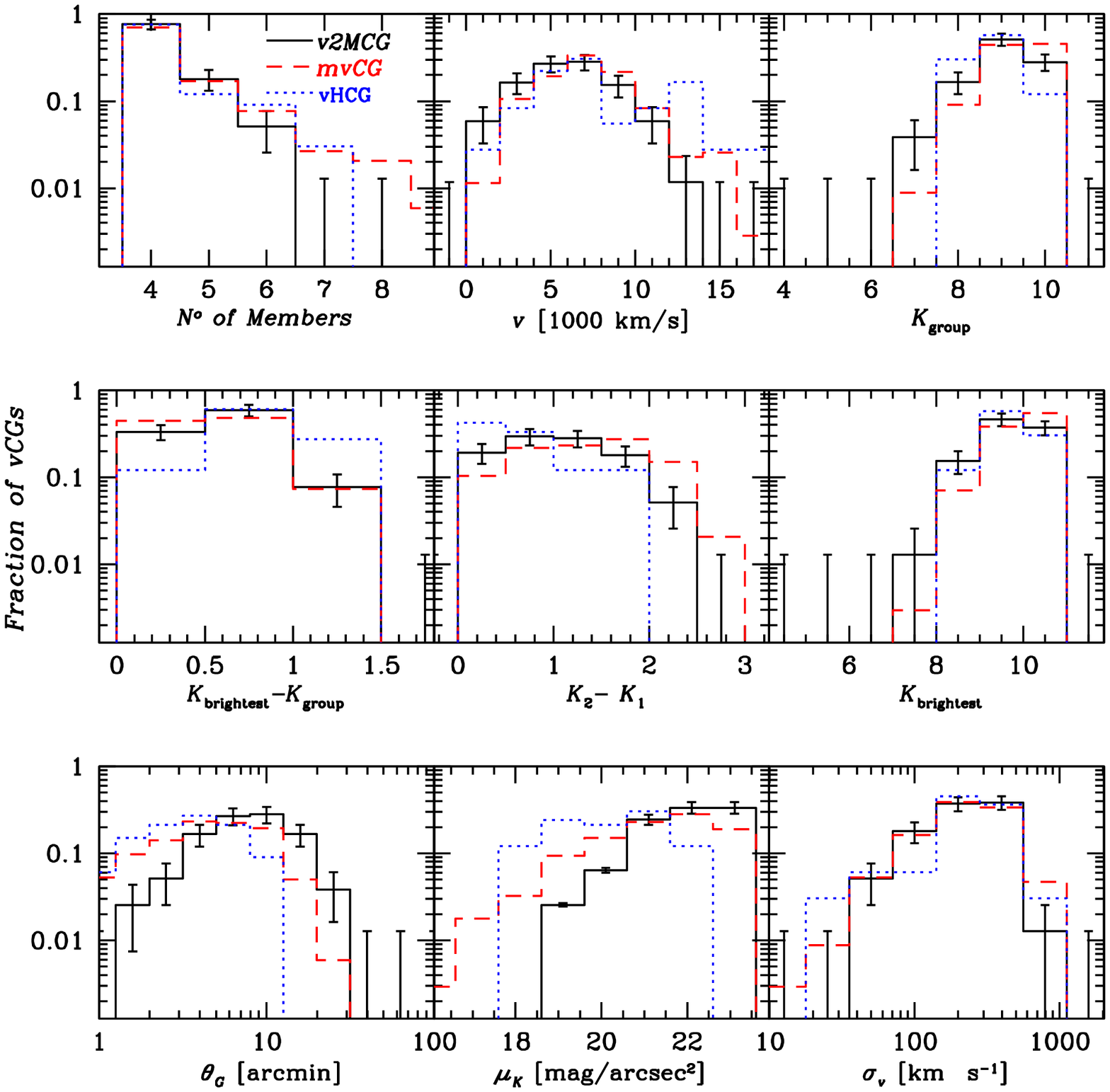}
\vskip -0.65cm
\includegraphics[width=0.93\hsize]{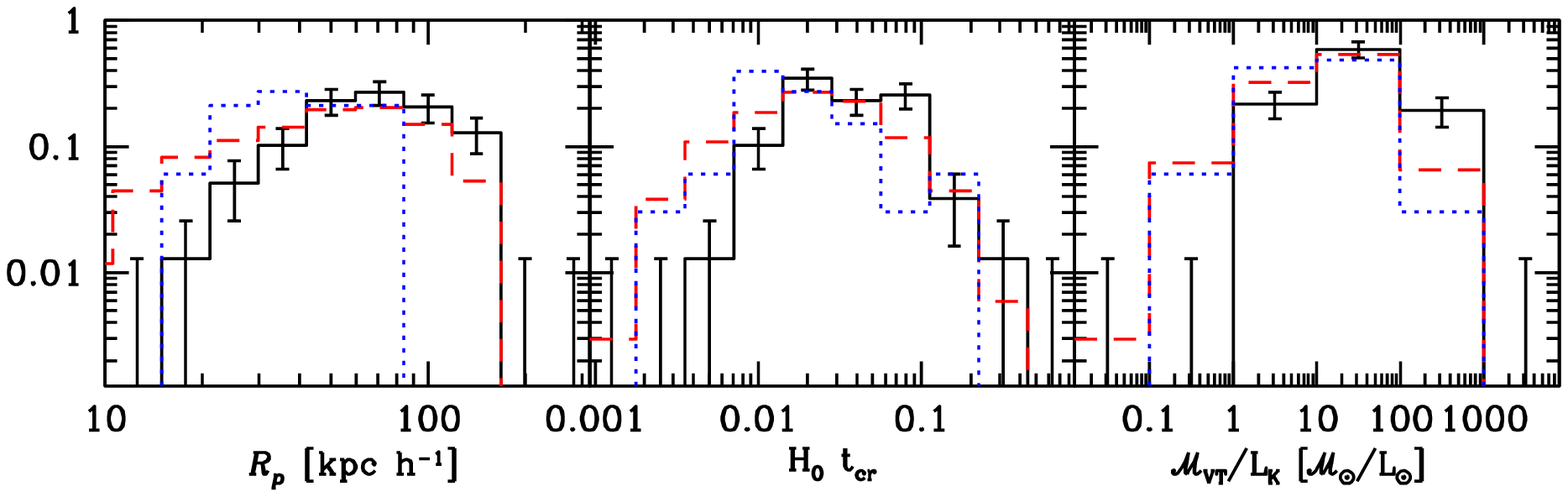}
\caption{Distributions of properties of the Compact groups after the velocity filtering. 
All panels are restricted to groups with $\langle v \rangle > 3000 \, {\rm km \,
  s^{-1}}$ except for the 
median group velocities. 
\emph{Thick black solid histograms}: \emph{v2MCG}s; \emph{thin red dashed histograms}: 
velocity-filtered mock Compact Groups (\emph{mvCG}s) 
identified in the $r$-band from the semi-analytical model of Guo et
al. (2011) run on the halos of the MS-II dark matter simulation, and
converted to the $K$-band using
$K=R-2.4=r-2.73$ (Table~\ref{median_v});
\emph{thin blue dotted histograms}: vHCG are the velocity-filtered HCGs
restricted to the limits used in this work and
converted to the $K$-band 
(sample Hick92/2, Table~\ref{median_v}).
Error bars correspond to Poisson errors.
} 
\label{v2MCG}
\end{figure*}

Our procedure led us to 85 
\emph{v2MCG}s that survive the velocity filtering, 
and are thus less likely to be contaminated by galaxies in chance
projections. The angular distribution of these groups is shown in 
Fig.~\ref{aitoff} (\emph{filled circles}).

Figure~\ref{v2MCG} shows the properties (see Sect.~\ref{measureprops} below) of the
\emph{v2MCG}s (solid black histograms).
One sees that the sample of \emph{v2MCG}s appears to be complete up to $K_{\rm
  group} \simeq 9$, close to the theoretical limits for quartets ($10.57- 2.5
\log 4 = 9.06$) and the rarer quintets ($10.57-2.5 \log 5 = 8.82$).
The \emph{v2MCG} sample also appears to be fairly complete\footnote{Of
  course, a flux-limited catalogue is never complete in terms of volume, since
galaxies are sampled with increasing minimum luminosity as one goes out to
increasing distances, hence the space density of galaxies always decreases
(if it were not for fluctuations from the large-scale structure).}
 to radial velocity of
$\sim 6000 \, \rm km \, s^{-1}$.

 Fig.~\ref{p2MCG} displays a comparison between the sample of projected and
 filtered CGs.
One sees 
that groups with higher multiplicity, or very large angular size
are more prone to be chance alignments along the line of sight.

\subsection{Cross-identification}
We found that $46\%$ of the \emph{v2MCG}s were previously (completely or partially) 
identified by other authors (see last column of Table~\ref{v2MASSCG}).
In particular, the \emph{v2MCG}s include 16 Hickson compact groups: HCG~7, 10,
15, 16, 21, 23, 25, 40, 42, 51, 58, 86, 88, 93, 97, and 99.
Moreover, $52\%$ of the \emph{v2MCG}s lie in the SDSS area.
The median of the properties are quoted in the second column of Table~\ref{median_v}.

\subsection{Measurement of group properties}
\label{measureprops}
The main properties of the \emph{v2MCG}s are quoted in Table~\ref{v2MASSCG}. 
They are:\\
{\it Column 1}: Group ID\\
{\it Column 2}: Right Ascension of the CG centre\\
{\it Column 3}: Declination of the CG centre\\
{\it Column 4}: Median radial velocity\\
{\it Column 5}: Number of galaxy members in the CG in the range of 3 magnitudes from the brightest member\\
{\it Column 6}: Extinction-corrected $K$-band apparent magnitude of the brightest galaxy\\
{\it Column 7}: Extinction-corrected $K$-band group surface brightness\\
{\it Column 8}: Angular diameter of the smallest circumscribed circle\\
{\it Column 9}: Median projected separation among galaxies\\
{\it Column 10}: Apparent group elongation\\
{\it Column 11}: Radial velocity dispersion of the galaxies in the CG\\
{\it Column 12}: Dimensionless crossing time\\
{\it Column 13}: Mass-to-light ratio in the $K$-band\\
{\it Column 14}: Cross-identification with other group catalogues\\

The group velocity dispersions, $\sigma_v$, are computed using the 
gapper algorithm following \cite{BFG90}, who found it to be more efficient
than standard estimators of dispersion for small samples.\footnote{One of us
  (G.A.M., unpublished) also found the gapper estimate of dispersion to be
  much less biased for small samples than are other measures.} Our values of $\sigma_v$
are corrected (in quadrature) for the velocity  errors.

The extinction corrections in Table~\ref{v2MASSCG} refer to the galactic
extinction, deduced from \cite{schlegel98}. We did not correct for internal
extinction, because the corrections are usually negligible, except for
edge-on spirals where it is probably of order of 0.2 mag. Moreover, we expect
that internal extinction increases not only with the inclination of the disk,
but also with disk luminosity (with increasing column density of dust at
increasing luminosity for given dust/stars ratio) and metallicity (which
controls the dust/stars ratio), as well as on the bulge/disk ratio.
Since we lack bulge/disk decomposition for our galaxies, we could have used
the internal extinction formulae for 2MASS wavebands 
of \cite{MGH03}, given as a function of
inclination and luminosity. However, their modulation of internal extinction 
by the luminosity saturates at luminosities of about $0.2\,L*$, whereas it
should keep rising because, at increasing luminosity, the column density
should increase and the metallicity, hence dust/stars ratio, should also
increase. We thus prefer to leave the internal extinction to further analysis.

The physical radii and luminosities assume distances obtained from the
redshifts, i.e. we neglect the peculiar velocities of the galaxies relative
to the Sun.
We could have included a Virgocentric infall model to correct for
the peculiar velocity of the Local Group (as given in
HyperLEDA\footnote{http://leda.univ-lyon1.fr/}, see 
\citealp{TPE02}), but this would not have included 
the peculiar motions of \emph{v2MCG} galaxies.
The attractor model of \cite{Mould+00} does include the peculiar motions of 
both the Local Group and the other galaxies, but it misses all the repelling voids. 
The peculiar velocity flow model of \cite{LTMC10} does include the full matter 
distribution and not just the attractors. 
But it was built from redshift data and lacks accuracy at distances less 
than $30 \,h^{-1}\,\rm Mpc$ because it is not calibrated with available 
quality distance estimators (Cepheids or surface brightness fluctuations).
Since none of the models available satisfied our expectations
taking into account all the main velocity components, 
we decided not to correct for peculiar motions, 
and we leave this open to further analysis. For peculiar velocities
of $300 \, \rm km \, s^{-1}$, the effects of peculiar motions on distances
are less than 10\% for galaxies with $c\,z > 3000 \, \rm km \, s^{-1}$
(leading to physical size and luminosity errors less than 10\% and 20\%
respectively). For the presentation and analysis of statistical \emph{v2MCG}
properties, we thus restrict our sample to the 78 \emph{v2MCG}s 
for which the median group velocity is greater than $3000 \, \rm km \, s^{-1}$. 

We compute the absolute magnitudes of the individual galaxies,
assuming that their luminosity distances are all based upon the
median group redshift.
The cosmology adopted for computing the luminosity distances is the standard 
cosmology also used in the MS ($\Omega_m=0.25$, $\Omega_\Lambda=0.75$). 
Note that the luminosities are not only corrected for galactic extinction, but are
also $k$-corrected. For computing the $k$-corrections, we have used
the polynomial expressions in terms of redshift and colour $H$--$K$ given by \cite{CMZ10}.
The results of the present work depend very little on the
details of the $k$-corrections,
since the galaxy samples studied here are from shallow flux-limited surveys, hence limited
to low redshifts. 

Our dimensionless crossing times are obtained with
\begin{equation}
H_0 \, t_{\rm cr} = H_0 \, \frac{\langle d_{ij}^{3D}\rangle}{\sigma_{3D}} =
\frac{100 \, \pi}{2 \sqrt{3}} \,h\, \frac{ \langle
  d_{ij}\rangle}{\sigma_v }  \ , 
\label{tcross}
\end{equation}
where $\langle d_{ij}\rangle$ is the median of
the inter-galaxy projected separations in $h^{-1} \,\rm Mpc$.
Our mass-to-light ratios are obtained from an 
application of the virial theorem: 
\begin{equation}
{ {\cal M}_{\rm VT}\over L} = {3\pi \over 2}\,{(2 \, R_{\rm h})\,
  \sigma_v^2\over G\, L} \ ,  
\label{moverl}
\end{equation}
where  
$R_{\rm h} = \left \langle 1/d_{ij} \right\rangle^{-1}$ is the harmonic mean
projected separation, given the projected separations
$d_{ij}$ (see eq. [10--23] of \citealp{BT87}). 

The distributions
of the main properties of the \emph{v2MCG}s are shown as {solid black histograms} in Fig.~\ref{v2MCG}.

\subsection{Mock velocity-filtered compact groups}
\label{mock_sample}

It is instructive to compare the distribution of the properties of
\emph{v2MCG}s with the mock, velocity-filtered compact groups (\emph{mvCG}s)
selected on mock galaxy catalogues with the exact same criteria as those described at the top of
Sect.~\ref{2mcgc}. We have done this following the prescriptions of
\cite{diaz-mamon10}, who had analysed the $z=0$ outputs of three different
semi-analytical models (SAMs) of galaxy formation
(B06, \citealp{Bower+06}; C06, \citealp{Croton+06}; and DLB07, \citealp{dLB07}). 

However, since this work, a new SAM has been developed by \cite{Guo11}
(hereafter G11)
that reproduces much better, among other things,
the $z=0$ stellar mass function of galaxies. Moreover,
\citeauthor{Guo11} have run their SAM not only on the Millennium dark matter
simulation, but also on the Millennium-II simulation (MS-II, see
\citealp{BoylanKolchin+09}), which has 5 times 
better space resolution and 125 times better mass resolution. Since CGs are
small, space and mass resolution are crucial in producing realistic mock CG
catalogues.
We have therefore primarily used the outputs of the \citeauthor{Guo11} SAM
run on top of the MS-II to build realistic mock light cones to a magnitude
limit of $r=16.3$ (the limit of the 2MASS converted to the SDSS $r$ band)
and identify CGs.
We have also reanalysed the $z=0$ output of the SAM by \cite{Croton+06}, this
time using their $K$-band magnitudes instead of the $R$-band ones to a
magnitude limit of $K=13.57$ (the 2MASS completeness limit used throughout
this work). We refer
to these \emph{mvCG}s as C06K.
Note that, as in \cite{diaz-mamon10}, we assumed for all SAMs that mock
galaxies that are close in projection on
the plane of the mock sky would be blended by observers if their angular
separation is less than the sum of their angular half-light radii.

\begin{table*}
\begin{center}
\caption{Mock velocity-filtered compact groups 
\label{tabSAMs}
}
\begin{tabular}{lccccc}
\hline
Parent $\Lambda$CDM 
simulation & MS & MS & MS & MS & MS-II \\
SAM & B06 & C06 & DLB07 & C06K &   G11 \\
Selection band & $R$ & $R$ & $R$ & $K$ & $r$ \\
\hline
$R_{\rm brightest} \le 14.44$ \& $\mu_R \le 26$ \\
Number of \emph{mvCG}s & 1952 & 2011 & 1251 & \nodata & 1782 \\
Fraction of physically dense & 0.77 & 0.73 & 0.58 & \nodata & 0.69 \\
$K_{\rm brightest} \le 10.57$ \& $\mu_K \le 23.6$ \\
Number of \emph{mvCG}s & 379 & 360 & 288 & 486 & 340 \\
Fraction of physically dense & 0.73 & 0.65 & 0.53 & 0.66 & 0.66 \\
\hline
\end{tabular}
\end{center}
\parbox{0.9\hsize}{
\small
{\bf Notes.} MS: Millennium Simulation \citep{Springel+05};
MS-II: Millennium-II Simulation \citep{BoylanKolchin+09};
B06: \cite{Bower+06};
C06: \cite{Croton+06};
DLB07: \cite{dLB07};
G11: \cite{Guo11}.
}
\end{table*}

Table~\ref{tabSAMs} shows the fraction of \emph{mvCG}s, selected in redshift
space as the observed catalogues,  
that are physically dense in real space with the criteria adopted by
\cite{diaz-mamon10}, who call $s$ the maximum pair separation in real space among the
closest subsample of 4 galaxies of the CG or the CG
itself for quartets, while $S_\perp$ and $S_\parallel$ denote the 
maximum line-of-sight
and projected separations of the subsample,
respectively.
With these notations, the criterion for physically dense groups is that they
be physically very small or that they be physically small and not elongated
along the line-of-sight:
($s < 100 \, h^{-1} \, \rm kpc$) OR ($s < 200 \, h^{-1} \, \rm kpc$ AND
$S_\parallel/S_\perp < 2$).
Assuming that the predictions from the SAM can be directly applied to the 
\emph{v2MCG}s, we predict that between $\sim 53\%$ and $\sim 73\%$ of the 
sample can be considered as physically dense systems, which means that between 
$\sim 45$ and $\sim 62$ \emph{v2MCG}s may be truly 
dense systems.
The remaining $37\pm10\%$ of the \emph{mvCG}s are caused by chance
alignments of galaxies along the line-of-sight, usually originating from
larger virialised groups (see also \citealp{diaz-mamon10}).

In particular, for the more realistic G11 SAM run on the much better resolved
MS-II cosmological dark matter simulation, \emph{two-thirds of the mock
  velocity-filtered compact groups are physically dense}, while one-third is
  caused by chance alignments of galaxies along the line-of-sight, mostly
  within larger virialised groups.
So, this better SAM produces a fraction of mock velocity-filtered compact
groups that are physically dense that is similar to what \cite{diaz-mamon10}
had found for the three other SAMs. Comparing the upper and lower rows of
Table~\ref{tabSAMs}, it is encouraging that the fraction of
physically dense \emph{mvCG}s depends little on the waveband (red or $K$) used.

\section{General Properties}
\label{props}

It is interesting to compare the properties of the CGs presented in this work to 
those found in the literature for other CG samples.
We downloaded several CG catalogues available at the VizieR
service\footnote{http://vizier.u-strasbg.fr} 
service of the Centre de Donn\'ees
astronomiques of Strasbourg (CDS),
and computed the properties of these groups in the same way we did for 
our sample of CGs. 

\subsection{Comparison with photometric catalogues}
\label{compareproj}
\begin{table*}
\begin{center}
\caption{Median Properties of CGs identified in projection. For a fair comparison, 
all photometric properties have been translated to the $R$-band }
\label{projected}
\tabcolsep 2pt
\begin{tabular}[t]{ccccccccc}
\hline
\hline
& \emph{p2MCG} & HCG &
DPOSSCG/03 & DPOSSCG/05 &
SDSSCG/04 & SDSSCG-A/09 &
SDSSCG-B/09 & HCG\\ 
\hline
ref. & & Hick82 & Iov03 & deCarv05 & Lee04 & McCon09 & McCon9 & Hick82/2\\ 
colour eq. & $K=R$--$2.4$ & $E=R$ & $r=R$+$0.33$ & $R$ & $r=R$+$0.33$ & $r=R$+$0.33$ &
$r=R$+$0.33$ &  $E=R$ \\
$\#$ & 230  & 100 & 84 & 459 & 177 & 2297 & 74791 & 40\\
$\theta_G$ [arcmin] & $\ \, 7.6\pm2.8$  & $\ \, 3.1\pm1.4$ &  $\ \, 0.7\pm0.1$ & $\ \, 0.7\pm0.1$ & $\ \, 0.4\pm0.1$ & $\ \,  1.5\pm0.4$ & $\ \, 0.4\pm0.1$ &  $\ \, 4.4\pm1.8$ \\
$R_{\rm brightest}$ & $12.5\pm0.4$      & $12.7\pm0.6$     &  $16.2\pm0.2$     & $16.5\pm0.3$     & $16.8\pm0.6$     & $15.8\pm0.5$      & $18.4\pm0.7$ & $12.3\pm0.6$ \\
$R_{\rm G}$         & $11.9\pm0.5$     & $11.9\pm0.6$     &  $15.3\pm0.3$     & $15.6\pm0.3$     & $16.2\pm0.5$     & $15.0\pm0.4$      & $17.6\pm 0.5$ & $11.3\pm0.6$\\
$\mu_G$ [$R \, \rm mag \ arcsec^{-2}$] & $25.0\pm0.7$  & $22.8\pm0.7$ & $23.1\pm0.3$ & $23.6\pm0.3$ & $23.2\pm0.3$ & $24.5\pm0.6$ & $24.5\pm0.5$ & $23.2\pm0.8$ \\
$R_{\rm faintest}$--$R_{\rm brightest}$ & $ \ \, 2.7\pm0.2$  & $\ \, 2.4\pm0.5$ & $\ \, 1.6\pm0.3$ & $\ \, 1.6\pm0.3$ & $\ \, 2.7\pm0.2$ & $\ \, 1.6\pm 0.5$ & $\ \, 1.9\pm0.6$ & $\ \, 2.2\pm0.4$\\
 $R_{\rm brightest}$--$R_G$& $\ \, 0.5\pm0.2$ & $\ \, 0.9\pm0.2$ & $\ \, 0.9\pm0.2$ & $\ \, 0.9\pm0.1$ & $\ \, 0.5\pm0.2$ & $\ \, 0.9\pm0.2$ & $\ \, 0.8 \pm 0.2$ & $\ \, 0.9 \pm 0.2$  \\
\hline
\end{tabular}
\end{center}

\parbox{0.9\hsize}{
\small
{\bf Notes.} $\theta_G$: group angular diameter, $R_{\rm brightest}$: apparent magnitude
of the brightest galaxy member in the $R$-band, $R_{\rm G}$: total apparent
magnitude of the group (i.e., sum of all members), $\mu_G$: group mean surface brightness, $R_{\rm faintest}-R_{\rm
  brightest}$: difference of apparent magnitudes between the faintest and the
brightest galaxy members,  $R_{\rm brightest} - R_G$: difference between the
brightest galaxy and the total apparent magnitudes of the groups.
Errors are the
semi-interquartile ranges. 
References: 
Hick82: \cite{Hickson82};
Iov03: \cite{Iovino03};
deCarv05: \cite{decarvalho05};
Lee04: \cite{Lee+04};
McCon09: \cite{McCPES09};
Hick82/2: \cite{Hickson82}, restricted to 
$ R_{\rm brightest} \leq 10.57+2.4=12.97$ and $R_{\rm faintest}-R_{\rm
  brightest} \leq 3$ and with non-isolated groups \citep{Sulentic97}
removed.
\normalsize
}
\end{table*}
%
We
retrieved data from VizieR for the following six catalogues: 
HCG \citep{Hickson82}, DPOSSCG/03 \citep{Iovino03}, DPOSSCG/05 \citep{decarvalho05},
SDSSCG/04 \citep{Lee+04}, SDSSCG-A/09 \citep{McCPES09}, SDSSCG-B/09 \citep{McCPES09}.
It is important to note that, although we used the membership
information from the authors (angular positions and magnitudes), 
we recomputed all the properties using our own algorithms, 
to ensure that they were  all estimated in the same manner.
In this way, we can compare
our projected \emph{p2MCG} catalogue with others in the literature to highlight 
differences in the searching algorithms and selection criteria.

Table~\ref{projected} shows the median observable properties and their inter-quartile ranges 
for the photometric samples of CGs. 
For a fair comparison among these samples, 
it is necessary to take into account the 
different bands in which CGs have been identified in each different catalogue. 
While our \emph{p2MCG}s are based upon $K$-band magnitudes, HCGs have been
first identified on POSS-I $E$ plates, whose spectral response is close to the
$R$ band, for which the galaxy magnitudes are available. DPOSSCG/03 have SDSS-$r$ band magnitudes, DPOSSCG/05 have $R$-Gunn magnitudes, and SDSSCG/04/09 have SDSS-$r$ band magnitudes. 
We assumed that $R=K+2.4$ (appendix~\ref{mu_k})
 and $R=r - 0.33$ \citep{diaz-mamon10} to compare 
magnitudes in the $K$ and $r$-band to those in the $R$-band. Therefore, 
in Table~\ref{projected}, all magnitudes are converted to the $R$-band.
Table~\ref{projected} also includes
a cleaner subsample of the HCG (Hick82/2) that 
meets equivalent criteria as the used in this work 
($R_{\rm brightest} \leq 10.57+2.4=12.97$ and $R_{\rm faintest}-R_{\rm brightest} 
\leq 3$) and for which we omitted 6 HCGs that fail to meet the isolation criterion
\citep{Sulentic97} (see also \emph{thin dotted blue lines} in Fig.~\ref{v2MCG}).

Table~\ref{projected} indicates that the \emph{p2MCG}s  have brighter
group and first-ranked galaxy apparent magnitudes than those of the other photometric
catalogues.
This is
a consequence of the shallower magnitude limit of the 2MRS spectroscopic survey
used here. Restricting the HCG sample to the magnitude limits used here (`Hick82/2'), 
the median first-ranked galaxy magnitude is slightly brighter than that of the \emph{p2MCG}s. 
The differences remaining between the \emph{p2MCG} and Hick82/2 samples
arise from the differences between automatic and visual identifications, 
since the latter by \cite{Hickson82} were biased 
(e.g., \citealp{diaz-mamon10}) towards identifying groups with 
similar galaxies (lower values of $R_{\rm faintest}-R_{\rm brightest} $, 
and higher values of $R_{\rm brightest}$--$R_G$), 
and missing groups close to the compactness limit (lower values of $\mu_G$).
On the other hand, the median angular diameter of the \emph{p2MCG}s 
is larger than for the other samples, 
making the surface brightness of our sample the faintest. 
Moreover, not all the CG catalogues were constructed taking into account 
our fourth criterion that ensures that group members can be found in a 3 
magnitude range from the first-ranked galaxy. 
In several of the catalogues, the magnitude limit of the sample
is sometimes just one or two magnitudes fainter than that of the first-ranked galaxy. 
It is clear that this leads to a bias towards identification of smaller differences between the first-ranked and 
faintest member of the group, as can be seen, e.g., in the average values of 
$R_{\rm faintest} - R_{\rm brightest}$ of 1.6 for the DPOSS/03/05 and SDSSCG-A/B
catalogues.

It is interesting to compare the number of \emph{p2MCG}s 
with $\delta > -33^\circ$ and $K_{\rm brightest}<10.57$ 
with the number of HCGs in the same range of magnitudes. 
We find $193$ 
\emph{p2MCG}s vs. $40$ 
\emph{pHCG}s that meet the criteria 
used in this work  translated to the $R$-band 
($K_{\rm brightest} \le 10.57$ and $K_i-K_{\rm brightest} \leq 3$), 
which means that the completeness
of the visually identified HCGs is  $\sim 21\pm2\%$  
(binomial errors)
This result is higher than the $14\%$ 
predicted by \cite{diaz-mamon10} from the semi-analytic models of galaxy
formation.
Moreover, in our analysis of the SAM of \cite{Croton+06}, 
identifying in the deeper $R$-band mock catalogue and then translating
to the $K$ band produces a sample of \emph{mvCG}s that is only 74\% 
of the size of the \emph{mvCG} sample directly selected in $K$.
Thus, the incompleteness of the HCG catalogue relative to the 2MCG one is
$0.21/0.74=0.28$, 
even higher than the prediction from the SAMs in the $R$ band.

Given that the properties shown in Table~\ref{projected} are dependent on 
the distances
to the CGs, which are not included in the analysis above, it is also interesting 
to compare catalogues for which velocity information is available. 
This is done in the following subsection.

\subsection{Comparison with observed and mock spectroscopic catalogues}

%
\begin{table*}
\begin{center} 
\caption{Median properties of compact groups after velocity filtering, with
  radial velocity larger than $3000 \, \rm km \, s^{-1}$
\label{median_v}}
\tabcolsep 2.9pt
\begin{tabular}{lccccccc}
\hline
\hline
& \emph{v2MCG} & HCG & UZC-CG
& LCCG & DPOSSII-CG & HCG & \emph{mvCG} \\
\hline
ref. & & Hick92 & Foc02 & Allam00 & Pom12 & Hick92/2 & G11\\ 
colour eq. & $K=R$--2.4 & $E=R$ & $B=R$+1.7 &$R$ & $B=R$+1.7 & $E=R$ & $r=R+0.33$ \\
$\#$ & 78 & 67 & 49 & 17 & 33 & 33 & 326 \\ 

$\theta_G$ [arcmin] 
& $ \ \,7.7 \pm 3.1 $ 
& $\ \, 2.5 \pm 1.3$ 
& $     11.8\pm 4.2$ 
& $\ \, 1.0 \pm 0.3$ 
& --- 
& $\ \, 3.6 \pm 1.5$ 
& $\ \, 4.8 \pm 2.8$ \\
 
$R_{\rm brightest}$ 
& $12.2 \pm 0.4 $ 
& $12.7 \pm 0.6$ 
& $12.6 \pm 0.5$ 
& $16.1 \pm 0.4$ 
& --- 
& $12.2 \pm 0.5$ 
& $12.4 \pm 0.4$ \\

$R_G$ 
& $ 11.5 \pm 0.5 $ 
& $11.9\pm0.7$ 
& $11.3\pm0.5$ 
& $15.1\pm0.3$ 
& --- 
& $11.3\pm0.5$ 
& $11.8 \pm 0.4$ \\

$\mu_G$ [$\rm mag \ arcsec^{-2}$] 
& $ 24.5 \pm 0.7 $ 
& $22.7 \pm 0.6$ 
& $25.5 \pm 0.6$ 
& $23.9 \pm 0.6$ 
& --- 
& $22.7 \pm 0.9$ 
& $23.9 \pm 1.1$ \\

$R_{\rm faintest}$--$R_{\rm brightest}$ 
& $\ \, 2.5 \pm 0.3 $ 
& $\ \, 2.1 \pm 0.5$ 
& $\ \, 1.2 \pm 0.4$ 
& $\ \, 1.2 \pm 0.5$ 
& --- 
& $\ \, 2.2 \pm 0.4$ 
& $\ \, 2.7 \pm 0.2$ \\

$R_{\rm brightest}$--$R_G$
& $\ \, 0.6 \pm 0.2 $ 
& $\ \, 0.9 \pm 0.2$ 
& $\ \, 1.2 \pm 0.4$ 
& $\ \, 1.1 \pm 0.2$ 
& --- 
& $\ \, 0.8 \pm 0.2$ 
& $\ \, 0.6 \pm 0.2$ \\

$R_2$--$R_1$ 
& $\ \, 1.0 \pm 0.4$ 
& $\ \, 0.6 \pm 0.4$ 
& $\ \, 0.4 \pm 0.4$ 
& $\ \, 0.5 \pm 0.3$ 
& --- 
& $\ \, 0.6 \pm 0.4$ 
& $\ \, 1.3 \pm 0.5$ \\

$L_R/10^{10} [h^{-2} L_{\odot}] $
& $ \ \,6.7 \pm 1.7$ 
& $    11.5 \pm 5.1$ 
& $\ \, 7.1 \pm 2.6$ 
& $\ \, 3.0 \pm 0.7$ 
& $\ \, 4.2 \pm 1.0$ 
& $    12.6 \pm 5.1$ 
& $\ \, 6.7 \pm 2.2$ \\

$v$  [$\rm km \  s^{-1}$]  
& $ \ \, 6361 \pm 1680 $ 
& $ \ \, 9248 \pm 2976$
& $\ \, 6287 \pm 1380$ 
& $23599\pm3715$ 
& $     32321 \pm 5522$ 
& $ \ \, 7042 \pm 3191$ 
& $ \ \, 7023 \pm 1471$ \\

$\sigma_v$ [$\rm km \  s^{-1}$]  
& $\ \ \ 237  \pm \ \, 105$ 
& $\ \ \ 262  \pm \ \ \ 93$ 
& $\ \ \ 298  \pm \ \ \ 99$ 
& $\ \ \ 243  \pm \ \, 103$ 
& $\ \ \ 194  \pm \ \ \ 55$ 
& $\ \ \ 271  \pm \ \ \ 78$ 
& $\ \ \ 248  \pm \ \, 115$ \\

$\langle d_{ij} \rangle$ [$h^{-1} \rm kpc$] 
& $\ \ \ \ \, 86\pm \ \ \ 24$ 
& $\ \ \ \ \, 43\pm \ \ \ 15$ 
& $ \ \ \ 132\pm \ \ \ 34$ 
& $\ \ \ \ \, 42 \pm\ \ \ \ \,5$ 
& $\ \ \ \ \, 31 \pm \ \ \ \ \, 6$ 
& $\ \ \ \ \, 47\pm \ \ \ 15$ 
& $\ \ \ \ \, 59\pm \ \ \ 28$ \\

$r_p$ [$h^{-1} \rm kpc$] 
& $\ \ \ \ \, 65\pm \ \ \ 25$ 
& $\ \ \ \ \, 34\pm \ \ \ 12$
& $\ \ \ 108\pm\ \ \ 27$ 
& $\ \ \ \ \, 35\pm \ \ \ \ \,6$ 
& --- 
& $\ \ \ \ \, 36\pm \ \ \ 12$ 
& $\ \ \ \ \, 48\pm \ \ \ 23$ \\

$b/a$ 
& $\ \, 0.43 \pm 0.17$ 
& $\ \, 0.37 \pm 0.17$ 
& $\ \, 0.47 \pm 0.19$ 
& $\ \, 0.37 \pm 0.16$ 
& --- 
& $\ \, 0.37 \pm 0.15$ 
& $\ \, 0.40 \pm 0.18$\\

$H_0\,t_{\rm cr}$ 
& $\ \, 0.032 \pm 0.024$ 
& $\ \, 0.013 \pm 0.008$ 
& $\ \, 0.039 \pm 0.024$ 
& $\ \, 0.014 \pm 0.008$  
& $\ \, 0.018 \pm 0.005$ 
& $\ \, 0.016 \pm 0.008$ 
& $\ \, 0.020 \pm 0.017$  \\

${\cal M}_{\rm VT}/ L_R$ [$h \, {\cal M}_\odot/L_\odot$] 
& $     116 \pm 42$ 
& $\ \,  42 \pm 28$ 
& $\ \, 235 \pm 193$ 
& $     117 \pm 76$ 
& $\ \,  94 \pm 34$ 
& $\ \,  39 \pm 20$ 
& $\ \,  53^{+91}_{-32}$ \\

$T_1$ 
& $\ \, 0.51 \pm 0.06$ 
& $\ \, 1.27 \pm 0.17$ 
& $\ \, 1.04 \pm 0.15$
& $\ \, 1.10 \pm 0.27$ 
& --- 
& $\ \, 1.15 \pm 0.22$ 
& $\ \, 0.46 \pm 0.02$ \\

$T_2$ 
& $\ \, 0.70 \pm 0.06$ 
& $\ \, 1.01 \pm 0.10$ 
& $\ \, 1.13 \pm 0.11$
& $\ \, 1.10 \pm 0.19$ 
& --- 
& $\ \, 0.98 \pm 0.13$ 
& $\ \, 0.59 \pm 0.02$ \\

$P_{\rm S}$ 
& 3$\times$$10^{-4}$  
& 0.19
& 0.53
& 0.86
& --- 
& 0.31
& 0  \\

$P_{\rm KS}^{1-2}$ 
& 9$\times$$10^{-4}$ 
& (0.40) 
& 0.09
& (0.93)
& --- 
& (0.08)
& 5$\times$$10^{-10}$   \\

$P_{\rm KS}^{2-3}$ 
& (0.90) 
& (0.69) 
& (0.23) 
& (0.93) 
& --- 
& 0.25 
& (0.38)  \\
\hline
\end{tabular}
\end{center}

\parbox{0.95\hsize}{
\small
{\bf Notes.} 
All the photometric properties have been translated to the $R$-band to allow comparison among catalogues.
$\#$: number of  CGs with 4 of more concordant members; 
$\theta_G$: group angular diameter; 
$R_{\rm brightest}$: 
apparent magnitude of the brightest galaxy member in the $R$-band; 
$R_{\rm G}$: total apparent magnitude; 
$\mu_G$: group mean surface brightness; 
$R_{\rm faintest}-R_{\rm brightest}$: difference of apparent magnitudes between the faintest 
and the brightest galaxy members; 
$R_{\rm brightest} - R_G$: difference between the brightest 
galaxy and the total apparent magnitudes; 
$R_2-R_1$: difference of absolute magnitudes between the brightest and the second brightest 
galaxy of the group (same statistics for difference in absolute $R$-band 
magnitudes,
after including $k$-corrections from \citealp{CMZ10} and \citealp{Poggianti97});
$L_G$: total luminosity of the CG; 
$v$: group median radial velocity; 
$\sigma_v$: group gapper \citep{WT76} velocity dispersion, corrected for
galaxy velocity errors 
(assumed to be $40 \, \rm km \, s^{-1}$ when unavailable);
$\langle d_{ij} \rangle$: median inter-galaxy separation;
$r_{\rm p}$: group radius (of smallest circumscribed circle);
$b/a$: apparent elongation of the group (1=round);
$H_0 \, t_{\rm cr}$: dimensionless crossing time (eq.~[\ref{tcross}]); 
${\cal M}_{\rm VT}/L_R$: mass-to-$R$-light ratio from the virial theorem 
(eq.~[\ref{moverl}]);
$T_1$ and $T_2$: Tremaine-Richstone statistics (\citealp{Tremaine77}, eq.~[\ref{trstats}]); 
$P_{\rm S}$: probability of greater anti-correlation of luminosity with
position occurring by chance (Spearman rank correlation test);
$P_{\rm KS}^{1-2}$: probability of greater difference in distributions of
positions between 1st and 2nd ranked galaxies, occurring by chance
(Kolmogorov-Smirnov test);
$P_{\rm KS}^{2-3}$: same for difference in distribution of positions between
2nd and 3rd ranked galaxies.
Numbers in parentheses for these three quantities indicates reverse
luminosity segregation (brighter galaxies further out).
Errors are the semi-interquartile ranges, except for $T_1$ and $T_2$, where
they are standard deviations computed with 10$\,$000 bootstraps. 
{\bf References:}
Hick92: \cite{HMdOHP92};
Foc02: \cite{FK02};
Allam00: \cite{AT00};
Pom12: \cite{PI12}, restricted to isolated (classes A, CH and CP) with at
least 4 accordant redshifts;
Hick92/2: \cite{HMdOHP92}, restricted to isolated groups (following
\citealp{Sulentic97})
and restricted to
$ R_{\rm brightest} \leq 10.57+2.4=12.97$ and $R_{\rm faintest}-R_{\rm brightest} \leq
3$;
G11: mock compact groups extracted (following the method of
\citealp{diaz-mamon10}) from the
mock galaxy catalogue of 
\cite{Guo11} applied to the MS-II \citep{BoylanKolchin+09} cosmological $N$ body simulation.
}
\normalsize

\end{table*}

We retrieved galaxy data from VizieR for the following compact group
catalogues with velocity information:
HCG \citep{HMdOHP92}, UZC-CG \citep{FK02}, and LCCG \citep{AT00}.
We also extracted the group information from the new DPOSSII-CG catalogue
\citep{PI12}.
We, then, proceeded to compare those samples to our \emph{v2MCG} sample,
after transforming again all samples to the $R$ band with colours $R$--$K=2.4$
(Appendix~\ref{mu_k}), $B=R+1.7$ \citep{PIM94} and $r = R+0.33$
\citep{diaz-mamon10}.
We have applied $k$-corrections to the different catalogues using the
morphology-based corrections of \cite{Poggianti97} (UZC-CG and LCCG) or
colour-based corrections of 
\cite{CMZ10} (2MCG, HCG).\footnote{We did not apply $k$-corrections to
  DPOSSII-CG because of lack of galaxy information, and we corrected their
  crossing time definition to ours ($\pi^2/9 \simeq 1.2$ times greater).}
We have included the cleaner Hick92/2 sub-sample (see Sect.~\ref{compareproj})
now velocity-filtered,
and also the sample of \emph{mvCG}s that we extracted 
from G11's SAM. 
The median values of the properties of the velocity-filtered CGs 
are quoted in Table~\ref{median_v} as well as their semi-interquartile ranges.

\subsubsection{Space density}
We computed the space density within the median distance of the sample ($60 h^{-1} \ Mpc$) 
as $\eta_{60}={3 \, N(<60)}/(60^3 \Omega )$. 
For the \emph{v2MCG}s the space density is $8.0 \times 10^{-5} h^{3}\rm Mpc^{-3}$.
In comparison, the space density for the Hick92/2 sample is $1.86 \times 10^{-5} h^{3}\rm Mpc^{-3}$, i.e. that the space density of the \emph{v2MCG}s is $\sim 4.3$ times larger.
From the G11 SAM, the space density of the \emph{mvCG}s is $12.7 \times 10^{-5} h^{3}\rm Mpc^{-3}$, which means that it almost doubles ($\sim 1.6$) that of the \emph{v2MCG}s.

\subsubsection{Distribution of group properties}
In Fig.~\ref{v2MCG}, we show the distribution of group properties for our
\emph{v2MCG}s (solid histograms),
\emph{vHCG}s (blue dotted histograms), 
and the \emph{mvCG}s from G11
(thin dashed red histograms).
The comparison with other SAMs can be found in the
Appendix~\ref{props_SAMs}\footnote{Given that the numbers of \emph{mvCG}s
  from the different SAMs are typically 4 times the number of \emph{v2MCG}s,
  the relative uncertainties on their differential distribution is half of
  those of the \emph{v2MCG}s, and are not shown in the Figure, for clarity.}.
Table~\ref{median_v} shows 
that the nearest samples are the UZC-CG and \emph{v2MCG} samples, 
although the HCG sample restricted
to the criteria used in this work also includes only the nearest groups. 
The two nearby CG samples present the largest projected radii, 
median inter-particle distances, dimensionless crossing times and mass-to-light ratios. 
The five CG samples have similar median properties (to within the
semi-interquartile ranges), except for $T_1$ and $T_2$ (see below).
In particular, the median velocity dispersions 
for the different catalogues are fairly similar, ranging 
from $194$ to $295 \, {\rm km \, s^{-1}}$.

Our sample has the highest median crossing time of all samples, while the HCG
has the lowest crossing time. The latter result is probably caused by the
lack of HCGs near the surface brightness limit
\citep{WM89,PIM94,diaz-mamon10}.


There is a good general agreement between the
predictions from the SAM and the observations from 2MASS. But some
differences stand out: in comparison with the \emph{mvCG}s, the \emph{v2MCG}
sample lacks groups of high-multiplicity (as checked with a KS test on the full
$N$ distributions), very low velocity dispersion, small angular and physical sizes, high
surface brightness, and low ${\cal M}_{\rm VT}/L_K$, but
has too many groups that lie at low redshifts, or that are globally bright
($K_{\rm group}$) or with bright 1st-ranked galaxy ($K_{\rm brightest}$).

Our identification of more \emph{v2MCG}s at low redshifts than
predicted by the SAM might be a sign that our local neighbourhood ($c\,z<2000
\, \rm km \, s^{-1}$) is denser than on average, perhaps thanks to the presence of the Virgo
and Fornax clusters, or conversely that the observer we placed in a random
position in the cosmological box turned out to be in an underdense region
(for small volumes one might expect that cosmic variance is than Poisson
variance). 
This excess of nearby CGs would explain our excess of \emph{v2MCG}s with
large angular size and of low surface brightness.
However, we also find an excess in physical radii, which suggests that we
suffer more from galaxy blending than we accounted for in our \emph{mvCG}s.
%

\subsubsection{Apparent group elongations}

Using projected Cartesian coordinates on the plane of the sky, we calculated
the 2-dimensional shape tensor, whose eigenvalues are related to the major
($a$) and minor ($b$) semi-axes.
We measure the elongations of the groups in the plane of the sky as the ratio
between the major and minor semi-axes ($b/a$).  Lower values of $b/a$ imply
more elongated systems on the plane of the sky.
Table~\ref{median_v} indicates that the apparent group elongations are
similar between all catalogues.

Using a different technique to measure group apparent elongations, 
\cite{TMT99} found that group velocity dispersions were significantly 
smaller (by 28\%, with
large uncertainty) 
in chain-like
groups than in rounder ones, which we hereafter denote the Tovmassian effect.
Now, geometrical considerations imply that
the distribution of group shapes depends on the number of
its members (e.g. \citealp{HNHM84}),
with low multiplicity groups being on average more elongated.
Since velocity dispersion scales as mass, which scales as number, one
would then expect from the geometrical considerations that
high velocity dispersion groups will be rounder, as found
by \citeauthor{TMT99}. However, these authors also noticed trends for
triplets, quartets and quintets separately,
and while none were statistically significant, they argued
that the probability that all three trends were present
(albeit weak) was significant.\footnote{\cite{TMT99} did not
present any statistical tests for the separate multiplicities, nor for the
combination of the larger mean velocity dispersions for the triplets,
quartets and quintets.}

\begin{table}
\begin{center}
\caption{Group (quartets) velocity dispersion vs. apparent elongation\label{sigvbova}}
\tabcolsep=3pt
\begin{tabular}{lccccc}
\hline
\hline
Catalogue &  $r$ & $P_{\rm S}$ & $\left\langle\sigma_v^{\rm
  chain}\right\rangle$ & $\left\langle\sigma_v^{\rm   round}\right\rangle$ &
$P_{\rm KS}$ \\
\cline{4-5}
& & & \multicolumn{2}{c}{($\rm km\,s^{-1}$)} & \\
\hline
\emph{v2MCG} & 0.01 & 0.95 &  204 &  188 & 0.50 \\
HCG (Hick92/2) & 0.20 & 0.33 & 149 &  284 & 0.19 \\
\emph{mvCG}-G11 & 0.11 & 0.11 &  208 &  272 & 0.09\\
\emph{mvCG}-C06 & 0.01 & 0.85 &  240 &  209 & 0.94\\
\emph{mvCG}-B06 & 0.04 & 0.54 &  227 &  240 & 0.75\\
\emph{mvCG}-DLB07 & 0.02 & 0.81 &  290 &  274 & 0.99\\
\emph{mvCG}-C06K  & 0.10 & 0.10 &  240 &  280 & 0.10\\
\hline
\end{tabular}
\end{center}
\parbox{\hsize}{
\small
{\bf Notes.} The samples are those listed in Table~\ref{median_v}, hence limited to 
$v > 3000 \, \rm km \, s^{-1}$, but also restricted to quartets ($N=4$). 
The columns are
$r$: Spearman rank correlation coefficient;
$P_{\rm S}$: probability of stronger correlation than $r$ occurring by
chance;
$\left\langle\sigma_v^{\rm chain}\right\rangle$ and
$\left\langle\sigma_v^{\rm round}\right\rangle$: median group velocity
dispersions for chain-like ($b/a<0.3$) and round ($b/a>0.5$) groups,
respectively;
$P_{\rm KS}$: probability of greater difference between of velocity
dispersion distributions for groups with $b/a < 0.3$ and $b/a>0.5$ occurring
by chance (Kolmogorov-Smirnov statistic).
}
\end{table}

Table~\ref{sigvbova} shows our analysis of the velocity dispersion and
apparent elongations of the \emph{quartets} (thus avoiding any geometrical
source for the Tovmassian effect).
In the \emph{v2MCG} sample, there is no
correlation between group apparent elongation and velocity dispersion, while
in the 
cleaned HCG sample (Hick92/2) and the mock (from G11's SAM) CGs, 
there are weak correlations between $\sigma_v$ and $b/a$, 
but they are not statistically
significant.
However, for the Hick92/2 and G11 samples, 
the median velocity dispersion of the chain-like ($b/a<0.3$) groups is much
smaller than that of the round ($b/a>0.5$) groups, while the opposite 
behaviour is observed for the \emph{v2MCG}s. 
Yet, the effect is not significant in the
\emph{v2MCG}, only marginally significant in the mock sample, 
while the Hick92/2 sample, with
only 13 quartets, is too small to provide a statistically significant
difference in the distributions of velocity dispersions between chain-like
and round groups. We note that if we increase the Hick92/2 sample to groups
with galaxy magnitudes  
$R < 14.97$ (instead of $R<12.97$), we end up with 41 HCG quartets, for which the
rank correlation coefficient between apparent elongation and velocity
dispersion is now $r=0.29$, yielding a correlation with
97\% significance. However, for this deeper Hick92/2 sample,
the difference  in 
the distributions of velocity dispersions for chain-like and round quartets is
still not statistically significant.
\subsubsection{Bright end of the luminosity function}

\cite{Tremaine77} devised two simple, yet powerful statistics, based on
Cauchy-Schwarz inequalities, 
to test whether the first-ranked galaxies in groups and clusters were
consistent with one or several arbitrary luminosity functions.
They defined $T_1$ and $T_2$ as follows:
\begin{equation}
T_1= \frac{\sigma(M_1)}{\langle M_2-M_1\rangle} 
\ \ \ \ 
T_2={1\over \sqrt{0.677}}\,\frac{\sigma(M_2-M_1)}{\langle M_2-M_1 \rangle}
\label{trstats}
\end{equation}
where the averages are means and where
$\sigma(M_1)$ and $\sigma(M_2-M_1)$ are the standard deviations of the 
absolute magnitude of the brightest galaxy ($M_1$) and 
difference in absolute magnitude ($M_2-M_1$) between the second- and first-ranked
galaxies.  
Values of $T_1$ and $T_2$ lower 
than unity imply that the first-ranked group galaxies are abnormally
bright at the expense of the second-ranked galaxy.
$N$-body simulations indicate that galaxy mergers within physically
dense groups 
rapidly reduce the values of $T_1$ and $T_2$ below 0.7 \citep{Mamon87}.
$T_1$ and $T_2$ are biased low for samples with small number of groups, $N<50$ \citep{Mamon87_baas}.

In Table~\ref{median_v}, we find that the \emph{v2MCG} sample 
displays $T_1$ and $T_2$ significantly lower than unity:
$T_1=0.51\pm0.06$ and $T_2=0.70\pm0.06$ ($1\,\sigma$ errors from $10\,000$ 
bootstraps).
We also find such low values in our mock \emph{mvCG} sample from G11 as
well as in our four other \emph{mvCG} samples.

However, none of the other observed CG samples display low values of $T_1$
and $T_2$. 
In particular, the HCG samples show $T_1 \approx 1.2$ and $\simeq T_2 \simeq
1.0$. It appears that \cite{Hickson82} missed CGs with very
dominant galaxies \citep{PIM94,diaz-mamon10}, thus creating a spuriously
high $T_1$.
Indeed, Table~\ref{median_v} shows that $R_2-R_1$ (hence, the difference in
absolute magnitudes) has a median value
of 1.0 for the \emph{v2MCG} sample (1.3 for the \emph{mvCG}s),
but only 0.6 for the HCG samples (the means are similar). Still, part of the
difference in $T_1$ 
values is caused by the larger standard deviations of
1st-ranked absolute magnitudes in the HCG samples (0.8) in comparison with the 
\emph{v2MCG} (0.53) and G11-\emph{mvCG} (0.58) samples.

In comparison, \cite{LS06} found $T_1=0.75\pm0.1$ and
$T_2=0.86\pm0.1$ in nearby rich SDSS clusters dominated by Luminous Red Galaxies 
(LRGs), while \cite{LOM10}
recently found $T_1=0.70\pm0.01$ and 
$T_2=0.96\pm0.01$ in luminous SDSS clusters, but
$T_1=0.84\pm0.01$ and 
$T_2=0.94\pm0.01$ for low luminosity ones.

\begin{figure}
{\includegraphics[width=\hsize]{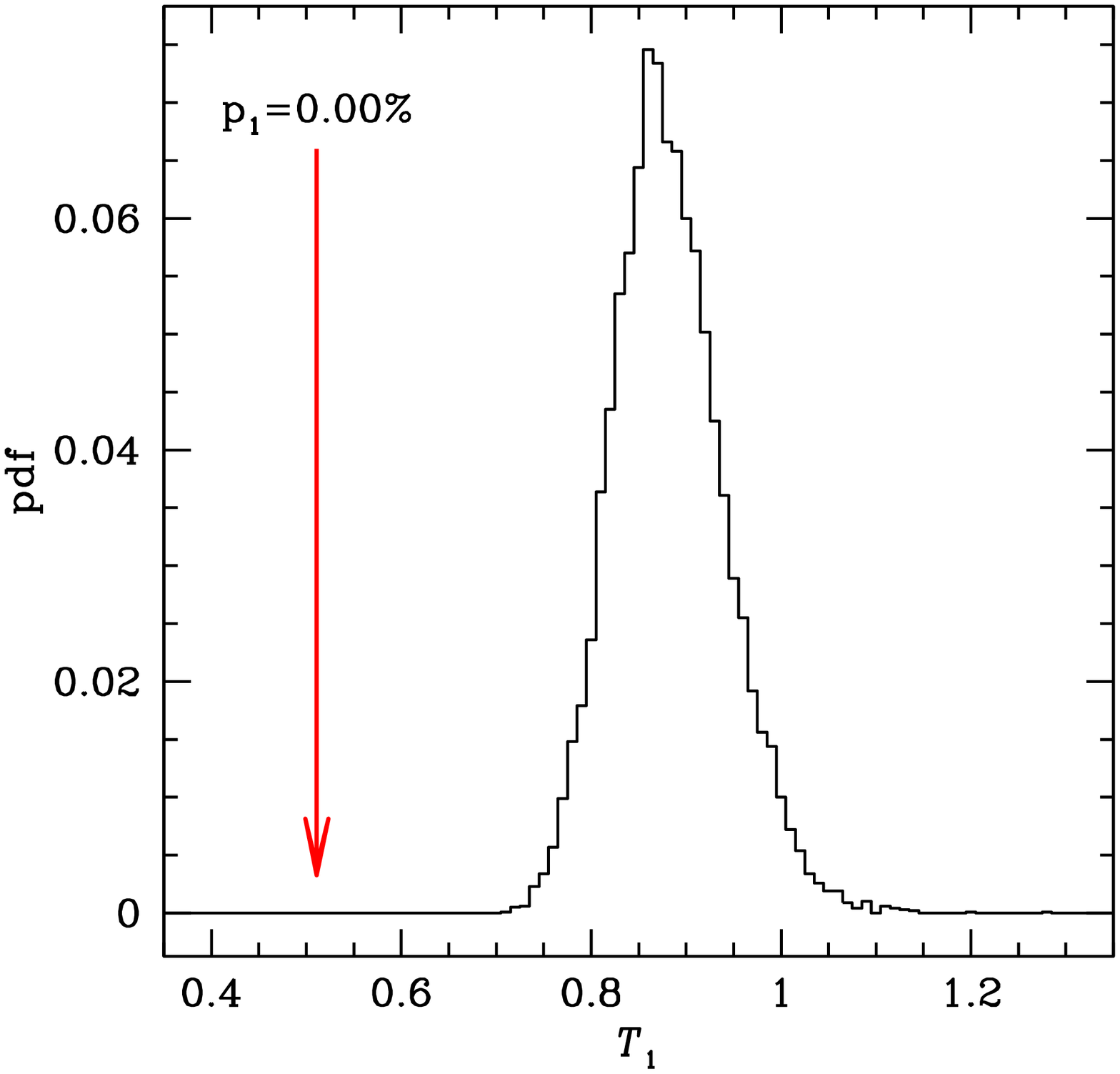}
\includegraphics[width=\hsize]{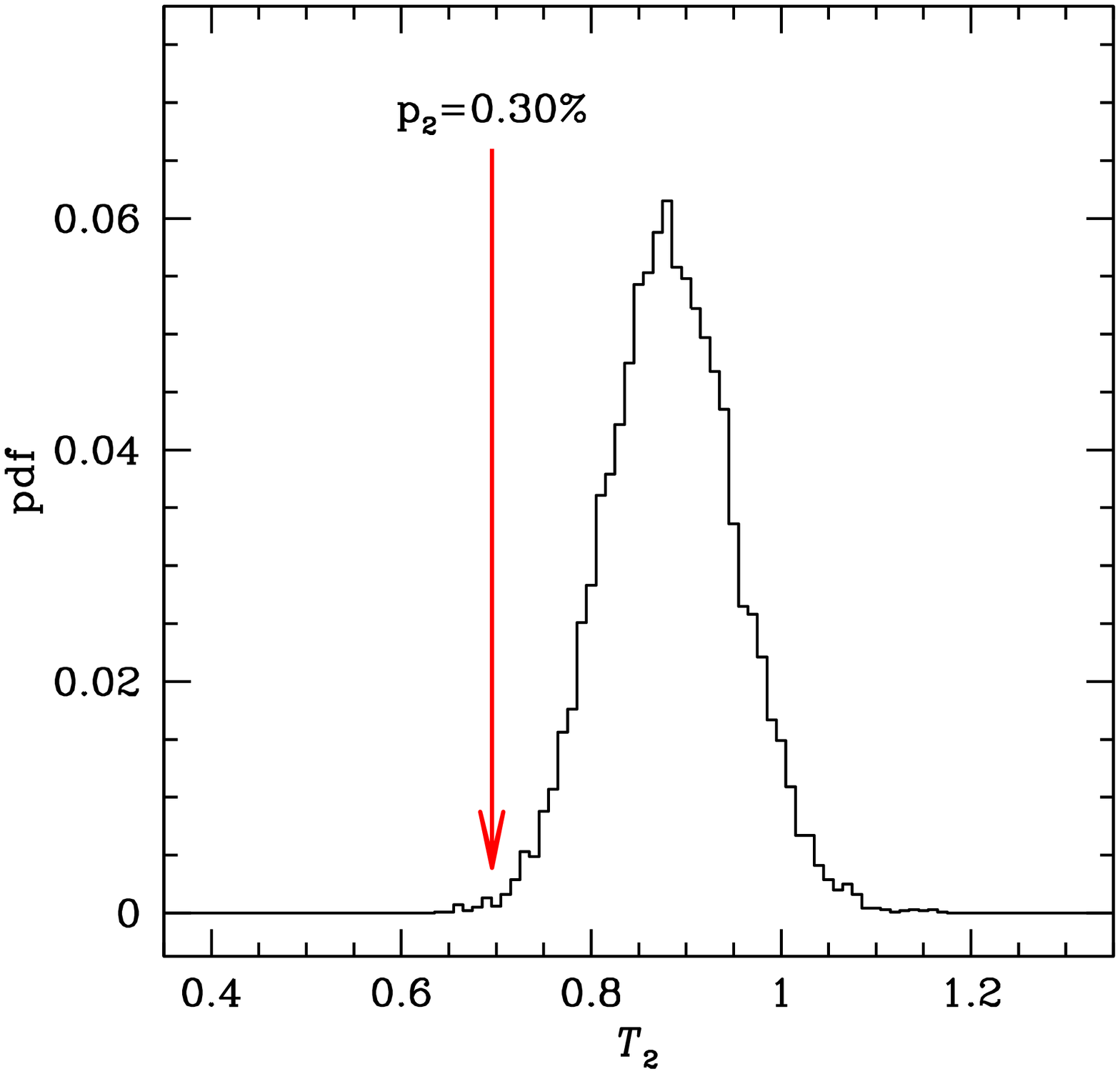}}
\caption{Distributions of Tremaine-Richstone statistics for $10\,000$
  Monte-Carlo realisations. Vertical arrows indicate the values
  observed in the \emph{v2MCG} catalogue ($v > 3000 \, \rm km \, s^{-1}$).
\label{TR_MC}
}
\end{figure}

We can also quantify how significant are the deviations of $T_1$ and $T_2$
from unity using a Monte-Carlo technique (see also \citealp{LOM10}).
We built mock CGs by adopting the absolute magnitudes of the first-ranked
\emph{v2MCG}s and adding to them the absolute magnitudes of galaxies chosen
at random from the 2MRS catalogue, but with 3 limitations: 
1) in the same range of redshifts
(velocities $1000 \, \rm km \, s^{-1}$ from that of the first-ranked);
2) absolute magnitudes, $M_K$, in the range of the group: $M_K^{\rm grp-1}\leq M_K \leq + M_K^{\rm
  grp-1}+3$, where $M_K^{\rm grp-1}$ is the absolute magnitude of the
first-ranked group member;
3) positions more than 5 degrees from the group (in Declination only for a
faster run).
The velocity criterion ensures that the flux limit of 2MRS is
properly handled, while the position criterion ensures that a
first-ranked galaxy is not duplicated in its mock group. 
In the end we thus generate mock CGs with the same multiplicity function and
distribution of most luminous absolute magnitudes.
We did not consider a surface brightness threshold on our Monte-Carlo groups (assuming that the galaxies are located in the same positions as in the observed sample), because this would increase the discrepancy between the observed values of $T_1$ and $T_2$ with those from our Monte-Carlo samples. Indeed, since we start with the brightest group galaxy, if we enforced a minimum group surface brightness, we would tend to reject groups with only one luminous member,
hence leading us to lower differences between second and first-ranked absolute magnitudes, and therefore larger values of $T_1$ and $T_2$.
We compute $T_1$ and $T_2$
for this mock sample of CGs and iterate to build a total of $10\,000$ samples.

The distribution of $T_1$ and $T_2$ for the $10\,000$ mock catalogues can
be seen in Fig.~\ref{TR_MC}. Let $p_i$ be 
the
fraction of Monte-Carlo realisations that have $T_i$ as low
as the observed value.
We found
$p_1=0$ (i.e. $p_1<0.01\%$) and $p_2=0.3\pm0.06\%$, i.e. none of our mocks reached
values of $T_1$ and $T_2$ both as low as
observed in \emph{v2MCG}s\footnote{Note that \cite{LOM10} also considered the compatibility
  of the distribution of 
observed values of $T_1$ and $T_2$ of clusters using bootstrap resampling with the
predicted distribution obtained with mocks. This incorrectly accounts twice
for the finite sample. One should either use bootstraps to provide error bars
on the observed $T_1$ and $T_2$ and compare to the mean prediction of the
mocks or conversely compare the observed  $T_1$ and $T_2$ without error bars
to the distribution of the mocks, but not do both.}.

All this confirms that \emph{the \emph{v2MCG} is the only observed compact group
  sample that has differences between first- and second-ranked absolute
magnitudes that are inconsistent with random sampling of luminosity
functions}, in agreement with the expectations from cosmological simulations.
 
\subsubsection{Luminosity segregation}

In the standard galaxy formation model used for SAMs, the brightest group
galaxies are centrally located (see \citealp{Skibba+11} for the
quantification and limits of
this idea). Indeed, N-body simulations of virialised dense groups show that
such luminosity segregation rapidly sets in \citep{Mamon87}. Moreover, the
two-body relaxation times in dense groups of galaxies, of order of the number
of galaxies times the orbital time, both of which are small, are expected to
be much smaller than the age of the Universe, hence galaxies should exchange
their orbital energies and reach equipartition on short timescales.
If CGs are caused by chance alignments, then one does not expect
to witness such luminosity segregation. \cite{Mamon86} measured luminosity 
segregation in HCGs, using the exact same technique as he used in the simulations: 
stacking the groups and searching for a correlation (with the Spearman rank test) 
between the fraction of group luminosity in the galaxy (hereafter, fractional luminosity) 
versus the projected distance relative to the group centroid 
(unweighted barycentre)
in units of the median of 
these distances 
per group
(hereafter, normalised radial coordinate).
The absence of luminosity segregation in HCGs, measured in this fashion, 
produced for him another argument that HCGs were heavily
contaminated by chance alignments.

Here, we performed the same analysis on the different observed and mock samples of CGs.
We first found that \emph{v2MCG}s show significant anti-correlation between
fractional luminosity and normalised distance: 
Spearman rank correlation $r=-0.19$. 
According to the Spearman rank correlation test, 
an anti-correlation at least as 
strong as this observed one has less than 0.1\% probability of arising 
by chance (see Table~\ref{median_v}). 
This is also the case in the mock \emph{mvCG} sample. 
One may argue that SAMs have
luminosity segregation within them by construction, since in SAMs, galaxies
form at the centre of a halo. But none of the other observed CG samples show
any significant sign of luminosity segregation, and this is not just a case
of poorer statistics, as the correlation coefficient between
fractional luminosities and normalised distances in the \emph{v2MCG} is much
more negative than in all other observed samples.

\begin{figure}
\includegraphics[width=\hsize]{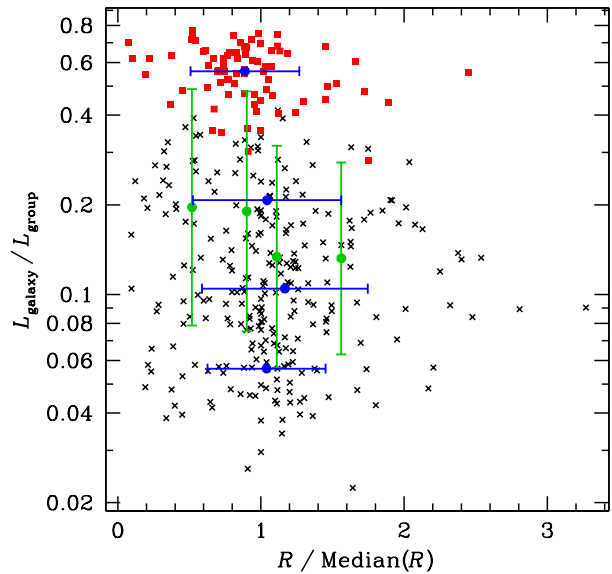}
\caption{Luminosity segregation (fractional luminosity in group vs. normalised radial coordinate relative to unweighted barycentre)
in the \emph{v2MCG} sample (restricted to $v
 > 3000 \, \rm km \, s^{-1}$).
The \emph{red squares} indicate the first-ranked galaxies, while the \emph{black crosses} show the other group galaxies.
The 
horizontal
and 
vertical
error bars show the standard deviations in equal number
subsamples of normalised radius and fractional luminosity, respectively.
Despite the large scatter, the rank correlation coefficient is $r=-0.19$ and
has only a probability $3\times10^{-4}$ of occurring by chance.
\label{lsegplot}}
\end{figure}
We also compared the stacked distributions of normalised distances between
first- and second-ranked galaxies, as well as between second- and
third-ranked galaxies, using the Kolmogorov-Smirnov (KS) test.
As seen in the last two rows in Table~\ref{median_v}, in \emph{v2MCG}s, 
the first-ranked galaxy is significantly more centrally located than the
second-ranked galaxies: according to the KS test, the probability that this
would occur by chance is again less than 0.1\%.
This is also seen in the \emph{mvCG}s, while
no such significant trends are seen in the other
group catalogues (including the HCG). 
Hence, \emph{The \emph{v2MCG} is the only CG sample to show that the most
  luminous galaxies are significantly more centrally located, in accordance
  with the mock CGs from SAMs, and contrary to what has been observed in all
  other CG sample.}

Interestingly, there are no statistically significant signs for different
distributions of normalised distances between second- and third-ranked
galaxies. In other words, while the first-ranked galaxies are in general 
more centrally located than the second-ranked, 
the latter are not more centrally located than the third-ranked.
This lack of luminosity segregation among the non-brightest galaxies 
may occur because tides from the parent group potential, may
limit the sizes hence luminosities of the galaxies as they approach the
central one (whose central location renders it immune to such tides). Then,
the second most luminous galaxy will preferentially lie at the outskirts,
while the third ranked one, will tend to lie closer because of this tidal
limitation. Therefore, for the non-brightest galaxies, 
group tides may cancel the effect of luminosity
segregation.

Figure~\ref{lsegplot} illustrates the luminosity segregation in the
\emph{v2MCG} sample.  
The fractional group luminosities appear to be enhanced within the median projected distance 
from the group centroid (i.e. abscissae smaller than unity).
 However, the normalised distances are only smaller in the galaxies in the 25\% highest 
quartile of fractional luminosity, which roughly corresponds to the first-ranked galaxies.

\section{Conclusions and Discussion}
\label{conclude}

In this work we have catalogued a new sample of CGs from the
2MASS survey and we have compared them with existing CG samples.

Following the criteria defined by \cite{Hickson82}, we have identified 230 CGs in projection 
in the $K$-band covering  $23\,844 \, {\rm deg}^2$. 
This catalogue has well defined criteria which produced an homogeneous 
sample useful to perform statistical analyses on it.
25\% of them (57 CGs) were previously identified in other catalogues as 
compact groups, triplets of galaxies or interacting galaxies. 
A total of $144$ \emph{p2MCG}s have \emph{all} their members
with redshifts available in the literature, and among them $85$ groups have 4 or more accordant 
galaxies, which makes this catalogue the largest sample of CGs with 4 or 
more spectroscopically confirmed members. 
The percentage of groups with accordant galaxies (59\%) is slightly lower than
that obtained from the HCG sample (67\%), 
and very similar to the predicted by \cite{diaz-mamon10} 
from the semi-analytical models of \cite{Bower+06} and
\cite{dLB07}.\footnote{For the SAM of \cite{Guo11}, we find that 70\% of mock
  CGs found in projected space survive the velocity filtering.}

As a side note, we have now built additional mock CG catalogues using the
\cite{Croton+06} SAM in the $K$-band and the \cite{Guo11} SAM in the $r$
band, where the latter was run on the Millennium-II Simulation, which has 512
times the mass resolution of the Millennium Simulation. For both samples, we
found that two-thirds of the mock CGs were physically dense systems of at
least 4 galaxies of accordant magnitudes, while the
remaining third was caused by chance alignments of galaxies along the
line-of-sight, mostly within larger virialised groups, confirming similar
conclusions of \cite{diaz-mamon10}.

In comparison with other CG catalogues, the \emph{v2MCG} 
catalogue presented in this work is 
one of the nearest and brightest samples of CGs, 
although these CGs have larger projected radii and 
interparticle separations. 

The \emph{v2MCG} does not show any significant correlation for quartets
between apparent elongation and velocity dispersion nor significantly larger
velocity dispersion in round groups relative to chain-like groups, contrary to what
\cite{TMT99} claimed in HCGs.

The \emph{v2MCG} is the only CG sample to display significantly large
differences between second- and first-ranked absolute magnitudes (from
Tremaine-Richstone statistics) as well as
centrally located first-ranked galaxies, both in agreement with mock
\emph{mvCG}s, but in sharp contrast with all other observed velocity-filtered
CG samples.

Galaxy mergers are an obvious way to decrease $T_1$ and $T_2$
\citep{Mamon87}, and we cannot think of any other physical mechanism that may
cause both $T_1$ and $T_2$ to be significantly smaller than unity in a group catalogue.

One major difference of our sample with others is that ours has many more
groups with dominant galaxies accounting for over half the total
luminosity. While this increases the gap between first and
2nd-ranked magnitudes, we found that our sample also has a small standard
deviation of first-ranked absolute magnitudes, which enhances the
significance of the Tremaine-Richstone $T_1$ statistic.

Why don't we find significant magnitude gaps and luminosity segregation in the
other CG samples? It is clear that in his visually search for CGs, \cite{Hickson82}
missed groups with dominant galaxies (\citealp{PIM94,diaz-mamon10} and
Table~\ref{median_v}). 
Could those \emph{v2MCG}s in common with \emph{vHCG}s have weaker signs of
magnitude gaps and luminosity segregation? One expects that if mergers cause
the magnitude gap, the masses, i.e. stellar masses, of the galaxies are the
crucial variable. Similarly, if luminosity segregation is produced by dynamical
friction or by energy equipartition from two-body relaxation, the galaxy
(stellar) masses should be the important variable.
Therefore, the magnitude gaps and
luminosity segregation should be weaker in the $R$ band, where the luminosity is
less a measure of stellar mass than in the $K$-band.

Unfortunately, we have only 14 groups in common,
among which 10 (HCG 7, 10, 23, 25, 40, 58, 86, 88, 93, 99)\footnote{and with
  slightly variations: HCG 15, 16, 51, 97.}
have exactly the same galaxies. 
For these 10 groups, we find $T_1=0.68\pm0.24$
and $T_2=0.87\pm0.24$ in the $K$ band, while in the $R$ band we find values greater than
unity: $T_1 = 1.75\pm1.69$ and $T_2=1.33\pm0.39$. So, indeed, the $K$-band
luminosities are more sensitive than their $R$-band counterparts 
to the magnitude gap, but given the bootstrap
errors, the large differences in $T_1$ and $T_2$ between $R$ and $K$-based
absolute magnitudes are not statistically significant (for $T_2$ the
difference is roughly $1\,\sigma$, while it is much less for $T_1$).
On the other hand, luminosity segregation is not seen in either waveband:
worse it is inverted, with
the brightest galaxy on average further away from the group centroid than the
2nd-brightest galaxy.

The other two CG samples, UZC-CG and LCCG, are based upon Friends-of-Friends
(LCCG) or similar (UZC-CG) algorithms, both with velocity linking length of $1000
\, \rm km \, s^{-1}$. Such a velocity link is much more liberal than imposing
that the velocities all lie within $1000 \, \rm km \, s^{-1}$ from the median
as done in \cite{HMdOHP92} and here. Indeed, according to
Table~\ref{median_v}, the median velocity dispersion
of UZC-CG groups of 4 or more galaxies is $295 \, \rm km \, s^{-1}$,
i.e. 25\% greater than in our sample. This suggests that the UZC-CG sample
is more contaminated by chance alignments of galaxies along the line-of-sight (as
\citealp{Mamon86} had suggested for the HCG sample) than is our sample. 
Moreover, the UZC-CG has a liberal linking length on projected distances of $200 \, h^{-1} \,
\rm kpc$, making these groups not so compact (as can be checked by their low
mean group surface brightness as seen in Table~\ref{median_v}).
Finally, the UZC-CG is based upon Zwicky's visually estimated magnitudes,
which may carry rms errors as large as 0.5 mag, thus washing out in
part the effects of the magnitude gap and luminosity segregation.

On the other hand, the LCCG sample
(again, restricted to groups with at least 4 members) has
a very similar median velocity dispersion to our sample (and a similar median
mass-to-light ratio). Note that the linking length for projected distances of
the LCCG is
only $50 \, h^{-1} \, \rm kpc$, i.e. 4 times less than in UZC-CG.
The problem with the LCCG is that its parent catalogue (the LCRS,
\citealp{Shectman+96}) 
is a collection of two samples with $16.0 < R < 17.3$ and
$15.0 < R < 17.7$. Thus the magnitude range is very restricted. Hence, it is not
a surprise that $\langle M_2-M_1\rangle$ is half our value
(Table~\ref{median_v}), leading to $T_1>1$ and $T_2>1$.

What does this tell us on the nature of the groups in the different CG
samples?
Over 25 years ago, \cite{Mamon86}  found $T_1=1.16$ and no signs of luminosity
segregation in the largest sample then available of 41
velocity-filtered HCGs with at least 4 members. This was in sharp 
contrast with the low values of $T_1$ and
significant luminosity segregation  he was
finding in coalescing dense groups \citep{Mamon87}. This provided him 
with two arguments (among several) 
to conclude that most HCGs were caused by chance
alignments of galaxies within larger groups \citep{Mamon86}.
As confirmed here with the SAM by \cite{Guo11}, roughly two-thirds of mock
CGs are physically dense (\citealp{diaz-mamon10} and Table~\ref{tabSAMs}
above). 
The statistically large magnitude gaps and luminosity segregation in both the
observed \emph{v2MCG}s and the mock \emph{mvCG}s suggest that
\cite{Mamon86} was misled by the bias of the HCG sample against large gaps
into concluding that most of them were not physically dense.

So the \emph{v2MCG} appear to be mostly \emph{bona fide} physically dense
groups. But can we conclude that the other CG samples are dominated by
chance alignments? \cite{diaz-mamon10} attempted to build sample of mock HCGs
that include the same biases as they had measured by comparing with the three
SAMs that they had built mock CGs from. They found that the same fraction
(if not slightly higher) of the mock (biased) HCGs were physically
dense. The nature of the groups in other CG samples could be studied in
similar ways, using mock CG samples from cosmological galaxy formation
simulations, mimicking their selection criteria
and observational select effects.

Could the lack of HCGs with strongly dominant brightest galaxies prevent the
visibility of luminosity segregation? 
We performed KS tests to compare the distribution of relative positions of
1st and 2nd-ranked group members for subsamples split between those dominated
by 1st-ranked members (`Dom') and those with galaxies of more comparable
luminosities (`Non-Dom'), making our splits at the median magnitude difference $\langle
M_2-M_1\rangle$.
We performed this analysis for the HCG and v2MCG samples as well as for the
C06K and G11 \emph{mvCG} samples.
\begin{table}
\begin{center} 
\caption{Luminosity segregation split by magnitude gap\label{lsegsplit}}
\begin{tabular}{llrc}
\hline
\hline
Catalogue & subsample & $N$ & $P_{\rm KS}$ \\
\hline
\emph{v2MCG} & Dom & 39 & $2.9\times10^{-7}$ \\
\emph{v2MCG} & Non-Dom & 39 & (3.9\%) \\
\emph{vHCG} & Dom & 17 & 67\% \\
\emph{vHCG} & Non-Dom & 16 & (6.6\%) \\
\emph{mvCG}-G11 & Dom & 163 & $1.2\times 10^{-11}$ \\
\emph{mvCG}-G11 & Non-Dom & 163 & 1.0\% \\
\emph{mvCG}-C06K & Dom & 223 & $4.0\times 10^{-9}$ \\
\emph{mvCG}-C06K & Non-Dom & 225 & 0.4\% \\
\hline
\end{tabular}
\end{center}
\parbox{\hsize}{
\small
{\bf Notes.} Dom and Non-Dom subsamples are those with $M_2-M_1$ above and
below the median value of the full sample, respectively.
$N$ is the number of groups in the subsample.
$P_{\rm KS}$ is the KS probability that a difference in the distributions of
normalised distances to the non-weighted group centre is greater than
`observed' by chance. Values of $P_{\rm KS}$ given in parentheses denote
reverse luminosity segregation: the 2nd-ranked galaxy is more centrally
located than the first-ranked.}
\end{table}
Table~\ref{lsegsplit} shows that indeed luminosity segregation is much
stronger for all catalogues in the Dom subsamples, and statistically
significant in all of them except the \emph{vHCG}. Surprisingly, while the
Non-Dom subsamples of the 
two mock CG samples display much weaker, but still statistically significant
luminosity segregation, the Non-Dom subsamples of both the \emph{v2MCG} and
\emph{vHCG} catalogues display \emph{reversed luminosity segregation}: the
2nd-ranked galaxy is more centrally located than the (slightly more luminous)
first-ranked galaxy. We can only see one explanation for this reverse luminosity segregation, if it occurs in wavebands bluer than $K$:
late-type galaxies that are second-ranked in stellar mass, hence not centrally located, can end up more luminous (thanks to their efficient star formation) than early-type galaxies of slightly higher stellar mass.
However the effect is also present in the $K$-selected \emph{v2MCG}, and with
even greater statistical significance
(96.1\% vs. 93.7\% confidence for the Non-Dom CGs of the
\emph{v2MCG} and \emph{vHCG} catalogues, respectively). So, we can only
explain this marginal effect as a statistical fluke.

Nevertheless, the absence of luminosity segregation in the HCG catalogue could be
consistent with their physical reality because the sample is too small to
detect the weak luminosity segregation expected from the mocks. Moreover, if
the reverse luminosity segregation for Non-Dom groups is real, then the lack
of groups with very dominant galaxies in the HCG (caused by the visual
selection bias) would cause the Non-Dom groups to cancel the luminosity
segregation of the Dom groups.

In conclusion,
the \emph{v2MCG} sample has numerous advantages over other CG samples:
\vspace{-0.5\baselineskip}
\begin{enumerate}
\itemindent 0pt
\item It is the largest available sample of velocity-filtered groups of at least
4 members of comparable luminosity (3 mags, i.e. factor 16).
\item It has an isolation criterion (in contrast with other CG samples except
for the HCG).
\item It is automatically extracted (contrary to the HCG).
\item It has a well-defined magnitude limit (which the HCG sample does not).
\item It is deep enough (which some may find surprising given the shallowness of
its parent 2MASS catalogue) to have a selection on brightest galaxy
magnitude, so as to ensure that all groups can span the maximum allowed
magnitude gap of 3.
\item It is selected by stellar mass ($K$ band), which is expected to be 
a better tracer for magnitude gaps and luminosity segregation (among other things).
\item It is the only sample to show statistical signs of mergers (magnitude gaps) and
luminosity segregation, expected in physically dense groups (in contrast with
all other CG samples).
\end{enumerate}
The last point implies that the \emph{v2MCG} is the only CG
sample for which one is reasonably sure that it is dominated by physically
dense groups.
For all these reasons, the \emph{v2MCG} appears to be the sample one ought to
study in depth to probe the effects on galaxies of this unique environment of 4 galaxies
of comparable luminosity lying close together in real space.

As a next step in this project, 
we are in the process of measuring redshifts for the members 
for which no spectra are available, and we are continuing our statistical
studies of the \emph{v2MCG}s and their galaxies.

\section{Acknowledgments}
We thank the anonymous referee for helpful comments that improved this work.

We dedicate this article to John P. Huchra (1948--2010), who among his numerous contributions to
astronomy, played a crucial role in obtaining redshifts for the galaxies in groups
in general and HCGs in particular.
%
%
We thank Roya Mohayaee and Guilhem Lavaux for useful discussions on peculiar
velocities and Igor Chilingarian on $k$-corrections and internal extinctions.

This publication makes use of data products from the Two Micron All Sky Survey, 
which is a joint project of the University of Massachusetts and the Infrared 
Processing and Analysis Centre/California Institute of Technology, 
funded by the National Aeronautics and Space Administration and the National Science Foundation. 
This research has made use of VizieR and Aladin at the Centre de Donn\'ees
astronomiques de Strasbourg (CDS).
This research made use of the ``K-corrections calculator'' service available
at http://kcor.sai.msu.ru/. An SM version ({\tt kcorr\_Chilingarian}) is available at ftp://ftp.iap.fr/from\_users/gam/SOFT/gam\_macros.
The Millennium and Millennium II Simulation databases used in this paper and the web application providing online access to them were constructed as part of the activities of the German Astrophysical Virtual Observatory
This work was partially supported by Consejo de Investigaciones Cient\'{\i}ficas y
T\'ecnicas de la Rep\'ublica Argentina (CONICET) and Funda\c c\~ao de Amparo \`a Pesquisa do Estado do S\~ao Paulo (FAPESP).
CMdO acknowledges financial support from FAPESP and CNPq.

\bibliography{cgs}

\appendix

\section{Transformation from $\mu_R$ to $\mu_K$}

\label{mu_k} In the visual search performed by \cite{Hickson82} on the photographic plates of POSS-I, 
he established a cutoff in surface brightness of $\mu_E = 26  \ \rm mag \ arcsec^{-2} $.
The POSS-I $E$ band roughly corresponds to the more familiar Cousins $R$ band.
As the galaxy database used for our search is selected in the $K_s$ band, we converted 
the original limit of \cite{Hickson82} to a corresponding one for $K$ magnitudes.
The $R$--$K$ colours\footnote{we drop the `s' subscript on the $K_s$ band for clarity.}  of galaxies depend on their luminosity (colour-luminosity relation) and morphology (e.g. Red Sequence vs. Blue Cloud).

We cross-identified the SDSS DR7 model $g$ and $r$ magnitudes, $A_g$ and $A_r$ extinctions, and redshifts with the 2MASS XSC $K$20 isophotal $J$ and $K$ magnitudes, with a maximum separation of $2''$ between the positions of the galaxies in the two catalogues. We corrected the 2MASS magnitudes for galactic extinction using the $A_g$ values of SDSS, assuming $A_g/A_V=1.256$, $A_r/A_V=0.798$, and $A_{K_s}/A_V=1.16$ from spline fits of $\log A_\lambda$ vs. $\log \lambda$ tabulated by \cite{CCM89} and $A_J/A_V=0.282$ directly from their table.
We then $k$-corrected the SDSS $r$ and 2MASS $K_s$ extinction-corrected magnitudes using the redshifts and extinction-corrected $g$--$r$ (SDSS) and the $J$--$K_s$ (2MASS) colours using the transformations of \cite{CMZ10}. This enabled us to derive extinction- and $k$-corrected $(r$--$K)^0$ colours.

In a first pass, we adopt the conservative $(r$--$K)^0$=2.33, which, with
$(r$--$R)^0$=0.33 \citep{diaz-mamon10}  yields $\mu_K \leq 24\,\rm
mag\,arcsec^{-2}$. Once we extract the \emph{p2MCG}s and then the
velocity-filtered \emph{v2MCG}s with this compactness limit, we find that the
mean galaxy luminosity in our \emph{v2MCG}s is $M_K=-23.40+5\,\log h$ 
and our median \emph{v2MCG} mean velocity is $5927 \, \rm km \, s^{-1}$, corresponding to $z=0.020$. 
\begin{figure}
\includegraphics[width=\hsize]{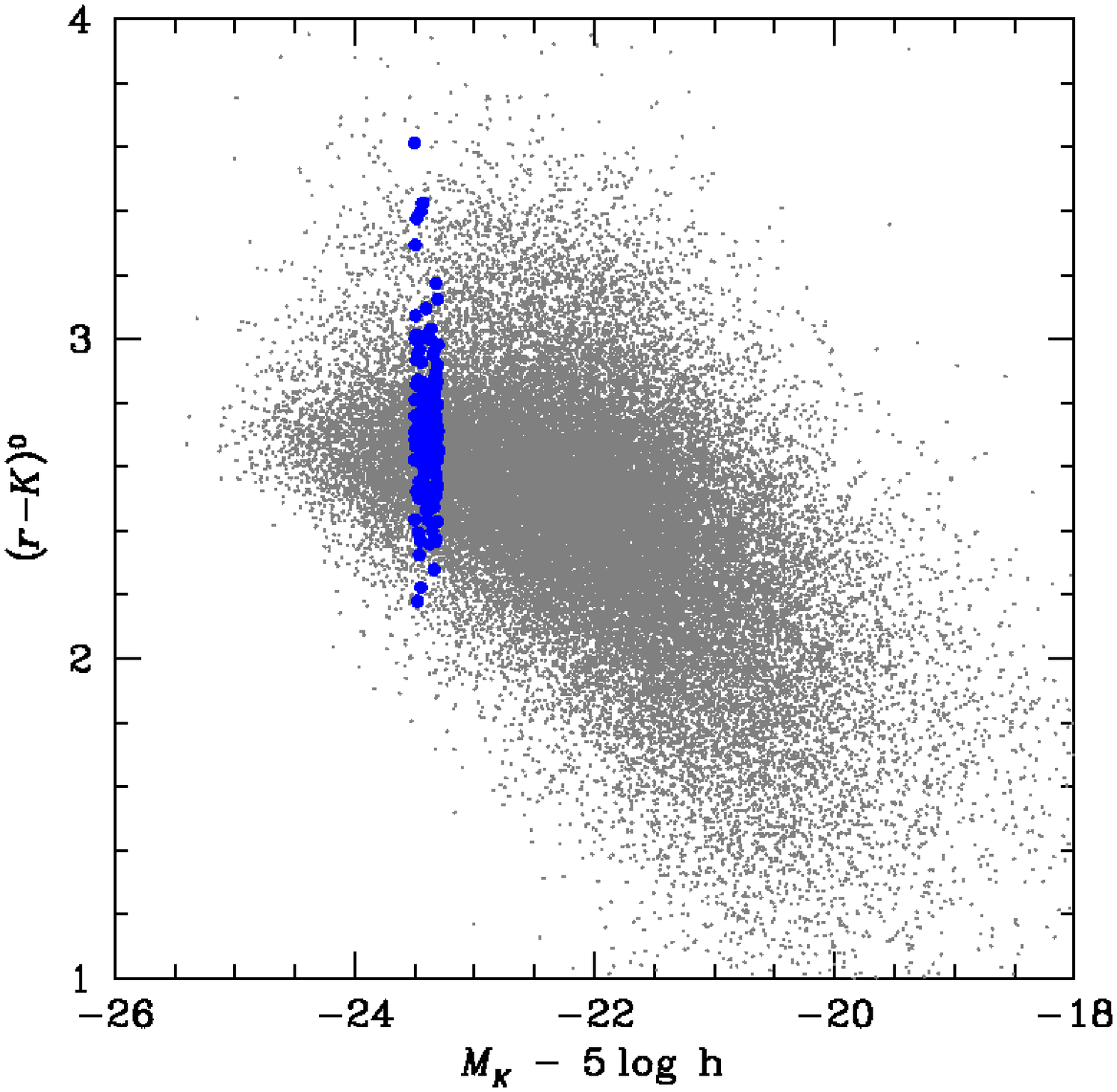}
\caption{Colour-luminosity relation for SDSS-2MASS matches. The \emph{grey
    points} show all 45974 non-flagged galaxies with $0.005 < z < 0.05$, $13
  < r<17.77$, $K<13.57$, while the \emph{blue points} use the additional
  criteria $M_K-5\,\log h=-23.40\pm0.10$ 
and $z=0.020\pm0.005$.
\label{colmag}}
\end{figure}
 In a second pass, we consider the $(r$--$K)^0$ colours for the 45974 galaxies among the 326320 SDSS-2MASS matches, with SDSS {\tt zWarn}=0, $13 < r < 17.77$, and 2MASS artifact flag {\tt cc\_flg}=0 and both $J$ and $K$ confusion flags, respectively {\tt j\_flg\_k20fe} and {\tt k\_flg\_k20fe}, equal to 0
(grey points of Fig.~\ref{colmag}). Restricting these matched galaxies to
those with $M_K-5\,\log h= -23.40\pm0.10$ and $z=0.020\pm0.005$ (large blue
points in Fig.~\ref{colmag}) yields a median $(r-K)^0$ of $2.72\pm0.04$
(assuming that the error on the median is $1.253\, \sigma/\sqrt{N}$, valid
for large Gaussian distributions) for 164 galaxies. 

With $(r$--$R)^0$=0.33, this yields $(R$--$K)^0$=2.39. For clarity, we therefore
assume $R=K+2.4$ and adopt a compactness limit of $\mu_K=23.6$. 
The
\emph{v2MCG}s obtained with this new compactness limit have very similar 
median redshifts (now 1.5\% larger), although the median $R$-band group luminosities
are now one-third lower (mainly because of the additional 0.4 magnitude correction
from $K$ to $R$).

\section{Comparison with different SAMs}
\label{props_SAMs}
\begin{figure*}
\centering
{\includegraphics[scale=0.43]{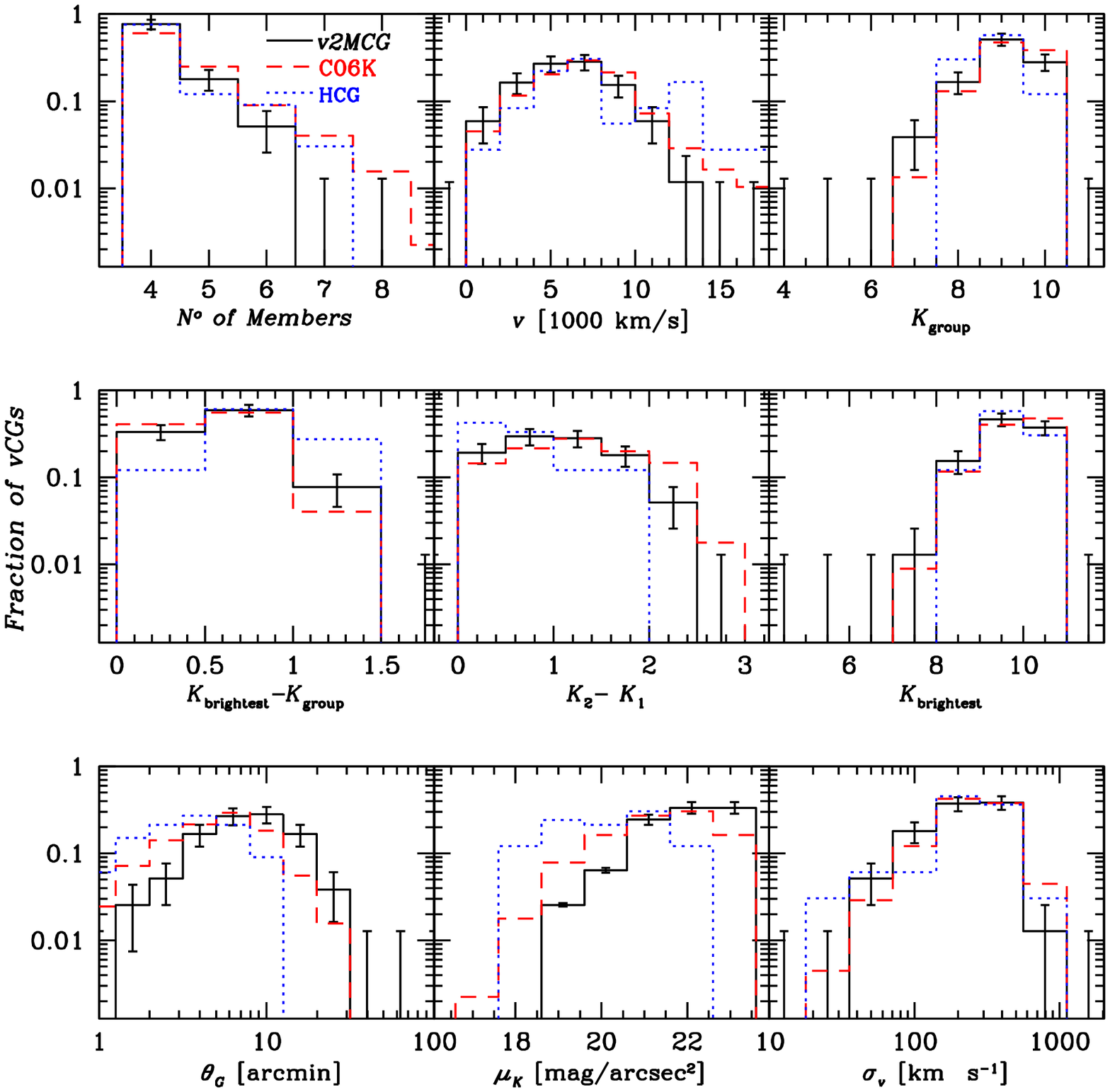}
\includegraphics[scale=0.43]{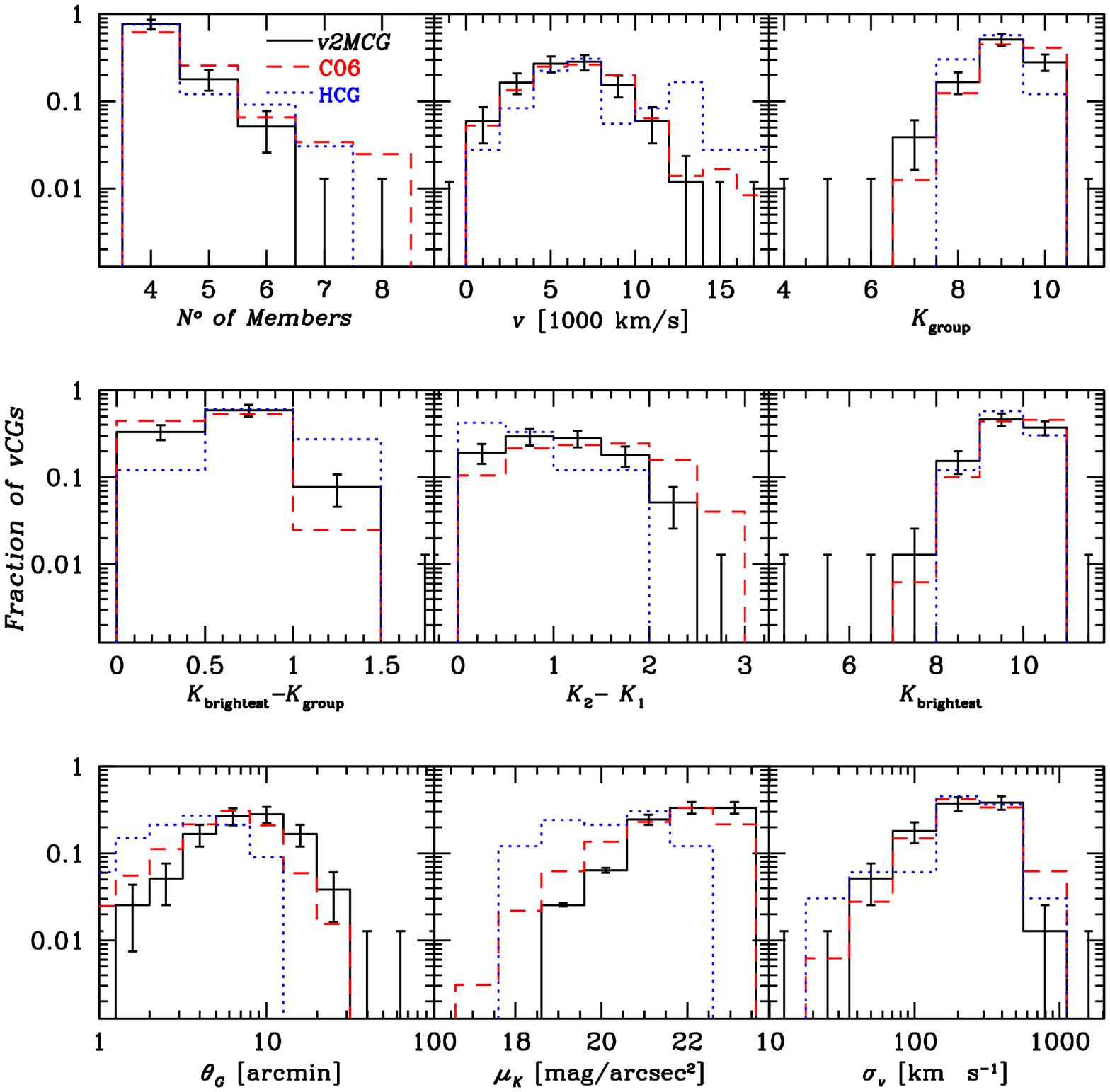}}
\vskip -0.39cm
{\includegraphics[scale=0.43]{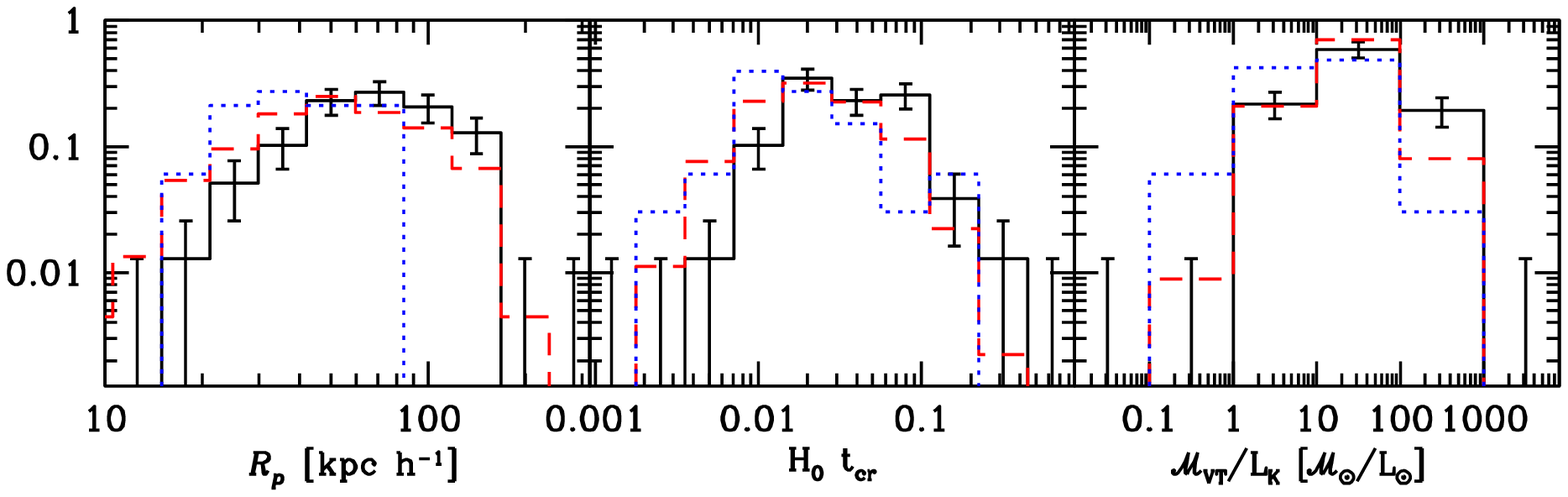}
\includegraphics[scale=0.43]{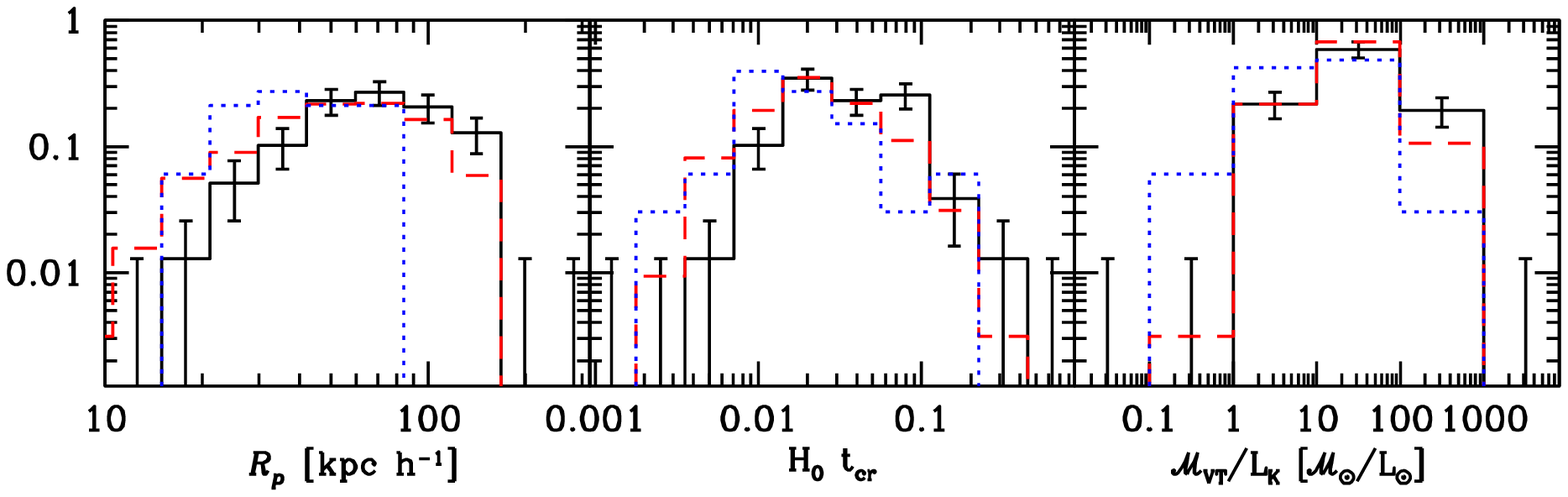}}
{\includegraphics[scale=0.43]{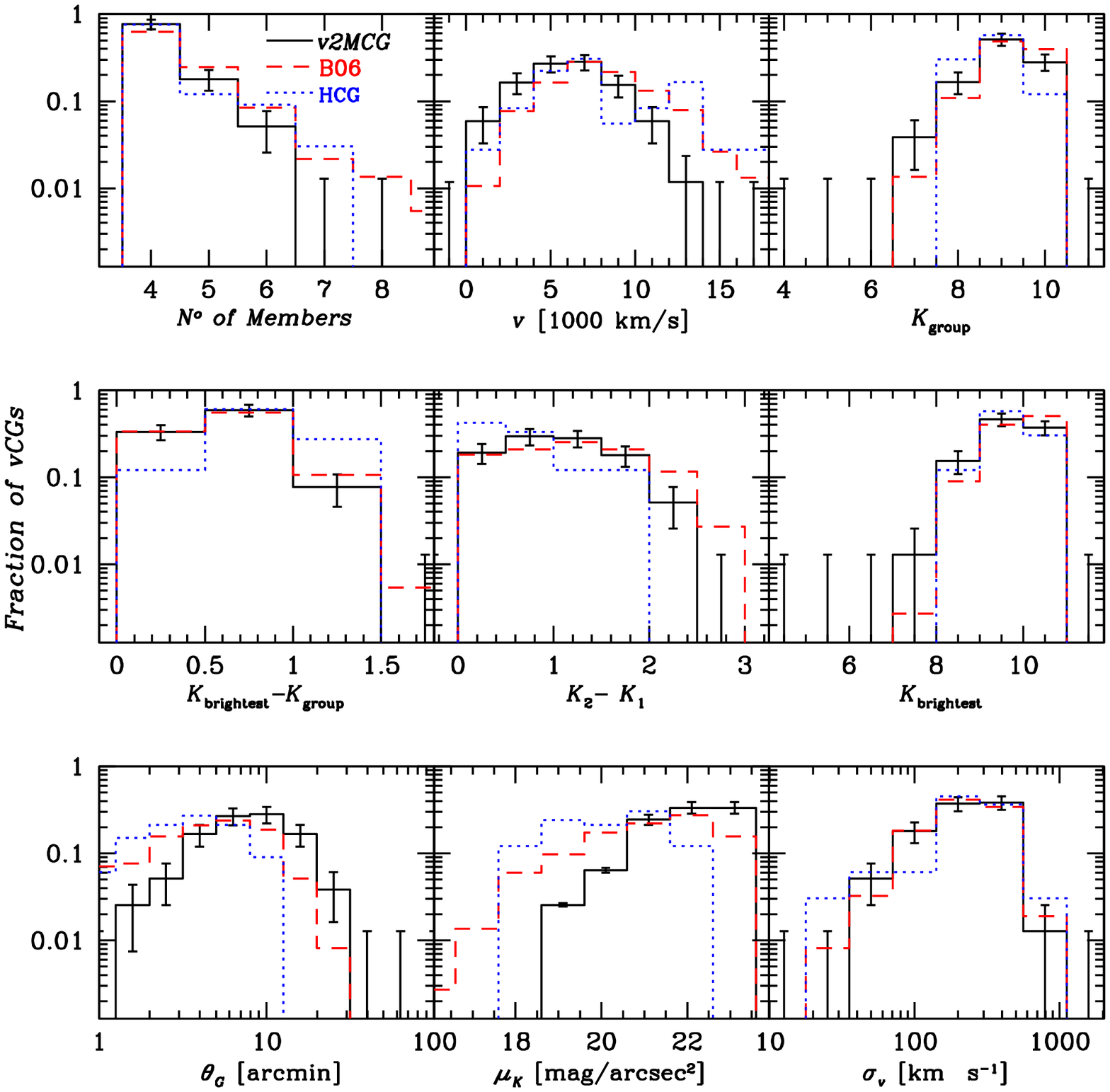}
\includegraphics[scale=0.43]{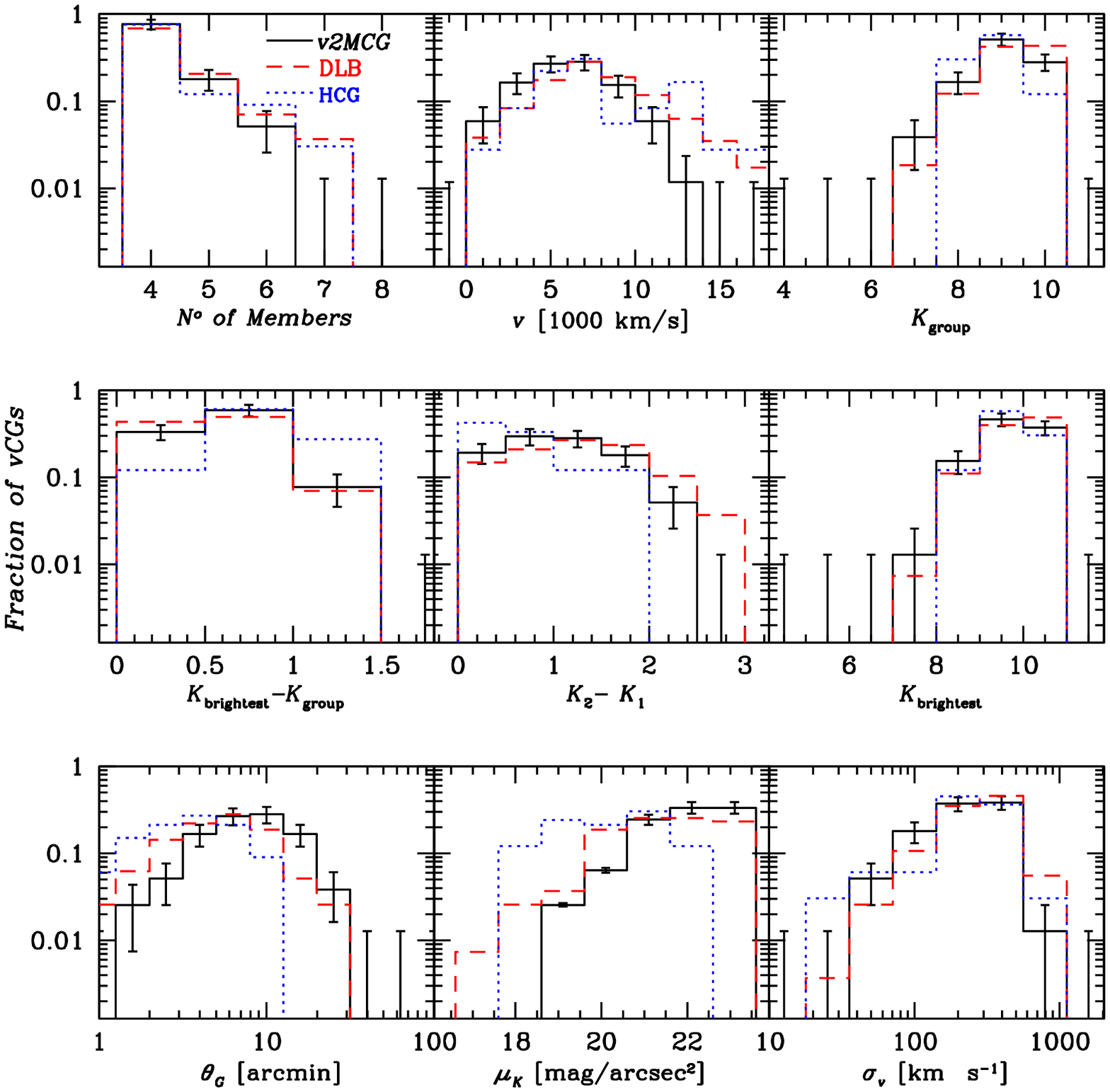}}
\vskip -0.39cm
{\includegraphics[scale=0.43]{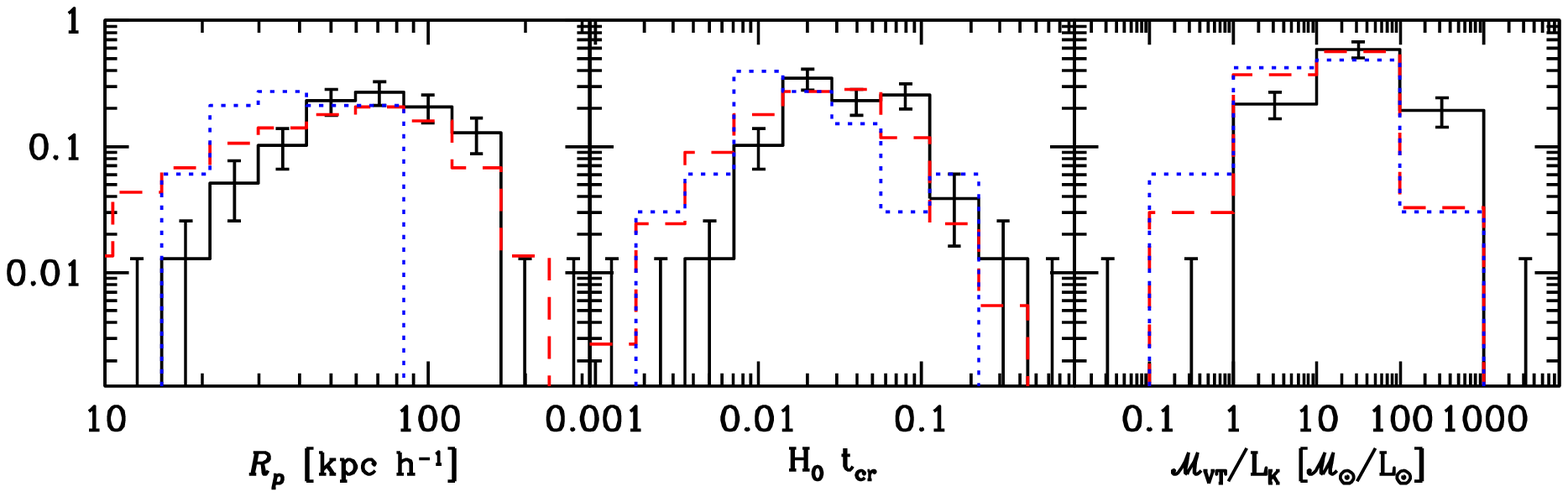}
\includegraphics[scale=0.43]{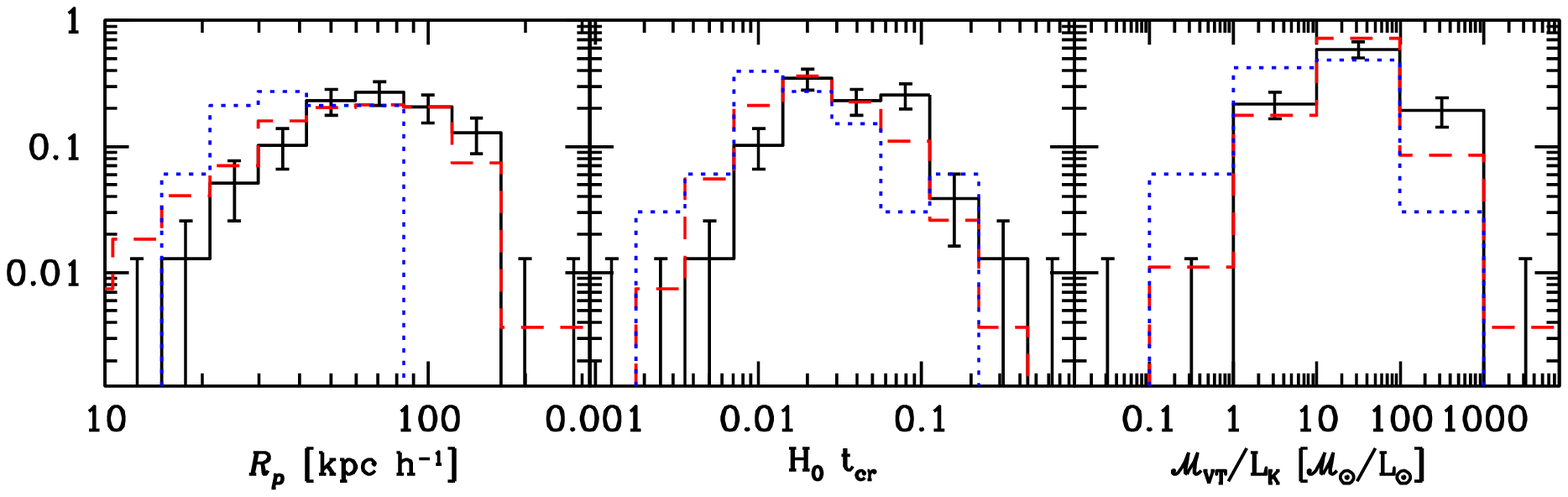}}
\caption{Distribution of properties of the \emph{v2MCG} (\emph{solid lines}) and HCG92/2 
(\emph{dotted lines}) samples, compared with different semi-analytical models (\emph{dashed lines}).
Error bars correspond to Poisson errors.
\label{2MCG_SAM}}
\end{figure*}

In Fig.~\ref{v2MCG}, we showed the comparison of the distributions of 
velocity-filtered CG properties between the \emph{v2MCG} and the \emph{mvCG} (from
\citealp{Guo11} run on the MS-II) samples. 
In Fig.~\ref{2MCG_SAM}, we show this comparison for several SAMs:
\emph{Left upper panels}: \emph{mvCG}s identified from \cite{Croton+06}'s SAM in the $K$-band, 
\emph{Right upper panels}: \emph{mvCG}s identified from \cite{Croton+06}'s SAM in the $R$-band, 
\emph{Left lower panels}: \emph{mvCG}s identified from \cite{Bower+06}'s SAM in the $R$-band, 
\emph{Right lower panels}: \emph{mvCG}s identified from \cite{dLB07}'s SAM in the $R$-band. 

\section{Properties of CGs after velocity-filtering in the 2MASS XSC}
\clearpage 

\begin{table*}
\caption{CGs after velocity filtering in the 2MASS XSC
} 
\label{v2MASSCG}
\tabcolsep 1pt
\begin{tabular}{r@{\hspace{3mm}}c@{\hspace{3mm}}r@{\hspace{3mm}}r@{\hspace{3mm}}crcr@{\hspace{4mm}}r@{\hspace{4mm}}cr@{\hspace{4mm}}cr@{\hspace{4mm}}l}
\hline
\hline
ID & RA & \multicolumn{1}{c}{Dec} & \multicolumn{1}{c}{$v$} & $N$ &  \multicolumn{1}{c}{$K_{\rm b}$} & $\mu_K$ &  \multicolumn{1}{c}{$\theta_{\rm G}$} & \multicolumn{1}{c}{$\langle R_{ij}\rangle  $} &
$b/a$ &  \multicolumn{1}{c}{$\sigma_v$} & 
$H_0 \, t_{\rm cr}$& \multicolumn{1}{c}{$M_{\rm VT}/L_K$} &  cross-ID \\
\cline{2-3}
  & \multicolumn{2}{c}{(J2000)\ \ \ \ }
& \multicolumn{1}{c}{[$\rm km
    \, s^{-1}$]} & & & [$\rm \frac{mag}{arcsec^{2}}$] 
& \multicolumn{1}{c}{[arcmin]} &\multicolumn{1}{c}{[kpc/$h$]} & &
\multicolumn{1}{c}{[$\rm km \, s^{-1}$]} & & \multicolumn{1}{c}{[$h \,\rm \cal{M}_\odot/L_\odot$]} \\
\hline
  1 & 00:00:43 &  28:23:18 &  8705 &  5 & 10.41 & 19.92 &  2.34 &  43.44 & 0.67 &  322 & 0.012 &   28 &	HCG 99, KTG 83,VV 854 \\  
  2 & 00:00:59 & -43:22:47 & 11525 &  4 & 10.14 & 22.23 &  6.07 & 121.73 & 0.30 &  479 & 0.023 &  148 &	\nodata \\     
  3 & 00:28:54 &  02:45:34 &  4241 &  5 &  8.58 & 23.18 & 18.04 &  50.06 & 0.08 &  175 & 0.026 &   11 &	UZC-CG 7, VV 894 \\    
  4 & 00:39:24 &  00:52:28 &  4209 &  4 &  9.33 & 21.00 &  5.58 &  45.56 & 0.53 &  130 & 0.032 &   10 &	HCG 7, UZC-CG 9, USGC U024 \\
  5 & 00:56:42 & -52:56:40 &  7561 &  4 &  9.50 & 21.29 &  5.71 &  92.15 & 0.37 &  534 & 0.016 &  108 &	\nodata \\     
  6 & 00:57:48 & -05:02:22 &  5345 &  6 & 10.20 & 22.81 & 11.38 &  85.80 & 0.46 &  294 & 0.026 &   90 &	\nodata \\     
  7 & 01:08:58 & -45:48:11 &  7598 &  4 &  9.87 & 21.05 &  4.18 &  44.85 & 0.11 &  291 & 0.014 &   29 &	\nodata \\     
  8 & 01:13:55 & -31:48:02 &  5687 &  4 &  8.79 & 21.15 &  7.30 &  85.77 & 0.78 &  293 & 0.027 &   39 &	SCG 1\\     
  9 & 01:26:05 &  34:41:39 &  4742 &  4 &  9.21 & 22.08 & 10.63 &  92.15 & 0.41 &  257 & 0.032 &   44 &	HCG 10, RSCG 12\\   
 10 & 01:43:24 & -34:14:40 &  3786 &  4 & 10.11 & 23.20 & 11.09 &  62.52 & 0.24 &  149 & 0.038 &   46 &	SCG 0141-3429 \\    
 11 & 02:07:39 &  02:08:30 &  7042 &  4 & 10.40 & 22.57 &  7.60 &  89.13 & 0.40 &  443 & 0.018 &  226 &	HCG 15 \\    
 12 & 02:09:49 & -10:13:37 &  3874 &  5 &  9.04 & 22.82 & 17.94 &  63.75 & 0.12 &  147 & 0.039 &   10 &	HCG 16, USGC S077, RSCG 19, VV 1007 \\
 13 & 02:36:55 &  07:22:28 &  6310 &  4 &  9.48 & 23.38 & 14.37 & 168.32 & 0.50 &  325 & 0.047 &  110 &	USGC U136 \\    
 14 & 02:42:05 & -15:04:36 &  7437 &  4 & 10.57 & 21.03 &  3.59 &  53.33 & 0.29 &   73 & 0.066 &    3 &	SCG 19, USGC S093\\
 15 & 02:45:07 & -17:42:28 &  7412 &  4 &  9.75 & 22.91 & 13.61 & 164.29 & 0.28 &  218 & 0.068 &   38 &	HCG 21\\     
 16 & 03:03:50 & -12:02:25 &  3648 &  4 &  8.73 & 21.52 &  8.45 &  67.71 & 0.32 &  396 & 0.016 &  122 &	SCG 3 \\    
 17 & 03:07:05 & -09:35:15 &  4851 &  4 &  9.87 & 21.75 &  7.04 &  65.73 & 0.69 &  330 & 0.018 &  106 &	HCG 23, SCG 11\\   
 18 & 03:07:29 & -66:48:22 &  5580 &  4 &  9.55 & 23.48 & 16.76 & 194.28 & 0.74 &  174 & 0.101 &   61 &	SCG 0306-6659 \\    
 19 & 03:17:44 & -10:19:56 &  8882 &  4 & 10.06 & 22.58 &  6.97 & 103.54 & 0.35 &  440 & 0.021 &  135 &	\nodata \\     
 20 & 03:20:40 & -01:04:18 &  6288 &  4 & 10.52 & 21.87 &  4.55 &  47.93 & 0.40 &   87 & 0.050 &    6 &	HCG 25\\     
 21 & 03:25:22 & -06:09:30 & 10335 &  4 & 10.44 & 22.12 &  5.85 & 116.75 & 0.36 &  157 & 0.067 &   17 &	\nodata \\     
 22 & 03:46:31 & -04:12:39 &  3917 &  5 &  8.18 & 23.38 & 29.15 & 140.46 & 0.40 &  145 & 0.088 &   15 &	\nodata \\     
 23 & 04:51:45 & -03:52:38 &  4633 &  4 & 10.14 & 22.83 &  8.34 &  74.54 & 0.57 &   83 & 0.082 &   13 &	\nodata \\     
 24 & 04:59:25 & -11:08:00 &  3823 &  6 &  9.03 & 22.60 & 18.32 &  99.91 & 0.56 &  357 & 0.025 &   59 &	KTS 28, VV 699 \\    
 25 & 06:03:55 & -32:06:38 &  9565 &  4 & 10.09 & 23.16 &  9.23 & 129.26 & 0.10 &  460 & 0.026 &   73 &	\nodata \\     
 26 & 06:43:41 & -74:14:45 &  6458 &  4 &  9.54 & 21.37 &  6.09 &  59.98 & 0.23 &  377 & 0.014 &   58 &	VV 785 \\     
 27 & 07:04:34 &  64:03:45 &  4468 &  4 &  9.55 & 22.99 & 11.27 &  94.36 & 0.25 &   66 & 0.129 &    8 &	\nodata \\     
 28 & 07:26:32 &  85:37:41 &  2177 &  4 &  7.74 & 22.71 & 22.30 &  83.25 & 0.46 &  262 & 0.029 &   98 &	VV 1189 \\     
 29 & 07:40:56 &  55:25:55 & 10752 &  4 &  9.90 & 19.38 &  1.73 &  32.21 & 0.42 &  436 & 0.007 &   32 &	\nodata \\     
 30 & 09:05:05 &  18:21:41 &  4188 &  5 &  9.01 & 23.51 & 17.54 & 142.04 & 0.79 &  391 & 0.033 &  317 &	VV 612 \\     
 31 & 09:16:30 &  30:52:02 &  6954 &  4 &  9.82 & 22.64 &  8.38 &  85.16 & 0.11 &   73 & 0.106 &    3 &	\nodata \\     
 32 & 09:27:57 &  30:00:35 &  7994 &  4 & 10.24 & 21.21 &  3.56 &  48.52 & 0.64 &  387 & 0.011 &  106 &	\nodata \\     
 33 & 09:34:23 &  10:09:24 &  3141 &  4 &  8.86 & 23.29 & 19.19 & 105.31 & 0.37 &  355 & 0.027 &  207 &	RSCG 33, VV 1290/1292 \\    
 34 & 09:38:55 & -04:51:05 &  6633 &  5 &  9.73 & 18.60 &  1.74 &  15.19 & 0.36 &  189 & 0.007 &    4 &	HCG 40, VV 116 \\     
 35 & 10:00:01 & -19:34:37 &  4020 &  4 &  8.18 & 22.74 & 18.18 & 105.27 & 0.08 &  119 & 0.081 &    6 &	HCG 42\\     
 36 & 10:25:07 &  28:03:09 &  6361 &  4 & 10.14 & 21.84 &  5.07 &  72.55 & 0.84 &  188 & 0.035 &   47 &	\nodata \\     
 37 & 10:37:15 & -26:14:56 &  3433 &  4 &  9.29 & 22.59 & 10.66 &  79.84 & 0.59 &  268 & 0.027 &  142 &	\nodata \\     
 38 & 10:39:43 & -23:49:47 &  3758 &  4 &  9.97 & 23.32 & 11.88 & 100.45 & 0.95 &  159 & 0.057 &   71 &	\nodata \\     
 39 & 10:51:52 &  50:56:02 &  7491 &  5 & 10.08 & 23.14 & 11.22 & 139.23 & 0.46 &  497 & 0.025 &   78 &	VV 1393\\     
 40 & 11:09:46 &  21:46:22 &  9552 &  4 & 10.01 & 21.52 &  4.81 &  64.52 & 0.08 &  103 & 0.057 &    3 &	\nodata \\     
 41 & 11:16:12 &  18:07:46 &   896 &  4 &  7.11 & 21.95 & 21.98 &  36.44 & 0.33 &  289 & 0.011 &   98 &	\nodata \\     
 42 & 11:22:21 &  24:18:31 &  7619 &  4 & 10.07 & 21.12 &  4.07 &  63.83 & 0.20 &  322 & 0.018 &   46 &	HCG 51, UZC-CG 138, VV 1435\\   
 43 & 11:35:11 &  51:12:03 &  8010 &  4 & 10.41 & 21.94 &  4.76 &  57.86 & 0.44 &  325 & 0.016 &  102 &	\nodata \\     
 44 & 11:42:10 &  10:18:52 &  6218 &  5 &  9.63 & 21.86 &  8.82 &  91.33 & 0.60 &  173 & 0.048 &   17 &	HCG 58, UZC-CG 144 \\  
 45 & 11:42:51 &  26:31:54 &  9054 &  4 & 10.06 & 22.87 &  9.91 & 126.33 & 0.11 &  355 & 0.032 &   68 &	\nodata \\     
 46 & 11:44:06 &  33:30:45 &  9381 &  4 & 10.16 & 21.25 &  3.78 &  67.96 & 0.43 &  144 & 0.043 &    8 &	\nodata \\     
 47 & 11:57:06 &  55:17:50 &   948 &  4 &  7.42 & 22.44 & 23.41 &  51.99 & 0.45 &  211 & 0.022 &   88 &	\nodata \\     
 48 & 12:19:22 &  05:57:42 &  2156 &  4 &  7.40 & 21.67 & 16.44 &  73.71 & 0.39 &  348 & 0.019 &   89 &	\nodata \\     
 49 & 12:24:31 &  07:09:38 &   994 &  5 &  6.79 & 22.83 & 36.02 &  64.63 & 0.46 &  215 & 0.027 &   91 &	\nodata \\     
 50 & 12:27:47 &  12:11:30 &  1024 &  4 &  8.99 & 23.12 & 15.53 &  28.03 & 0.32 &  557 & 0.005 & 1560 &	\nodata \\     
 51 & 12:43:17 &  11:25:08 &  1117 &  5 &  5.81 & 22.16 & 45.77 &  75.83 & 0.38 &  363 & 0.019 &   56 &	M60 CG, VV 206/1558\\     
 52 & 13:01:58 &  27:37:34 &  7473 &  4 & 10.03 & 21.90 &  5.07 &  61.27 & 0.21 &  704 & 0.008 &  401 &	\nodata \\     
 53 & 13:06:50 & -40:22:49 &  4768 &  4 &  8.83 & 22.20 & 10.29 & 118.12 & 0.55 &  116 & 0.092 &   14 &	HDCE 0761 \\    
 54 & 13:08:15 &  34:01:53 & 10120 &  4 & 10.50 & 23.41 & 10.77 & 180.87 & 0.54 &  185 & 0.089 &   47 &	\nodata \\     
 55 & 13:22:02 & -17:21:16 &  6909 &  5 &  9.95 & 22.11 &  6.15 &  71.02 & 0.44 &  241 & 0.027 &   50 &	\nodata \\     
 56 & 13:24:31 &  14:01:36 &  7087 &  4 &  9.37 & 22.75 & 11.94 & 177.44 & 0.47 &  278 & 0.058 &   76 &	\nodata \\     
 57 & 13:52:16 &  02:20:05 &  7095 &  4 &  9.80 & 20.74 &  3.42 &  51.30 & 0.52 &   50 & 0.094 &    1 &	\nodata \\     
 58 & 13:53:01 & -28:27:48 &  4713 &  5 &  8.54 & 21.08 &  7.27 &  43.75 & 0.13 &  261 & 0.015 &   20 &	\nodata \\     
 59 & 14:00:33 & -02:51:35 &  7325 &  6 &  9.82 & 21.73 &  6.12 &  60.26 & 0.31 &  322 & 0.017 &   37 &	\nodata \\     
 60 & 14:19:14 &  35:08:15 &  8534 &  4 & 10.34 & 22.52 &  6.12 &  74.71 & 0.28 &   52 & 0.131 &    3 &	\nodata \\     
 61 & 14:27:27 &  11:19:18 &  8078 &  4 & 10.45 & 22.51 &  6.47 & 108.69 & 0.70 &  424 & 0.023 &  245 &	\nodata \\     
 62 & 14:28:02 &  25:53:33 &  4388 &  4 &  9.33 & 21.86 &  8.31 &  67.15 & 0.58 &  216 & 0.028 &   42 &	\nodata \\     
 63 & 14:58:09 & -19:10:04 &  3387 &  4 &  8.48 & 23.41 & 23.18 & 158.56 & 0.52 &  107 & 0.134 &   14 &	\nodata \\     
  \hline
\end{tabular}
\end{table*}

\begin{table*}
\contcaption{--- CGs after velocity filtering in the 2MASS XSC} 
\tabcolsep 1pt
\begin{tabular}{r@{\hspace{3mm}}c@{\hspace{3mm}}r@{\hspace{3mm}}r@{\hspace{3mm}}crcr@{\hspace{4mm}}r@{\hspace{4mm}}cr@{\hspace{4mm}}cr@{\hspace{4mm}}l}
\hline
\hline
ID & RA & \multicolumn{1}{c}{Dec} & \multicolumn{1}{c}{$v$} & $N$ &  \multicolumn{1}{c}{$K_{\rm b}$} & $\mu_K$ &  \multicolumn{1}{c}{$\theta_{\rm G}$} & \multicolumn{1}{c}{$\langle R_{ij}\rangle  $} &
$b/a$ &  \multicolumn{1}{c}{$\sigma_v$} & 
$H_0 \, t_{\rm cr}$& \multicolumn{1}{c}{$M_{\rm VT}/L_K$} &  cross-ID \\
  & \multicolumn{2}{c}{(J2000)\ \ \ \ }
& \multicolumn{1}{c}{[$\rm km
    \, s^{-1}$]} & & & [$\rm \frac{mag}{arcsec^{2}}$] 
& \multicolumn{1}{c}{[arcmin]} &\multicolumn{1}{c}{[kpc/$h$]} & &
\multicolumn{1}{c}{[$\rm km \, s^{-1}$]} & & \multicolumn{1}{c}{[$h \,\rm \cal{M}_\odot/L_\odot$]} \\
\hline
      64 & 15:36:22 &  43:29:29 &  5660 &  5 &  9.62 & 22.19 &  9.19 & 102.35 & 0.71 &  197 & 0.047 &   28 &	\nodata \\     
 65 & 16:12:50 &  33:02:06 &  9429 &  4 & 10.51 & 21.73 &  4.32 &  91.69 & 0.20 &  108 & 0.077 &    4 &	VV 1801 \\     
 66 & 16:37:53 &  36:03:18 &  9614 &  4 &  9.91 & 22.45 &  7.62 & 105.18 & 0.17 &  204 & 0.047 &   23 &	\nodata \\     
 67 & 19:14:47 & -54:36:26 &  5400 &  4 &  9.48 & 23.15 & 13.64 & 110.24 & 0.13 &  224 & 0.045 &   33 &	\nodata \\     
 68 & 19:51:59 & -30:49:31 &  5891 &  4 &  9.47 & 20.22 &  3.90 &  46.60 & 0.80 &  368 & 0.011 &   50 &	HCG 86 \\    
 69 & 20:00:59 & -47:04:38 &  6804 &  4 & 10.02 & 20.19 &  2.92 &  29.28 & 0.16 &  317 & 0.008 &   30 &	KTS 61, Rose 38, NGC6845, VV 1880\\  
 70 & 20:03:14 & -56:00:10 &  4413 &  4 &  8.01 & 21.62 & 12.97 &  93.18 & 0.30 &   84 & 0.100 &    3 &	\nodata \\     
 71 & 20:17:25 & -70:41:58 &  3969 &  4 &  7.77 & 22.94 & 28.03 & 167.79 & 0.20 &  488 & 0.031 &  148 &	VV 297 \\     
 72 & 20:43:41 & -26:34:43 & 12406 &  4 & 10.45 & 21.45 &  3.81 &  95.20 & 0.68 &  268 & 0.032 &   38 &	\nodata \\     
 73 & 20:47:22 &  00:23:32 &  3779 &  6 &  8.89 & 21.83 & 11.03 &  71.64 & 0.58 &  254 & 0.026 &   43 &	\nodata \\     
 74 & 20:52:24 & -05:45:16 &  6028 &  4 &  9.85 & 22.19 &  7.70 &  68.51 & 0.16 &   80 & 0.077 &    5 &	HCG 88\\     
 75 & 21:08:25 & -29:45:32 &  5935 &  4 &  9.57 & 23.48 & 14.46 & 166.54 & 0.82 &   53 & 0.285 &    6 &	SCG 2105-2957 \\    
 76 & 21:17:01 & -42:19:38 &  5337 &  4 &  9.46 & 22.32 &  9.77 & 123.47 & 0.81 &  112 & 0.100 &   11 &	SCG 2113-4235 \\    
 77 & 22:03:28 &  12:38:56 &  8113 &  4 &  9.83 & 20.15 &  2.66 &  39.48 & 0.51 &  465 & 0.008 &   69 &	WBL 677\\     
 78 & 22:36:24 & -24:18:35 & 10314 &  4 & 10.30 & 21.57 &  4.21 &  95.89 & 0.12 &  103 & 0.084 &    4 &	\nodata \\     
 79 & 22:55:22 & -33:54:22 &  8761 &  4 & 10.21 & 20.21 &  2.39 &  41.13 & 0.65 &  175 & 0.021 &   10 &	VV 1957\\      
 80 & 22:58:09 &  26:07:30 &  7588 &  5 &  9.25 & 22.17 &  8.95 & 137.84 & 0.72 &  201 & 0.062 &   22 &	UZC-CG 282, VV 84 \\    
 81 & 23:15:18 &  19:00:38 &  4922 &  4 &  9.00 & 21.58 &  8.31 &  70.59 & 0.41 &  237 & 0.027 &   37 &	HCG 93, USGC U837, Arp 99 \\
 82 & 23:28:03 & -67:47:18 &  3904 &  4 &  9.82 & 21.54 &  6.04 &  42.88 & 0.46 &  176 & 0.022 &   25 &	\nodata \\     
 83 & 23:28:18 &  32:25:09 &  5066 &  4 &  9.18 & 21.58 &  7.15 &  68.01 & 0.59 &  344 & 0.018 &   58 &	\nodata \\     
 84 & 23:47:27 & -02:18:37 &  6665 &  5 & 10.07 & 21.38 &  5.23 &  64.89 & 0.73 &  415 & 0.014 &   94 &	HCG 97, RSCG 87\\   
 85 & 23:53:35 &  07:59:10 &  5218 &  5 &  9.15 & 22.68 & 15.93 & 142.35 & 0.52 &  233 & 0.055 &   33 &	\nodata \\     

\hline
\end{tabular}
\parbox{16cm}{
\small
{\bf Notes.} ID: Group ID, RA: Right Ascension of the CG centre, Dec: Declination of the CG centre, 
$v$: Median velocity, 
$N$: Number of galaxy members in the CG in the range of 3 magnitudes from the brightest member,
$K_{\rm b}$: Galactic Extinction-corrected $K$-band apparent magnitude of the brightest galaxy,
$\mu_K$: Galactic Extinction-corrected $K$-band group surface brightness,
$\theta_{\rm G}$: Angular diameter of the smallest circumscribed circle,
$\langle R_{ij}\rangle$: Median projected separation among galaxies,
$b/a$: Apparent group elongation,
$\sigma_v$: Radial velocity dispersion of the galaxies in the CG computed using individual galaxy errors,
$H_0 \, t_{\rm cr}$: Dimensionless crossing time,
$M_{\rm VT}/L_K$: Mass to Light ratio in the $K$-band,
cross-ID: Cross-identification with other group catalogues\\
{\bf References for cross-ID:} 
AM: Arp+Madore Southern Peculiar Galaxies and Associations \citep{Arp87};
Arp: Arp Peculiar Galaxies \citep{Arp66};
HCG: Hickson Compact Group \citep{Hickson82};
HDCE: High-density-contrast groups - Erratum version \citep{Crook08};
KPG: Karachentsev Isolated Pairs of Galaxies Catalogue \citep{K72};
KTG: Karachentsev Isolated Triplets of Galaxies Catalogue \citep{KKL88}; 
KTS: Karachentseva Triple System \citep{KK00};
M60: \citep{Mamon89,Mamon08};
Rose: Rose Compact Groups of Galaxies \citep{Rose77};
RSCG: Redshift Survey Compact Group \citep{Barton+96}; 
SCG: Southern Compact Group \citep{PIM94,Iovino02};
UZC-CG:Updated Zwicky Catalogue-Compact Group \citep{FK02};
USGC: UZC/SSRS2 Group Catalogue \citep{Ramella02};
VV: Interacting galaxies catalogue \citep{VV01}
WBL: White+Bliton+Bhavsar groups \citep{WBL99};
}

\end{table*}
	
\begin{table*}
\caption{Table of galaxy members (\emph{v2MCG}s)}
\tabcolsep 1pt
\centering
\begin{tabular}{r@{\hspace{3mm}}cc@{\hspace{2mm}}r@{\hspace{3mm}}r@{\hspace{3mm}}r@{\hspace{3mm}}rr@{\hspace{3mm}}cl@{\hspace{2mm}}l}
\hline
GroupID & GalID & RA & Dec\ \ \ \ \ \  & \multicolumn{1}{c}{$K$} &
\multicolumn{1}{c}{$k_K$} &  \multicolumn{1}{c}{\ \ \ $v$} &
\multicolumn{1}{c}{\ \ \ \ \ \ err($v$)} & $v$ & SDSS\_ID & 2MASS\_ID \\
\cline{3-4}
\cline{7-8}
& & \multicolumn{2}{c}{(J2000)} & & & \multicolumn{2}{c}{[$\rm km \,s^{-1}$]}
&  source & & \\
\hline
  1 &  1 & 00:00:46.97 &  28:24:07.28 & 10.41 & --0.06 &  8764 &   19 & 1 & 758874298530726152 & 00004696+2824071 \\
  1 &  2 & 00:00:37.94 &  28:23:04.34 & 10.46 & --0.05 &  8705 &    9 & 1 & 758874298530726255 & 00003794+2823041 \\
  1 &  3 & 00:00:44.00 &  28:24:05.22 & 11.53 & --0.04 &  8156 &    0 & 1 & 758874298530726153 & 00004401+2824051 \\
  1 &  4 & 00:00:42.41 &  28:22:08.43 & 13.26 & --0.03 &  9006 &    0 & 2 & 758874298530791466 & 00004242+2822081 \\
  1 &  5 & 00:00:45.09 &  28:22:18.18 & 13.40 & --0.08 &  8642 &    0 & 2 & 758874298530791773 & 00004507+2822181 \\
  2 &  1 & 00:01:02.89 & --43:19:49.57 & 10.14 & --0.07 & 11627 &   45 & 1 & 000000000000000000 & 00010289--4319496 \\
  2 &  2 & 00:00:53.00 & --43:23:31.43 & 11.59 & --0.05 & 11980 &   40 & 1 & 000000000000000000 & 00005298--4323316 \\
  2 &  3 & 00:00:57.05 & --43:25:47.60 & 12.27 & --0.06 & 10964 &   45 & 2 & 000000000000000000 & 00005702--4325476 \\
  2 &  4 & 00:00:52.18 & --43:20:02.38 & 12.47 & --0.06 & 11422 &    0 & 2 & 000000000000000000 & 00005216--4320026 \\
  3 &  1 & 00:29:15.06 &  02:51:50.58 &  8.58 & --0.02 &  4241 &   16 & 1 & 000000000000000000 & 00291506+0251505 \\
  3 &  2 & 00:28:29.78 &  02:38:54.98 & 10.88 & --0.03 &  4317 &   29 & 1 & 000000000000000000 & 00282976+0238549 \\
  3 &  3 & 00:29:08.10 &  02:48:39.97 & 11.36 & --0.03 &  4046 &   25 & 1 & 000000000000000000 & 00290809+0248400 \\
  3 &  4 & 00:29:18.55 &  02:52:13.56 & 11.51 & --0.03 &  4433 &    0 & 1 & 000000000000000000 & 00291854+0252135 \\
  3 &  5 & 00:29:12.40 &  02:52:21.42 & 11.53 & --0.02 &  4093 &    0 & 1 & 000000000000000000 & 00291239+0252215 \\
  4 &  1 & 00:39:13.39 &  00:51:50.87 &  9.33 & --0.03 &  4177 &    3 & 1 & 588015510347382842 & 00391339+0051508 \\
  4 &  2 & 00:39:17.84 &  00:54:45.88 & 10.14 & --0.03 &  4239 &    1 & 1 & 588015510347382865 & 00391786+0054458 \\
  4 &  3 & 00:39:34.85 &  00:51:35.68 & 10.51 & --0.03 &  4379 &    3 & 1 & 588015510347448339 & 00393485+0051355 \\
  4 &  4 & 00:39:18.79 &  00:53:31.01 & 12.07 & --0.02 &  4107 &    0 & 2 & 588015510347382868 & 00391879+0053308 \\
  5 &  1 & 00:56:41.65 & --52:58:33.32 &  9.50 & --0.05 &  7779 &   31 & 1 & 000000000000000000 & 00564165--5258332 \\
  5 &  2 & 00:56:57.57 & --52:55:26.10 & 10.08 & --0.04 &  7342 &   29 & 1 & 000000000000000000 & 00565758--5255262 \\
  5 &  3 & 00:56:43.64 & --52:53:49.16 & 11.85 & --0.04 &  7331 &   28 & 2 & 000000000000000000 & 00564365--5253492 \\
  5 &  4 & 00:56:42.17 & --52:59:31.44 & 12.29 & --0.06 &  8392 &   45 & 2 & 000000000000000000 & 00564220--5259312 \\
  6 &  1 & 00:58:01.60 & --05:04:16.44 & 10.20 & --0.03 &  5184 &   45 & 1 & 000000000000000000 & 00580158--0504165 \\
  6 &  2 & 00:57:55.36 & --05:07:49.53 & 10.34 & --0.03 &  5314 &   45 & 1 & 000000000000000000 & 00575536--0507495 \\
  6 &  3 & 00:57:42.39 & --04:56:55.11 & 10.96 & --0.03 &  5375 &   45 & 1 & 000000000000000000 & 00574239--0456548 \\
  6 &  4 & 00:57:47.87 & --05:06:43.52 & 11.16 & --0.03 &  5454 &   36 & 1 & 000000000000000000 & 00574786--0506435 \\
  6 &  5 & 00:57:39.24 & --05:05:09.92 & 11.60 & --0.03 &  4845 &   45 & 1 & 000000000000000000 & 00573925--0505098 \\
  6 &  6 & 00:57:35.09 & --05:00:09.00 & 11.73 & --0.04 &  5675 &   45 & 1 & 000000000000000000 & 00573510--0500088 \\
  7 &  1 & 01:09:04.55 & --45:46:24.73 &  9.87 & --0.05 &  7746 &   27 & 1 & 000000000000000000 & 01090456--4546246 \\
  7 &  2 & 01:08:51.85 & --45:49:57.50 & 10.60 & --0.05 &  7207 &   37 & 1 & 000000000000000000 & 01085185--4549577 \\
  7 &  3 & 01:08:57.40 & --45:48:15.63 & 12.60 & --0.05 &  7768 &    0 & 2 & 000000000000000000 & 01085740--4548157 \\
  7 &  4 & 01:09:01.03 & --45:48:04.83 & 12.86 & --0.07 &  7449 &    0 & 2 & 000000000000000000 & 01090103--4548047 \\
  8 &  1 & 01:13:47.26 & --31:44:50.00 &  8.79 & --0.04 &  5738 &   45 & 1 & 000000000000000000 & 01134725--3144500 \\
  8 &  2 & 01:13:51.26 & --31:47:18.04 &  9.62 & --0.04 &  5636 &   19 & 1 & 000000000000000000 & 01135125--3147180 \\
  8 &  3 & 01:14:10.95 & --31:49:38.06 & 10.89 & --0.04 &  5594 &   45 & 1 & 000000000000000000 & 01141094--3149382 \\
  8 &  4 & 01:13:43.18 & --31:50:35.21 & 11.25 & --0.05 &  6228 &   45 & 1 & 000000000000000000 & 01134317--3150350 \\
  9 &  1 & 01:25:40.29 &  34:42:46.68 &  9.21 & --0.04 &  4822 &   18 & 1 & 758877278694080872 & 01254030+3442465 \\
  9 &  2 & 01:26:21.78 &  34:42:10.89 &  9.34 & --0.03 &  5188 &    5 & 1 & 758877153601913092 & 01262177+3442107 \\
  9 &  3 & 01:26:18.83 &  34:45:14.93 & 10.62 & --0.03 &  4660 &   32 & 1 & 758877153601978381 & 01261884+3445147 \\
  9 &  4 & 01:26:30.85 &  34:40:31.76 & 11.64 & --0.02 &  4662 &    0 & 1 & 758877153601913329 & 01263085+3440318 \\
 10 &  1 & 01:43:09.68 & --34:14:30.85 & 10.11 & --0.03 &  3801 &   45 & 1 & 000000000000000000 & 01430966--3414305 \\
 10 &  2 & 01:43:18.35 & --34:12:21.62 & 10.72 & --0.01 &  3770 &   45 & 1 & 000000000000000000 & 01431837--3412216 \\
 10 &  3 & 01:43:45.06 & --34:18:07.33 & 11.30 & --0.03 &  3756 &   31 & 1 & 000000000000000000 & 01434505--3418070 \\
 10 &  4 & 01:43:02.96 & --34:11:14.44 & 12.47 & --0.02 &  4093 &   32 & 2 & 000000000000000000 & 01430297--3411143 \\
 11 &  1 & 02:07:53.08 &  02:10:03.37 & 10.40 & --0.05 &  6966 &   30 & 1 & 000000000000000000 & 02075306+0210034 \\
 11 &  2 & 02:07:34.13 &  02:06:55.17 & 11.01 & --0.05 &  7117 &   36 & 1 & 000000000000000000 & 02073411+0206554 \\
 11 &  3 & 02:07:37.52 &  02:10:50.62 & 11.50 & --0.04 &  6243 &   36 & 1 & 000000000000000000 & 02073751+0210504 \\
 11 &  4 & 02:07:25.29 &  02:06:58.08 & 11.76 & --0.05 &  7196 &    0 & 2 & 000000000000000000 & 02072527+0206579 \\
 12 &  1 & 02:09:24.60 & --10:08:09.15 &  9.04 & --0.02 &  4174 &   45 & 1 & 587727177926508587 & 02092458--1008091 \\
 12 &  2 & 02:09:20.86 & --10:07:59.15 &  9.54 & --0.02 &  3854 &    2 & 1 & 587727177926508588 & 02092086--1007591 \\
 12 &  3 & 02:09:38.53 & --10:08:46.57 &  9.81 & --0.01 &  3849 &    1 & 1 & 587727177926574089 & 02093853--1008466 \\
 12 &  4 & 02:09:42.74 & --10:11:01.80 &  9.92 & --0.01 &  3874 &    5 & 1 & 000000000000000000 & 02094273--1011016 \\
 12 &  5 & 02:10:17.55 & --10:19:15.72 & 10.61 & --0.02 &  4045 &   45 & 1 & 000000000000000000 & 02101756--1019157 \\
 13 &  1 & 02:37:14.48 &  07:18:20.27 &  9.48 & --0.04 &  6498 &   61 & 1 & 587744296044986466 & 02371447+0718201 \\
 13 &  2 & 02:36:31.62 &  07:18:34.23 & 10.77 & --0.02 &  6122 &    3 & 1 & 587744296044920962 & 02363162+0718342 \\
 13 &  3 & 02:37:16.50 &  07:20:08.95 & 11.46 & --0.05 &  6546 &   49 & 1 & 587744296044986609 & 02371649+0720091 \\
 13 &  4 & 02:37:20.26 &  07:26:23.32 & 11.52 & --0.03 &  5931 &    0 & 1 & 587744296044986628 & 02372026+0726231 \\
 14 &  1 & 02:42:06.28 & --15:05:29.24 & 10.57 & --0.05 &  7396 &   31 & 1 & 000000000000000000 & 02420629--1505289 \\
 14 &  2 & 02:42:05.74 & --15:02:48.78 & 10.58 & --0.04 &  7328 &   23 & 1 & 000000000000000000 & 02420573--1502489 \\
 14 &  3 & 02:42:04.98 & --15:06:23.88 & 12.03 & --0.05 &  7477 &  270 & 3 & 000000000000000000 & 02420497--1506239 \\
 14 &  4 & 02:42:01.38 & --15:03:23.91 & 12.74 & --0.06 &  7476 &    0 & 2 & 000000000000000000 & 02420138--1503239 \\
\end{tabular}
\end{table*}
\begin{table*}
\contcaption{--- Table of galaxy positions} 
\tabcolsep 1pt
\centering
\begin{tabular}{r@{\hspace{3mm}}cc@{\hspace{2mm}}r@{\hspace{3mm}}r@{\hspace{3mm}}r@{\hspace{3mm}}rr@{\hspace{3mm}}cl@{\hspace{2mm}}l}
\hline
GroupID & GalID & RA & Dec\ \ \ \ \ \  & \multicolumn{1}{c}{$K$} &
\multicolumn{1}{c}{$k_K$} &  \multicolumn{1}{c}{\ \ \ $v$} &
\multicolumn{1}{c}{\ \ \ \ \ \ err($v$)} & $v$ & SDSS\_ID & 2MASS\_ID \\
\cline{3-4}
\cline{7-8}
& & \multicolumn{2}{c}{(J2000)} & & & \multicolumn{2}{c}{[$\rm km \,s^{-1}$]}
&  source & &\\
\hline

 15 &  1 & 02:44:53.65 & --17:39:32.92 &  9.75 & --0.05 &  7272 &   45 & 1 & 000000000000000000 & 02445363--1739330 \\
 15 &  2 & 02:45:36.08 & --17:41:20.16 &  9.83 & --0.04 &  7550 &   45 & 1 & 000000000000000000 & 02453607--1741201 \\
 15 &  3 & 02:45:18.02 & --17:42:30.09 &  9.99 & --0.04 &  7682 &   45 & 1 & 000000000000000000 & 02451802--1742301 \\
 15 &  4 & 02:44:39.73 & --17:43:35.99 & 12.01 & --0.05 &  7272 &   45 & 2 & 000000000000000000 & 02443974--1743359 \\
 16 &  1 & 03:03:54.48 & --11:59:30.48 &  8.73 & --0.03 &  3986 &   45 & 1 & 000000000000000000 & 03035448--1159306 \\
 16 &  2 & 03:03:35.20 & --12:04:34.72 &  9.94 & --0.03 &  3384 &   22 & 1 & 000000000000000000 & 03033518--1204349 \\
 16 &  3 & 03:04:06.22 & --12:00:55.65 & 11.13 & --0.02 &  3265 &   35 & 1 & 000000000000000000 & 03040620--1200556 \\
 16 &  4 & 03:03:32.80 & --12:02:23.16 & 11.17 & --0.03 &  3911 &   36 & 1 & 000000000000000000 & 03033280--1202229 \\
 17 &  1 & 03:06:55.94 & --09:32:38.66 &  9.87 & --0.03 &  4825 &   45 & 1 & 000000000000000000 & 03065595--0932389 \\
 17 &  2 & 03:07:09.47 & --09:35:33.55 & 10.11 & --0.02 &  4876 &   11 & 1 & 000000000000000000 & 03070944--0935334 \\
 17 &  3 & 03:07:18.38 & --09:36:45.48 & 10.56 & --0.03 &  5199 &   45 & 1 & 000000000000000000 & 03071837--0936454 \\
 17 &  4 & 03:06:55.22 & --09:37:42.86 & 12.00 & --0.04 &  4466 &   21 & 2 & 000000000000000000 & 03065521--0937429 \\
 18 &  1 & 03:06:31.08 & --66:46:32.20 &  9.55 & --0.03 &  5521 &   10 & 1 & 000000000000000000 & 03063104--6646320 \\
 18 &  2 & 03:07:02.09 & --66:56:19.20 &  9.99 & --0.04 &  5639 &   45 & 1 & 000000000000000000 & 03070209--6656192 \\
 18 &  3 & 03:07:40.74 & --66:40:04.47 & 10.92 & --0.04 &  5762 &   45 & 1 & 000000000000000000 & 03074073--6640045 \\
 18 &  4 & 03:08:39.99 & --66:53:03.40 & 11.44 & --0.04 &  5398 &   45 & 1 & 000000000000000000 & 03083999--6653035 \\
 19 &  1 & 03:17:45.52 & --10:17:20.72 & 10.06 & --0.06 &  8984 &   39 & 1 & 000000000000000000 & 03174554--1017207 \\
 19 &  2 & 03:17:35.93 & --10:18:50.78 & 11.94 & --0.05 &  8778 &   45 & 2 & 000000000000000000 & 03173593--1018511 \\
 19 &  3 & 03:17:42.95 & --10:23:24.41 & 12.61 & --0.05 &  9562 &    0 & 2 & 000000000000000000 & 03174295--1023247 \\
 19 &  4 & 03:17:46.00 & --10:16:28.59 & 12.82 & --0.06 &  8633 &    0 & 2 & 000000000000000000 & 03174601--1016287 \\
 20 &  1 & 03:20:45.41 & --01:02:40.87 & 10.52 & --0.04 &  6413 &    3 & 1 & 587731511545757880 & 03204541--0102407 \\
 20 &  2 & 03:20:42.94 & --01:06:30.89 & 11.66 & --0.03 &  6268 &    2 & 1 & 588015507680723132 & 03204294--0106307 \\
 20 &  3 & 03:20:38.55 & --01:02:06.05 & 12.41 & --0.05 &  6307 &    2 & 2 & 587731511545757775 & 03203854--0102057 \\
 20 &  4 & 03:20:45.34 & --01:03:14.06 & 12.51 & --0.03 &  6229 &    0 & 2 & 588015507680723008 & 03204534--0103137 \\
 21 &  1 & 03:25:11.58 & --06:10:51.17 & 10.44 & --0.05 & 10092 &    2 & 1 & 587724242304565362 & 03251157--0610510 \\
 21 &  2 & 03:25:25.37 & --06:08:38.02 & 10.82 & --0.05 & 10316 &    3 & 1 & 587724242304630856 & 03252538--0608380 \\
 21 &  3 & 03:25:31.39 & --06:07:43.98 & 11.93 & --0.05 & 10435 &    2 & 2 & 587724242304630877 & 03253137--0607438 \\
 21 &  4 & 03:25:19.37 & --06:12:21.02 & 13.01 & --0.04 & 10353 &    0 & 2 & 587724242304630917 & 03251935--0612210 \\
 22 &  1 & 03:46:27.25 & --03:58:07.63 &  8.18 & --0.02 &  3908 &   45 & 1 & 000000000000000000 & 03462726--0358075 \\
 22 &  2 & 03:45:43.12 & --04:05:29.68 &  9.27 & --0.02 &  3972 &   45 & 1 & 000000000000000000 & 03454312--0405295 \\
 22 &  3 & 03:46:35.94 & --04:27:11.52 &  9.62 & --0.02 &  3740 &   45 & 1 & 000000000000000000 & 03463595--0427115 \\
 22 &  4 & 03:46:07.16 & --04:04:09.13 &  9.98 & --0.03 &  3916 &   45 & 1 & 000000000000000000 & 03460717--0404093 \\
 22 &  5 & 03:46:03.07 & --04:08:17.17 & 10.07 & --0.03 &  4136 &   45 & 1 & 000000000000000000 & 03460309--0408173 \\
 23 &  1 & 04:51:41.51 & --03:48:33.68 & 10.14 & --0.02 &  4764 &   30 & 1 & 758885351078232088 & 04514150--0348335 \\
 23 &  2 & 04:51:55.95 & --03:55:47.66 & 11.48 & --0.03 &  4607 &   23 & 1 & 758887369173827734 & 04515593--0355475 \\
 23 &  3 & 04:51:41.75 & --03:50:29.54 & 11.95 & --0.03 &  4578 &   45 & 2 & 758885351078232404 & 04514177--0350295 \\
 23 &  4 & 04:51:28.41 & --03:52:12.53 & 12.09 & --0.02 &  4659 &   45 & 2 & 758885351078166820 & 04512841--0352124 \\
 24 &  1 & 04:59:25.89 & --10:58:50.54 &  9.03 & --0.03 &  3740 &    7 & 1 & 000000000000000000 & 04592589--1058504 \\
 24 &  2 & 04:59:27.73 & --11:07:22.62 &  9.24 & --0.02 &  3762 &   11 & 1 & 000000000000000000 & 04592771--1107224 \\
 24 &  3 & 04:59:17.38 & --11:07:07.20 &  9.35 & --0.03 &  4500 &    9 & 1 & 000000000000000000 & 04591738--1107071 \\
 24 &  4 & 04:59:22.89 & --11:07:56.34 &  9.43 & --0.02 &  3883 &   42 & 1 & 000000000000000000 & 04592288--1107561 \\
 24 &  5 & 04:59:41.37 & --11:16:17.44 & 10.77 & --0.02 &  3700 &   45 & 1 & 000000000000000000 & 04594137--1116175 \\
 24 &  6 & 04:58:50.32 & --11:04:53.04 & 11.78 & --0.02 &  4390 &   45 & 2 & 000000000000000000 & 04585029--1104531 \\
 25 &  1 & 06:03:39.89 & --32:08:52.03 & 10.09 & --0.05 &  9233 &   38 & 1 & 000000000000000000 & 06033989--3208521 \\
 25 &  2 & 06:03:38.22 & --32:09:19.99 & 11.96 & --0.06 &  9912 &   45 & 2 & 000000000000000000 & 06033824--3209201 \\
 25 &  3 & 06:04:13.62 & --32:03:57.30 & 12.33 & --0.03 &  9090 &   45 & 2 & 000000000000000000 & 06041362--3203574 \\
 25 &  4 & 06:03:42.26 & --32:09:42.48 & 12.41 & --0.04 &  9896 &   45 & 2 & 000000000000000000 & 06034225--3209422 \\
 26 &  1 & 06:43:06.02 & --74:14:10.45 &  9.54 & --0.04 &  6504 &  120 & 1 & 000000000000000000 & 06430596--7414103 \\
 26 &  2 & 06:44:17.48 & --74:16:36.32 & 10.61 & --0.04 &  6153 &   28 & 1 & 000000000000000000 & 06441744--7416364 \\
 26 &  3 & 06:43:25.51 & --74:15:25.74 & 10.64 & --0.04 &  6411 &   33 & 1 & 000000000000000000 & 06432557--7415255 \\
 26 &  4 & 06:43:06.00 & --74:12:55.17 & 11.21 & --0.05 &  7000 &  150 & 1 & 000000000000000000 & 06430602--7412553 \\
 27 &  1 & 07:04:20.30 &  64:01:12.98 &  9.55 & --0.03 &  4497 &    6 & 1 & 000000000000000000 & 07042030+6401132 \\
 27 &  2 & 07:05:23.68 &  64:05:32.50 & 11.33 & --0.02 &  4530 &   48 & 1 & 000000000000000000 & 07052368+6405326 \\
 27 &  3 & 07:03:45.99 &  64:01:57.06 & 11.50 & --0.03 &  4380 &    0 & 1 & 000000000000000000 & 07034600+6401570 \\
 27 &  4 & 07:05:19.69 &  64:02:26.83 & 11.54 & --0.04 &  4438 &    0 & 1 & 000000000000000000 & 07051971+6402266 \\
 28 &  1 & 07:32:20.49 &  85:42:31.90 &  7.74 & --0.01 &  1905 &    7 & 1 & 000000000000000000 & 07322048+8542319 \\
 28 &  2 & 07:27:14.36 &  85:45:16.37 &  9.25 & --0.01 &  2415 &    2 & 1 & 000000000000000000 & 07271448+8545162 \\
 28 &  3 & 07:34:57.53 &  85:32:13.90 & 10.03 & --0.02 &  2050 &   51 & 1 & 000000000000000000 & 07345760+8532138 \\
 28 &  4 & 07:17:47.09 &  85:42:47.75 & 10.61 & --0.02 &  2303 &   22 & 1 & 000000000000000000 & 07174680+8542479 \\
 29 &  1 & 07:40:58.22 &  55:25:37.92 &  9.90 & --0.06 & 10229 &   27 & 1 & 000000000000000000 & 07405822+5525379 \\
 29 &  2 & 07:41:00.08 &  55:25:13.89 & 11.84 & --0.03 & 11037 &    0 & 2 & 000000000000000000 & 07410010+5525139 \\
 29 &  3 & 07:40:52.73 &  55:26:36.89 & 12.20 & --0.06 & 11018 &    0 & 2 & 000000000000000000 & 07405269+5526369 \\
 29 &  4 & 07:40:59.74 &  55:26:21.01 & 12.66 & --0.07 & 10485 &    0 & 2 & 000000000000000000 & 07405974+5526209 \\
\end{tabular}
\end{table*}
\begin{table*}
\contcaption{--- Table of galaxy positions} 
\tabcolsep 1pt
\centering
\begin{tabular}{r@{\hspace{3mm}}cc@{\hspace{2mm}}r@{\hspace{3mm}}r@{\hspace{3mm}}r@{\hspace{3mm}}rr@{\hspace{3mm}}cl@{\hspace{2mm}}l}
\hline
GroupID & GalID & RA & Dec\ \ \ \ \ \  & \multicolumn{1}{c}{$K$} &
\multicolumn{1}{c}{$k_K$} &  \multicolumn{1}{c}{\ \ \ $v$} &
\multicolumn{1}{c}{\ \ \ \ \ \ err($v$)} & $v$ & SDSS\_ID & 2MASS\_ID \\
\cline{3-4}
\cline{7-8}
& & \multicolumn{2}{c}{(J2000)} & & & \multicolumn{2}{c}{[$\rm km \,s^{-1}$]}
&  source & &\\
\hline

 30 &  1 & 09:05:21.32 &  18:18:47.18 &  9.01 & --0.03 &  4189 &   21 & 1 & 587741708330074118 & 09052131+1818472 \\
 30 &  2 & 09:04:39.02 &  18:27:51.92 & 11.08 & --0.02 &  3423 &    5 & 1 & 587741708330008623 & 09043901+1827521 \\
 30 &  3 & 09:05:32.39 &  18:15:44.45 & 11.45 & --0.04 &  4250 &    2 & 1 & 588023045866193011 & 09053239+1815445 \\
 30 &  4 & 09:04:39.32 &  18:15:26.33 & 11.94 & --0.03 &  3635 &    2 & 2 & 587741708329943218 & 09043929+1815261 \\
 30 &  5 & 09:05:18.35 &  18:26:32.15 & 11.95 & --0.03 &  4188 &    2 & 2 & 587741708330074178 & 09051836+1826322 \\
 31 &  1 & 09:16:41.85 &  30:54:55.33 &  9.82 & --0.04 &  6949 &    2 & 1 & 588017979413299278 & 09164185+3054551 \\
 31 &  2 & 09:16:15.64 &  30:49:26.32 & 11.60 & --0.05 &  6812 &    2 & 1 & 588017979413233753 & 09161561+3049261 \\
 31 &  3 & 09:16:37.66 &  30:54:19.33 & 11.65 & --0.05 &  6958 &    2 & 1 & 588017979413299279 & 09163765+3054191 \\
 31 &  4 & 09:16:46.19 &  30:54:39.25 & 12.32 & --0.02 &  6974 &    2 & 2 & 588017979413299280 & 09164620+3054391 \\
 32 &  1 & 09:27:52.82 &  29:59:08.64 & 10.24 & --0.04 &  7993 &    2 & 1 & 587739158725001301 & 09275281+2959085 \\
 32 &  2 & 09:28:00.92 &  30:02:13.33 & 11.81 & --0.06 &  7993 &    2 & 2 & 587739158725001338 & 09280090+3002135 \\
 32 &  3 & 09:28:04.06 &  29:59:29.31 & 12.30 & --0.05 &  8296 &    1 & 2 & 587739115781947490 & 09280405+2959294 \\
 32 &  4 & 09:27:57.72 &  30:00:39.63 & 12.97 & --0.01 &  7424 &    0 & 2 & 587739158725001442 & 09275774+3000395 \\
 33 &  1 & 09:33:46.09 &  10:09:09.09 &  8.86 & --0.02 &  3231 &    2 & 3 & 587735344799350868 & 09334609+1009093 \\
 33 &  2 & 09:34:02.78 &  10:06:31.49 &  9.99 & --0.02 &  3144 &    1 & 1 & 587735344799350940 & 09340276+1006315 \\
 33 &  3 & 09:34:47.54 &  10:17:01.42 & 10.15 & --0.01 &  2432 &    8 & 1 & 587735344799481904 & 09344754+1017014 \\
 33 &  4 & 09:34:00.18 &  10:01:46.51 & 11.37 & --0.02 &  3136 &    5 & 1 & 587734949661769909 & 09340019+1001465 \\
 34 &  1 & 09:38:53.47 & --04:50:55.32 &  9.73 & --0.05 &  6627 &   27 & 1 & 000000000000000000 & 09385347--0450553 \\
 34 &  2 & 09:38:53.62 & --04:51:36.58 & 10.32 & --0.03 &  6405 &    6 & 1 & 000000000000000000 & 09385360--0451365 \\
 34 &  3 & 09:38:54.96 & --04:51:57.31 & 10.71 & --0.05 &  6842 &   27 & 1 & 000000000000000000 & 09385496--0451573 \\
 34 &  4 & 09:38:55.76 & --04:50:13.45 & 10.92 & --0.04 &  6837 &   45 & 1 & 000000000000000000 & 09385576--0450134 \\
 34 &  5 & 09:38:55.29 & --04:51:28.47 & 12.17 & --0.04 &  6632 &   23 & 2 & 000000000000000000 & 09385529--0451284 \\
 35 &  1 & 10:00:14.14 & --19:38:11.21 &  8.18 & --0.03 &  3964 &   10 & 1 & 000000000000000000 & 10001412--1938113 \\
 35 &  2 & 09:59:29.19 & --19:29:32.18 & 10.12 & --0.03 &  4004 &   17 & 1 & 000000000000000000 & 09592917--1929323 \\
 35 &  3 & 10:00:10.29 & --19:37:18.53 & 10.40 & --0.03 &  4034 &   45 & 1 & 000000000000000000 & 10001030--1937183 \\
 35 &  4 & 10:00:33.11 & --19:39:43.13 & 10.63 & --0.03 &  4228 &   18 & 1 & 000000000000000000 & 10003309--1939433 \\
 36 &  1 & 10:24:59.32 &  28:01:26.00 & 10.14 & --0.05 &  6417 &   27 & 1 & 587741566050304065 & 10245932+2801259 \\
 36 &  2 & 10:25:07.28 &  28:05:41.04 & 11.60 & --0.04 &  6130 &    2 & 1 & 587741566050369539 & 10250725+2805409 \\
 36 &  3 & 10:25:13.58 &  28:00:58.31 & 12.26 & --0.04 &  6520 &    2 & 2 & 587741534390517974 & 10251359+2800584 \\
 36 &  4 & 10:25:18.72 &  28:03:20.38 & 12.43 & --0.04 &  6304 &    0 & 2 & 587741566050369557 & 10251872+2803204 \\
 37 &  1 & 10:37:29.09 & --26:19:01.22 &  9.29 & --0.02 &  3358 &   45 & 1 & 000000000000000000 & 10372908--2619014 \\
 37 &  2 & 10:36:56.64 & --26:11:39.32 & 10.88 & --0.02 &  3251 &   45 & 1 & 000000000000000000 & 10365664--2611391 \\
 37 &  3 & 10:37:37.87 & --26:16:38.32 & 10.97 & --0.02 &  3815 &   45 & 1 & 000000000000000000 & 10373785--2616384 \\
 37 &  4 & 10:37:00.43 & --26:19:05.25 & 11.78 & --0.02 &  3507 &   45 & 2 & 000000000000000000 & 10370041--2619051 \\
 38 &  1 & 10:39:34.06 & --23:55:21.69 &  9.97 & --0.03 &  3813 &   45 & 1 & 000000000000000000 & 10393405--2355217 \\
 38 &  2 & 10:40:08.94 & --23:49:13.31 & 10.69 & --0.03 &  3619 &   45 & 1 & 000000000000000000 & 10400892--2349132 \\
 38 &  3 & 10:39:26.11 & --23:45:17.50 & 11.56 & --0.01 &  3955 &   45 & 1 & 000000000000000000 & 10392612--2345176 \\
 38 &  4 & 10:39:40.26 & --23:53:15.73 & 12.83 & --0.02 &  3703 &    0 & 2 & 000000000000000000 & 10394025--2353157 \\
 39 &  1 & 10:51:43.67 &  51:01:19.76 & 10.08 & --0.05 &  7491 &    3 & 1 & 588013383811923987 & 10514368+5101195 \\
 39 &  2 & 10:51:44.49 &  51:01:30.32 & 10.44 & --0.04 &  7137 &    0 & 1 & 588013383811923988 & 10514450+5101303 \\
 39 &  3 & 10:51:53.71 &  51:00:23.47 & 12.01 & --0.05 &  8277 &    3 & 2 & 588013383811924105 & 10515373+5100234 \\
 39 &  4 & 10:51:23.23 &  50:55:51.54 & 12.24 & --0.03 &  7281 &    2 & 2 & 588013383811858538 & 10512323+5055516 \\
 39 &  5 & 10:52:00.77 &  50:50:34.70 & 12.55 & --0.05 &  7805 &    0 & 2 & 588013383811924012 & 10520073+5050344 \\
 40 &  1 & 11:09:44.43 &  21:45:31.88 & 10.01 & --0.06 &  9542 &    2 & 1 & 587742061602734111 & 11094441+2145316 \\
 40 &  2 & 11:09:50.31 &  21:48:36.88 & 11.40 & --0.06 &  9432 &    2 & 1 & 587742014350557248 & 11095029+2148366 \\
 40 &  3 & 11:09:41.19 &  21:44:25.54 & 11.62 & --0.05 &  9659 &    2 & 1 & 587742061602734109 & 11094118+2144256 \\
 40 &  4 & 11:09:42.82 &  21:44:07.82 & 12.37 & --0.07 &  9562 &    0 & 0 & 587742061602734112 & 11094283+2144076 \\
 41 &  1 & 11:16:54.66 &  18:03:06.51 &  7.11 & --0.01 &   959 &   20 & 1 & 587742865818779699 & 11165465+1803065 \\
 41 &  2 & 11:16:58.95 &  18:08:54.95 &  8.17 & --0.01 &  1253 &    5 & 1 & 587742571074355242 & 11165896+1808547 \\
 41 &  3 & 11:15:26.95 &  18:06:37.29 &  9.48 & --0.01 &   832 &    9 & 1 & 587742571074158634 & 11152695+1806373 \\
 41 &  4 & 11:16:46.61 &  18:01:01.74 &  9.78 & --0.01 &   642 &    1 & 1 & 587742865818779669 & 11164662+1801017 \\
 42 &  1 & 11:22:26.35 &  24:17:56.93 & 10.07 & --0.05 &  7539 &    2 & 1 & 587741829122555933 & 11222635+2417567 \\
 42 &  2 & 11:22:13.28 &  24:19:02.12 & 11.24 & --0.05 &  7699 &   23 & 1 & 587742190436352010 & 11221325+2419017 \\
 42 &  3 & 11:22:14.22 &  24:18:00.88 & 11.47 & --0.06 &  8200 &    3 & 1 & 587741829122556024 & 11221420+2418007 \\
 42 &  4 & 11:22:30.55 &  24:17:59.96 & 11.99 & --0.06 &  7527 &    2 & 2 & 587741829122555934 & 11223052+2417597 \\
 43 &  1 & 11:35:00.00 &  51:13:04.81 & 10.41 & --0.05 &  8085 &    2 & 1 & 588013382204194882 & 11345999+5113048 \\
 43 &  2 & 11:35:13.62 &  51:09:42.87 & 11.75 & --0.05 &  7949 &    2 & 1 & 588013382204194936 & 11351359+5109426 \\
 43 &  3 & 11:35:15.64 &  51:12:13.68 & 12.22 & --0.05 &  7393 &    2 & 2 & 588013382204194957 & 11351563+5112136 \\
 43 &  4 & 11:35:08.66 &  51:14:24.72 & 13.14 & --0.04 &  8071 &    0 & 2 & 588013382204194911 & 11350862+5114247 \\
 44 &  1 & 11:42:11.09 &  10:16:39.88 &  9.63 & --0.02 &  6267 &    6 & 1 & 587734893288095762 & 11421107+1016398 \\
 44 &  2 & 11:42:23.72 &  10:15:50.88 &  9.92 & --0.04 &  6475 &    2 & 1 & 587734893288095841 & 11422374+1015508 \\
 44 &  3 & 11:41:52.95 &  10:18:15.83 & 10.30 & --0.04 &  6074 &    2 & 1 & 587732772126261318 & 11415296+1018160 \\
 44 &  4 & 11:42:05.86 &  10:21:04.51 & 10.70 & --0.04 &  6217 &    2 & 1 & 587732772126261358 & 11420585+1021047 \\
 44 &  5 & 11:42:04.90 &  10:23:02.98 & 11.81 & --0.02 &  6093 &    4 & 2 & 587732772126261508 & 11420490+1023027 \\
\end{tabular}
\end{table*}
\begin{table*}
\contcaption{--- Table of galaxy positions} 
\tabcolsep 1pt
\centering
\begin{tabular}{r@{\hspace{3mm}}cc@{\hspace{2mm}}r@{\hspace{3mm}}r@{\hspace{3mm}}r@{\hspace{3mm}}rr@{\hspace{3mm}}cl@{\hspace{2mm}}l}
\hline
GroupID & GalID & RA & Dec\ \ \ \ \ \  & \multicolumn{1}{c}{$K$} &
\multicolumn{1}{c}{$k_K$} &  \multicolumn{1}{c}{\ \ \ $v$} &
\multicolumn{1}{c}{\ \ \ \ \ \ err($v$)} & $v$ & SDSS\_ID & 2MASS\_ID \\
\cline{3-4}
\cline{7-8}
& & \multicolumn{2}{c}{(J2000)} & & & \multicolumn{2}{c}{[$\rm km \,s^{-1}$]}
&  source & &\\
\hline

 45 &  1 & 11:42:32.86 &  26:29:19.89 & 10.06 & --0.05 &  9090 &    3 & 1 & 587741600954712088 & 11423285+2629198 \\
 45 &  2 & 11:42:52.24 &  26:32:26.21 & 10.93 & --0.05 &  8543 &    2 & 1 & 587741600954712170 & 11425225+2632260 \\
 45 &  3 & 11:43:11.85 &  26:33:32.17 & 11.05 & --0.06 &  9017 &    3 & 1 & 587741600954777644 & 11431185+2633320 \\
 45 &  4 & 11:43:10.74 &  26:34:28.25 & 11.67 & --0.06 &  9320 &    3 & 1 & 587741708346327131 & 11431073+2634280 \\
 46 &  1 & 11:44:13.99 &  33:30:52.21 & 10.16 & --0.06 &  9517 &    3 & 1 & 587739405705150546 & 11441398+3330521 \\
 46 &  2 & 11:44:03.35 &  33:32:06.14 & 11.73 & --0.05 &  9325 &    3 & 1 & 587739405705150521 & 11440335+3332060 \\
 46 &  3 & 11:44:09.32 &  33:28:56.16 & 12.33 & --0.05 &  9435 &    3 & 2 & 587739405705150653 & 11440934+3328561 \\
 46 &  4 & 11:44:04.32 &  33:32:33.97 & 12.65 & --0.04 &  9230 &    0 & 2 & 587739405705150522 & 11440431+3332340 \\
 47 &  1 & 11:57:56.15 &  55:27:12.79 &  7.42 & --0.01 &  1040 &   13 & 1 & 587733081348571206 & 11575616+5527128 \\
 47 &  2 & 11:56:28.15 &  55:07:31.15 &  8.92 & --0.01 &  1113 &    3 & 1 & 587731869633871916 & 11562816+5507313 \\
 47 &  3 & 11:55:45.11 &  55:19:14.29 &  9.63 & --0.01 &   855 &    4 & 1 & 587733081348440213 & 11554511+5519144 \\
 47 &  4 & 11:57:35.58 &  55:27:31.64 &  9.63 & --0.01 &   695 &   40 & 1 & 587733081348571153 & 11573559+5527318 \\
 48 &  1 & 12:19:23.26 &  05:49:28.96 &  7.40 & --0.02 &  2238 &    7 & 1 & 588010360162091053 & 12192326+0549289 \\
 48 &  2 & 12:19:22.24 &  06:05:55.53 &  8.64 & --0.01 &  1776 &    2 & 1 & 588010360698961934 & 12192224+0605556 \\
 48 &  3 & 12:19:35.78 &  05:50:48.35 &  9.98 & --0.02 &  2507 &    1 & 1 & 588010880371851309 & 12193577+0550484 \\
 48 &  4 & 12:19:49.19 &  06:00:53.93 & 10.02 & --0.02 &  2073 &   19 & 1 & 588010880371851341 & 12194918+0600541 \\
 49 &  1 & 12:24:28.23 &  07:19:03.07 &  6.79 & --0.01 &  1242 &    6 & 1 & 588017729225752606 & 12242822+0719030 \\
 49 &  2 & 12:23:39.00 &  07:03:14.33 &  9.04 & --0.01 &   751 &   24 & 1 & 588017728688750698 & 12233902+0703141 \\
 49 &  3 & 12:23:38.69 &  06:57:14.32 &  9.05 & --0.01 &   993 &    1 & 1 & 588017728688750678 & 12233869+0657141 \\
 49 &  4 & 12:24:54.92 &  07:26:40.24 &  9.41 &  0.00 &   771 &    3 & 1 & 588017729225818214 & 12245493+0726404 \\
 49 &  5 & 12:25:42.66 &  07:13:00.33 &  9.78 & --0.01 &   999 &    5 & 1 & 588017724937994262 & 12254263+0713001 \\
 50 &  1 & 12:27:53.57 &  12:17:35.62 &  8.99 & --0.01 &   723 &   18 & 1 & 588017703470366754 & 12275357+1217354 \\
 50 &  2 & 12:27:27.36 &  12:17:25.20 & 10.56 & --0.01 &   934 &   10 & 1 & 588017703470301335 & 12272735+1217252 \\
 50 &  3 & 12:27:41.22 &  12:18:57.41 & 10.95 & --0.01 &  1113 &    2 & 1 & 588017703470301247 & 12274122+1218574 \\
 50 &  4 & 12:28:08.59 &  12:05:35.74 & 11.25 & --0.02 &  1921 &    3 & 1 & 588017566027546723 & 12280859+1205356 \\
 51 &  1 & 12:43:40.00 &  11:33:09.40 &  5.81 & --0.01 &  1116 &    6 & 1 & 588017702398328841 & 12434000+1133093 \\
 51 &  2 & 12:42:02.32 &  11:38:48.95 &  6.86 &  0.00 &   409 &    6 & 1 & 588017569774370819 & 12420232+1138489 \\
 51 &  3 & 12:43:32.55 &  11:34:56.88 &  8.19 & --0.01 &  1394 &    5 & 1 & 588017702398263345 & 12433254+1134568 \\
 51 &  4 & 12:44:31.98 &  11:11:25.89 &  8.23 & --0.01 &  1083 &    4 & 1 & 588017569237762053 & 12443197+1111259 \\
 51 &  5 & 12:42:47.43 &  11:26:32.89 &  8.24 & --0.01 &  1164 &   10 & 1 & 588017702398197833 & 12424743+1126328 \\
 52 &  1 & 13:01:53.75 &  27:37:27.87 & 10.03 & --0.06 &  7856 &    2 & 1 & 587741602573058054 & 13015375+2737277 \\
 52 &  2 & 13:02:07.91 &  27:38:54.03 & 12.02 & --0.05 &  6898 &    2 & 2 & 587741602573058113 & 13020791+2738539 \\
 52 &  3 & 13:01:48.39 &  27:36:14.48 & 12.88 & --0.06 &  8229 &    0 & 2 & 587741602572992587 & 13014841+2736147 \\
 52 &  4 & 13:02:01.05 &  27:39:10.80 & 12.90 & --0.05 &  7089 &    0 & 2 & 587741602573058178 & 13020106+2739109 \\
 53 &  1 & 13:06:26.13 & --40:24:52.56 &  8.83 &  0.00 &  4500 &   14 & 1 & 000000000000000000 & 13062614--4024521 \\
 53 &  2 & 13:06:32.24 & --40:19:07.21 & 11.12 & --0.03 &  4772 &   45 & 1 & 000000000000000000 & 13063226--4019071 \\
 53 &  3 & 13:07:16.26 & --40:21:02.55 & 11.15 & --0.04 &  4778 &   45 & 1 & 000000000000000000 & 13071625--4021026 \\
 53 &  4 & 13:07:12.09 & --40:24:27.36 & 11.31 & --0.03 &  4764 &   59 & 1 & 000000000000000000 & 13071206--4024276 \\
 54 &  1 & 13:08:35.62 &  33:58:33.03 & 10.50 & --0.06 & 10151 &    3 & 1 & 587739406786560053 & 13083562+3358330 \\
 54 &  2 & 13:07:54.82 &  34:05:13.34 & 11.17 & --0.06 & 10087 &    3 & 1 & 587739406786494548 & 13075483+3405131 \\
 54 &  3 & 13:08:04.17 &  34:00:27.19 & 11.26 & --0.06 &  9943 &    3 & 1 & 587739406786494592 & 13080418+3400272 \\
 54 &  4 & 13:08:27.01 &  34:04:13.37 & 12.20 & --0.03 & 10339 &    0 & 2 & 587739406786560033 & 13082700+3404129 \\
 55 &  1 & 13:21:50.81 & --17:20:11.35 &  9.95 & --0.05 &  7064 &   39 & 1 & 000000000000000000 & 13215080--1720114 \\
 55 &  2 & 13:21:54.15 & --17:21:36.41 & 11.89 & --0.05 &  6908 &   45 & 2 & 000000000000000000 & 13215415--1721364 \\
 55 &  3 & 13:22:07.60 & --17:24:07.86 & 12.07 & --0.02 &  7370 &   45 & 2 & 000000000000000000 & 13220758--1724080 \\
 55 &  4 & 13:22:15.68 & --17:21:38.06 & 12.69 & --0.03 &  6888 &    0 & 2 & 000000000000000000 & 13221568--1721380 \\
 55 &  5 & 13:22:06.79 & --17:22:37.51 & 12.72 & --0.05 &  6767 &    0 & 2 & 000000000000000000 & 13220680--1722374 \\
 56 &  1 & 13:24:10.02 &  13:58:35.32 &  9.37 & --0.05 &  6885 &    2 & 1 & 587738568710357008 & 13241000+1358351 \\
 56 &  2 & 13:24:28.90 &  14:05:33.28 &  9.98 & --0.05 &  7320 &   27 & 1 & 587736808298774638 & 13242889+1405332 \\
 56 &  3 & 13:24:24.17 &  13:56:15.19 & 11.51 & --0.04 &  6826 &    2 & 1 & 587738568710357048 & 13242415+1356152 \\
 56 &  4 & 13:24:52.47 &  14:04:38.15 & 12.25 & --0.03 &  7288 &    5 & 2 & 587736808298840234 & 13245247+1404382 \\
 57 &  1 & 13:52:10.08 &  02:19:30.39 &  9.80 & --0.05 &  7109 &   31 & 1 & 587726032797696030 & 13521008+0219305 \\
 57 &  2 & 13:52:22.83 &  02:20:44.83 & 11.64 & --0.05 &  6997 &    2 & 1 & 587726032797696100 & 13522284+0220448 \\
 57 &  3 & 13:52:14.56 &  02:21:34.31 & 12.05 & --0.04 &  7107 &    3 & 2 & 587726032797696243 & 13521455+0221345 \\
 57 &  4 & 13:52:11.55 &  02:18:54.71 & 12.51 & --0.04 &  7082 &    0 & 2 & 587726032797696031 & 13521155+0218545 \\
 58 &  1 & 13:52:53.32 & --28:29:21.54 &  8.54 & --0.03 &  4713 &   45 & 1 & 000000000000000000 & 13525331--2829213 \\
 58 &  2 & 13:52:59.18 & --28:28:14.39 & 10.60 & --0.03 &  4783 &   45 & 1 & 000000000000000000 & 13525917--2828141 \\
 58 &  3 & 13:53:15.02 & --28:25:39.40 & 10.74 & --0.03 &  4427 &   35 & 1 & 000000000000000000 & 13531502--2825391 \\
 58 &  4 & 13:53:00.18 & --28:27:06.35 & 11.24 & --0.03 &  5047 &   25 & 1 & 000000000000000000 & 13530016--2827061 \\
 58 &  5 & 13:52:48.39 & --28:29:58.18 & 11.45 & --0.03 &  4531 &   45 & 1 & 000000000000000000 & 13524838--2829584 \\
 59 &  1 & 14:00:37.20 & --02:51:28.05 &  9.82 & --0.04 &  7446 &    2 & 1 & 587729776905486440 & 14003719--0251281 \\
 59 &  2 & 14:00:30.06 & --02:48:38.40 & 11.25 & --0.05 &  7856 &    3 & 1 & 587729776905486416 & 14003005--0248387 \\
 59 &  3 & 14:00:36.60 & --02:54:32.37 & 11.49 & --0.04 &  7204 &   16 & 1 & 587729772610125865 & 14003658--0254321 \\
 59 &  4 & 14:00:37.98 & --02:54:22.37 & 11.95 & --0.05 &  7581 &    2 & 2 & 587729772610125863 & 14003799--0254221 \\
 59 &  5 & 14:00:37.52 & --02:52:23.13 & 12.12 & --0.04 &  7064 &    2 & 2 & 587729776905486441 & 14003752--0252231 \\
 59 &  6 & 14:00:40.01 & --02:49:55.26 & 12.51 & --0.05 &  7165 &    0 & 2 & 587729776905486570 & 14003999--0249551 \\
\end{tabular}
\end{table*}
\begin{table*}
\contcaption{--- Table of galaxy positions} 
\tabcolsep 1pt
\centering
\begin{tabular}{r@{\hspace{3mm}}cc@{\hspace{2mm}}r@{\hspace{3mm}}r@{\hspace{3mm}}r@{\hspace{3mm}}rr@{\hspace{3mm}}cl@{\hspace{2mm}}l}
\hline
GroupID & GalID & RA & Dec\ \ \ \ \ \  & \multicolumn{1}{c}{$K$} &
\multicolumn{1}{c}{$k_K$} &  \multicolumn{1}{c}{\ \ \ $v$} &
\multicolumn{1}{c}{\ \ \ \ \ \ err($v$)} & $v$ & SDSS\_ID & 2MASS\_ID \\
\cline{3-4}
\cline{7-8}
& & \multicolumn{2}{c}{(J2000)} & & & \multicolumn{2}{c}{[$\rm km \,s^{-1}$]}
&  source & &\\
\hline

 60 &  1 & 14:19:17.62 &  35:08:16.58 & 10.34 & --0.05 &  8529 &   26 & 1 & 587736940371247137 & 14191759+3508168 \\
 60 &  2 & 14:19:21.31 &  35:05:31.69 & 11.86 & --0.05 &  8459 &    2 & 2 & 587736940371247186 & 14192133+3505318 \\
 60 &  3 & 14:19:07.73 &  35:10:58.80 & 12.86 & --0.03 &  8580 &    0 & 2 & 588017977829818403 & 14190772+3510588 \\
 60 &  4 & 14:19:24.77 &  35:07:42.73 & 13.06 & --0.01 &  8538 &    0 & 2 & 587736940371247301 & 14192476+3507427 \\
 61 &  1 & 14:27:34.68 &  11:19:50.06 & 10.45 & --0.06 &  7993 &    2 & 1 & 588017704556888172 & 14273466+1119498 \\
 61 &  2 & 14:27:14.54 &  11:20:21.00 & 11.12 & --0.05 &  8324 &    2 & 1 & 588017704556888080 & 14271453+1120212 \\
 61 &  3 & 14:27:22.36 &  11:16:17.32 & 12.78 & --0.06 &  8162 &    0 & 2 & 587736478664556756 & 14272235+1116172 \\
 61 &  4 & 14:27:33.02 &  11:22:12.03 & 12.81 & --0.06 &  7424 &    0 & 2 & 588017704556888164 & 14273302+1122118 \\
 62 &  1 & 14:28:16.36 &  25:50:55.61 &  9.33 & --0.03 &  4497 &   24 & 1 & 587739720299577472 & 14281635+2550557 \\
 62 &  2 & 14:28:07.23 &  25:52:07.60 & 10.18 & --0.03 &  4371 &    2 & 1 & 587739707410415672 & 14280724+2552077 \\
 62 &  3 & 14:28:13.47 &  25:56:50.76 & 10.72 & --0.02 &  4021 &    2 & 1 & 587739707410415654 & 14281346+2556507 \\
 62 &  4 & 14:27:50.79 &  25:50:17.11 & 11.27 & --0.03 &  4405 &    3 & 1 & 587739720299577369 & 14275077+2550172 \\
 63 &  1 & 14:58:46.22 & --19:16:00.69 &  8.48 & --0.02 &  3350 &   11 & 1 & 000000000000000000 & 14584622--1916006 \\
 63 &  2 & 14:57:21.72 & --19:12:49.32 &  9.67 & --0.02 &  3423 &   45 & 1 & 000000000000000000 & 14572172--1912491 \\
 63 &  3 & 14:58:55.06 & --19:14:20.70 & 10.44 & --0.02 &  3291 &   45 & 1 & 000000000000000000 & 14585504--1914206 \\
 63 &  4 & 14:58:50.11 & --19:03:35.76 & 10.82 & --0.03 &  3521 &   24 & 1 & 000000000000000000 & 14585010--1903356 \\
 64 &  1 & 15:35:57.01 &  43:29:35.42 &  9.62 & --0.04 &  5659 &   47 & 1 & 587733411526606950 & 15355700+4329351 \\
 64 &  2 & 15:36:42.17 &  43:32:21.68 & 10.31 & --0.04 &  5566 &    2 & 1 & 587733411526672449 & 15364216+4332217 \\
 64 &  3 & 15:36:27.08 &  43:31:07.59 & 10.53 & --0.03 &  6038 &    2 & 1 & 587733411526672426 & 15362709+4331075 \\
 64 &  4 & 15:36:02.79 &  43:30:12.37 & 11.45 & --0.04 &  5597 &    2 & 1 & 587733411526606951 & 15360279+4330122 \\
 64 &  5 & 15:36:35.19 &  43:25:32.24 & 12.30 & --0.05 &  5768 &    0 & 2 & 587733411526672511 & 15363521+4325326 \\
 65 &  1 & 16:12:40.14 &  33:02:05.55 & 10.51 & --0.06 &  9601 &    3 & 1 & 587736783607365679 & 16124014+3302054 \\
 65 &  2 & 16:12:41.33 &  33:02:15.64 & 11.19 & --0.05 &  9446 &    0 & 1 & 587736783607365680 & 16124129+3302158 \\
 65 &  3 & 16:13:00.77 &  33:02:07.08 & 12.99 & --0.04 &  9368 &    0 & 2 & 587736783607365887 & 16130077+3302069 \\
 65 &  4 & 16:12:56.70 &  33:03:05.93 & 13.26 & --0.06 &  9410 &    0 & 2 & 587736783607365861 & 16125671+3303059 \\
 66 &  1 & 16:37:53.93 &  36:04:23.05 &  9.91 & --0.06 &  9399 &    3 & 1 & 587733608555216981 & 16375395+3604233 \\
 66 &  2 & 16:37:59.89 &  35:59:43.85 & 11.18 & --0.06 &  9618 &    2 & 1 & 587733603732422719 & 16375990+3559438 \\
 66 &  3 & 16:37:47.18 &  36:06:54.12 & 11.68 & --0.05 &  9856 &    3 & 1 & 587733608555216924 & 16374718+3606543 \\
 66 &  4 & 16:37:48.44 &  36:03:39.14 & 12.71 & --0.09 &  9609 &    0 & 2 & 587733603732357553 & 16374842+3603393 \\
 67 &  1 & 19:15:34.12 & --54:37:36.11 &  9.48 & --0.04 &  5546 &   45 & 1 & 000000000000000000 & 19153413--5437359 \\
 67 &  2 & 19:14:22.02 & --54:33:56.18 & 10.31 & --0.04 &  5134 &   30 & 1 & 000000000000000000 & 19142202--5433561 \\
 67 &  3 & 19:14:12.81 & --54:34:26.91 & 10.97 & --0.04 &  5252 &   22 & 1 & 000000000000000000 & 19141282--5434270 \\
 67 &  4 & 19:14:01.32 & --54:35:15.60 & 12.25 & --0.04 &  5549 &   45 & 2 & 000000000000000000 & 19140128--5435155 \\
 68 &  1 & 19:52:08.76 & --30:49:30.98 &  9.47 & --0.04 &  6012 &   11 & 1 & 000000000000000000 & 19520876--3049311 \\
 68 &  2 & 19:51:59.05 & --30:48:57.17 & 10.07 & --0.03 &  5805 &   45 & 1 & 000000000000000000 & 19515906--3048572 \\
 68 &  3 & 19:51:57.33 & --30:51:23.99 & 10.64 & --0.03 &  5233 &   45 & 1 & 000000000000000000 & 19515735--3051241 \\
 68 &  4 & 19:51:51.94 & --30:48:30.38 & 11.07 & --0.04 &  5976 &   45 & 1 & 000000000000000000 & 19515192--3048301 \\
 69 &  1 & 20:00:58.42 & --47:04:12.85 & 10.02 & --0.04 &  6355 &   18 & 1 & 000000000000000000 & 20005841--4704130 \\
 69 &  2 & 20:00:56.84 & --47:05:03.87 & 10.40 & --0.05 &  7070 &   54 & 1 & 000000000000000000 & 20005684--4705040 \\
 69 &  3 & 20:01:05.26 & --47:03:33.97 & 11.80 & --0.06 &  6792 &   45 & 2 & 000000000000000000 & 20010526--4703340 \\
 69 &  4 & 20:00:53.58 & --47:05:42.51 & 11.96 & --0.03 &  6815 &   44 & 2 & 000000000000000000 & 20005354--4705424 \\
 70 &  1 & 20:03:27.03 & --55:56:49.81 &  8.01 & --0.03 &  4433 &   26 & 1 & 000000000000000000 & 20032702--5556498 \\
 70 &  2 & 20:02:47.25 & --56:05:25.10 &  8.82 & --0.03 &  4391 &   76 & 1 & 000000000000000000 & 20024726--5605250 \\
 70 &  3 & 20:03:01.06 & --55:56:52.88 &  9.97 & --0.03 &  4388 &   33 & 1 & 000000000000000000 & 20030107--5556528 \\
 70 &  4 & 20:03:41.79 & --55:54:55.68 & 10.95 & --0.03 &  4590 &   19 & 1 & 000000000000000000 & 20034177--5554557 \\
 71 &  1 & 20:18:19.15 & --70:51:31.77 &  7.77 & --0.03 &  4009 &   16 & 1 & 000000000000000000 & 20181914--7051317 \\
 71 &  2 & 20:16:56.50 & --70:46:05.67 &  8.37 & --0.03 &  4554 &   30 & 1 & 000000000000000000 & 20165648--7046057 \\
 71 &  3 & 20:19:29.62 & --70:51:35.56 &  9.29 & --0.02 &  3928 &   22 & 1 & 000000000000000000 & 20192966--7051356 \\
 71 &  4 & 20:15:23.88 & --70:32:15.87 & 10.50 & --0.03 &  3479 &   30 & 1 & 000000000000000000 & 20152387--7032159 \\
 72 &  1 & 20:43:45.71 & --26:33:00.99 & 10.45 & --0.07 & 12221 &   92 & 1 & 000000000000000000 & 20434570--2633011 \\
 72 &  2 & 20:43:49.67 & --26:35:32.19 & 11.76 & --0.06 & 12131 &   45 & 2 & 000000000000000000 & 20434965--2635321 \\
 72 &  3 & 20:43:33.67 & --26:35:10.11 & 12.15 & --0.06 & 12589 &   45 & 2 & 000000000000000000 & 20433370--2635101 \\
 72 &  4 & 20:43:38.11 & --26:34:17.42 & 12.66 & --0.07 & 12631 &    0 & 2 & 000000000000000000 & 20433810--2634171 \\
 73 &  1 & 20:47:19.06 &  00:19:14.94 &  8.89 & --0.02 &  4200 &    2 & 1 & 587731173842026987 & 20471908+0019150 \\
 73 &  2 & 20:47:24.29 &  00:18:02.99 &  9.65 & --0.02 &  3790 &    2 & 1 & 587731173842027210 & 20472428+0018030 \\
 73 &  3 & 20:47:34.09 &  00:24:41.89 &  9.97 & --0.03 &  3768 &   20 & 1 & 587731173842092228 & 20473408+0024420 \\
 73 &  4 & 20:47:07.24 &  00:25:48.66 & 10.65 & --0.02 &  3663 &    2 & 1 & 587730847960596651 & 20470724+0025486 \\
 73 &  5 & 20:47:10.50 &  00:21:47.90 & 11.13 & --0.03 &  3715 &    2 & 1 & 587731173842026895 & 20471048+0021479 \\
 73 &  6 & 20:47:20.35 &  00:29:02.36 & 11.35 & --0.03 &  4228 &    2 & 1 & 587730847960596912 & 20472035+0029020 \\
 74 &  1 & 20:52:35.46 & --05:42:40.06 &  9.85 & --0.04 &  5968 &    2 & 1 & 587727214416953538 & 20523547--0542399 \\
 74 &  2 & 20:52:29.69 & --05:44:45.97 & 10.22 & --0.04 &  6128 &    2 & 1 & 587727214416953446 & 20522971--0544459 \\
 74 &  3 & 20:52:26.02 & --05:46:19.86 & 12.06 & --0.05 &  5995 &    4 & 2 & 587727214416953420 & 20522602--0546198 \\
 74 &  4 & 20:52:12.74 & --05:47:53.63 & 12.25 & --0.02 &  6060 &    4 & 2 & 587726878873486063 & 20521275--0547538 \\
\end{tabular}
\end{table*}
\begin{table*}
\contcaption{--- Table of galaxy positions} 
\tabcolsep 1pt
\centering
\begin{tabular}{r@{\hspace{3mm}}cc@{\hspace{2mm}}r@{\hspace{3mm}}r@{\hspace{3mm}}r@{\hspace{3mm}}rr@{\hspace{3mm}}cl@{\hspace{2mm}}l}
\hline
GroupID & GalID & RA & Dec\ \ \ \ \ \  & \multicolumn{1}{c}{$K$} &
\multicolumn{1}{c}{$k_K$} &  \multicolumn{1}{c}{\ \ \ $v$} &
\multicolumn{1}{c}{\ \ \ \ \ \ err($v$)} & $v$ & SDSS\_ID & 2MASS\_ID \\
\cline{3-4}
\cline{7-8}
& & \multicolumn{2}{c}{(J2000)} & & & \multicolumn{2}{c}{[$\rm km \,s^{-1}$]}
&  source & &\\
\hline

 75 &  1 & 21:08:32.01 & --29:46:08.58 &  9.57 & --0.04 &  5914 &   45 & 1 & 000000000000000000 & 21083199--2946083 \\
 75 &  2 & 21:07:59.93 & --29:50:08.89 & 10.87 & --0.04 &  5954 &   31 & 1 & 000000000000000000 & 21075991--2950089 \\
 75 &  3 & 21:08:58.28 & --29:46:58.14 & 11.44 & --0.04 &  5855 &   31 & 1 & 000000000000000000 & 21085826--2946582 \\
 75 &  4 & 21:08:10.63 & --29:39:04.87 & 11.78 & --0.03 &  5985 &   31 & 2 & 000000000000000000 & 21081059--2939049 \\
 76 &  1 & 21:16:55.29 & --42:15:36.06 &  9.46 & --0.04 &  5325 &   19 & 1 & 000000000000000000 & 21165529--4215361 \\
 76 &  2 & 21:16:46.19 & --42:15:42.05 & 10.11 & --0.04 &  5151 &   45 & 1 & 000000000000000000 & 21164619--4215421 \\
 76 &  3 & 21:17:27.92 & --42:20:22.93 & 10.70 & --0.04 &  5348 &   33 & 1 & 000000000000000000 & 21172789--4220231 \\
 76 &  4 & 21:16:47.80 & --42:23:47.45 & 12.41 & --0.04 &  5409 &    0 & 2 & 000000000000000000 & 21164779--4223471 \\
 77 &  1 & 22:03:30.92 &  12:38:12.38 &  9.83 & --0.05 &  8041 &    7 & 1 & 587730774413213871 & 22033093+1238124 \\
 77 &  2 & 22:03:30.31 &  12:39:38.75 & 11.15 & --0.05 &  7895 &    2 & 1 & 587730774413213915 & 22033031+1239384 \\
 77 &  3 & 22:03:22.58 &  12:38:57.53 & 12.09 & --0.03 &  8184 &    0 & 2 & 587730774413214125 & 22032258+1238574 \\
 77 &  4 & 22:03:33.46 &  12:38:54.46 & 12.79 & --0.04 &  8897 &    0 & 2 & 587730774413213877 & 22033345+1238544 \\
 78 &  1 & 22:36:27.96 & --24:20:30.34 & 10.30 & --0.06 & 10215 &   37 & 1 & 000000000000000000 & 22362797--2420306 \\
 78 &  2 & 22:36:20.35 & --24:16:40.42 & 11.74 & --0.06 & 10392 &   45 & 1 & 000000000000000000 & 22362036--2416406 \\
 78 &  3 & 22:36:25.77 & --24:19:54.72 & 12.30 & --0.07 & 10414 &   45 & 2 & 000000000000000000 & 22362578--2419546 \\
 78 &  4 & 22:36:23.15 & --24:16:46.55 & 12.50 & --0.08 & 10235 &    0 & 2 & 000000000000000000 & 22362315--2416466 \\
 79 &  1 & 22:55:20.45 & --33:53:16.47 & 10.21 & --0.05 &  8507 &   45 & 1 & 000000000000000000 & 22552046--3353166 \\
 79 &  2 & 22:55:22.38 & --33:53:41.70 & 11.23 & --0.06 &  8734 &   38 & 1 & 000000000000000000 & 22552239--3353416 \\
 79 &  3 & 22:55:18.62 & --33:55:13.71 & 12.39 & --0.06 &  8787 &  150 & 2 & 000000000000000000 & 22551861--3355136 \\
 79 &  4 & 22:55:27.79 & --33:53:49.80 & 12.79 & --0.07 &  8938 &    0 & 2 & 000000000000000000 & 22552777--3353496 \\
 80 &  1 & 22:57:57.52 &  26:09:00.01 &  9.25 & --0.05 &  7375 &   21 & 1 & 758883828517699894 & 22575750+2609000 \\
 80 &  2 & 22:58:19.56 &  26:03:43.03 & 10.59 & --0.05 &  7662 &   23 & 1 & 758883828517765263 & 22581956+2603431 \\
 80 &  3 & 22:57:51.71 &  26:09:43.75 & 11.72 & --0.05 &  7587 &    0 & 1 & 758883828517699852 & 22575170+2609436 \\
 80 &  4 & 22:58:24.96 &  26:10:12.28 & 12.16 & --0.04 &  7233 &    0 & 2 & 758883828517765149 & 22582499+2610119 \\
 80 &  5 & 22:57:56.56 &  26:06:39.86 & 12.24 & --0.06 &  7654 &    0 & 2 & 758883828517765157 & 22575655+2606400 \\
 81 &  1 & 23:15:16.08 &  18:57:41.04 &  9.00 & --0.03 &  5072 &   20 & 1 & 758883880592736344 & 23151609+1857409 \\
 81 &  2 & 23:15:03.46 &  18:58:24.24 & 10.04 & --0.03 &  4772 &   10 & 1 & 758883828521304188 & 23150347+1858245 \\
 81 &  3 & 23:15:17.26 &  19:02:29.98 & 10.39 & --0.03 &  4736 &    3 & 1 & 758883880592670819 & 23151728+1902299 \\
 81 &  4 & 23:15:33.06 &  19:02:53.24 & 11.11 & --0.04 &  5173 &   34 & 1 & 758883880592736265 & 23153308+1902529 \\
 82 &  1 & 23:28:17.74 & --67:49:16.92 &  9.82 & --0.02 &  3903 &   40 & 1 & 000000000000000000 & 23281779--6749170 \\
 82 &  2 & 23:27:36.97 & --67:48:55.66 & 10.27 & --0.02 &  3905 &   45 & 1 & 000000000000000000 & 23273700--6748557 \\
 82 &  3 & 23:28:21.97 & --67:45:36.91 & 11.37 & --0.03 &  3882 &   45 & 1 & 000000000000000000 & 23282200--6745369 \\
 82 &  4 & 23:28:30.81 & --67:45:40.73 & 11.49 & --0.02 &  4290 &   45 & 1 & 000000000000000000 & 23283081--6745409 \\
 83 &  1 & 23:28:35.11 &  32:24:56.65 &  9.18 & --0.03 &  5131 &   22 & 1 & 000000000000000000 & 23283508+3224565 \\
 83 &  2 & 23:28:10.77 &  32:28:22.36 & 10.78 & --0.03 &  4544 &   36 & 1 & 000000000000000000 & 23281075+3228224 \\
 83 &  3 & 23:28:23.17 &  32:21:44.54 & 10.84 & --0.03 &  4999 &   29 & 1 & 000000000000000000 & 23282316+3221445 \\
 83 &  4 & 23:28:31.60 &  32:25:19.68 & 11.37 & --0.03 &  5277 &   26 & 1 & 000000000000000000 & 23283161+3225195 \\
 84 &  1 & 23:47:22.99 & --02:18:02.42 & 10.07 & --0.04 &  6932 &   22 & 1 & 000000000000000000 & 23472298--0218025 \\
 84 &  2 & 23:47:18.91 & --02:18:48.60 & 10.71 & --0.05 &  6327 &   30 & 1 & 000000000000000000 & 23471891--0218485 \\
 84 &  3 & 23:47:23.79 & --02:21:04.55 & 11.03 & --0.04 &  6003 &   26 & 1 & 000000000000000000 & 23472378--0221045 \\
 84 &  4 & 23:47:37.84 & --02:19:00.03 & 11.56 & --0.05 &  6817 &   45 & 1 & 000000000000000000 & 23473787--0218598 \\
 84 &  5 & 23:47:19.86 & --02:16:50.78 & 12.72 & --0.06 &  6664 &    0 & 0 & 000000000000000000 & 23471984--0216505 \\
 85 &  1 & 23:53:53.88 &  07:58:13.99 &  9.15 & --0.03 &  5373 &   22 & 1 & 587743797290008717 & 23535389+0758138 \\
 85 &  2 & 23:53:26.79 &  07:52:32.36 &  9.52 & --0.03 &  5106 &    7 & 1 & 587743797290008586 & 23532680+0752322 \\
 85 &  3 & 23:53:19.66 &  07:52:15.34 &  9.71 & --0.04 &  5218 &    7 & 1 & 587743797289943215 & 23531967+0752152 \\
 85 &  4 & 23:53:32.15 &  08:07:05.27 & 10.66 & --0.03 &  5154 &   58 & 1 & 587743960499159116 & 23533218+0807052 \\
 85 &  5 & 23:53:45.98 &  07:51:37.72 & 10.77 & --0.04 &  5661 &   29 & 1 & 587743959962288248 & 23534595+0751377 \\
\hline
\end{tabular}
\parbox{15cm}{
\small
{\bf Notes.} Group ID, Galaxy ID, RA: Right Ascension, Dec: Declination, 
$K_{\rm b}$: Galactic Extinction-corrected $K$-band apparent magnitude,
k-corr: $k$-correction in the $K$-band 
computed from \cite{CMZ10} as a function of redshift and colour $H$--$K$, 
$v_r$: radial velocity,
err$_{v_r}$: error in radial velocity,
$v_r$\_source: catalogue from which the $v_r$ and err$_{v_r}$ were extracted,
SDSS\_ID: Object ID in the SDSS-DR7 database,
2MASS\_ID: ID in the 2MASS database\\
{\bf References for redshift source:}\\
1= main 2MRS \citep{Huchra+12}\\
2= extra 2MRS \citep{Huchra+12}\\
3= 2M++ redshift catalogue \citep{2M++_11}\\
0= NED\\ 
}
\end{table*}

\end{document}